\documentclass{article}

\input{setup.sty}

\usepackage[fontsize=10.5pt]{scrextend}
\selectfont
\title{\boldmath Wormholes and Factorization in Exact Effective Theory}

\author{Sergio Hern\'andez-Cuenca \orcidlink{0000-0002-0765-6905}}

\affiliation{Center for Theoretical Physics, Massachusetts Institute of Technology, Cambridge, MA 02139, USA}

\emailAdd{sergiohc@mit.edu}

\preprint{MIT-CTP/5708}

\abstract{~

We study the general framework of effective theories obtained via exact path integration of a complete theory over some sector of its degrees of freedom.
Theories constructed this way contain multi-integrals which couple fields arbitrarily far apart, and in certain settings even on path-disconnected components of the space. These are not just entanglement, but genuine non-local interactions that we dub quantum wormholes.

Any state the path integral of such an effective theory prepares is shown to be a partial trace of a state of the complete theory over the integrated-out sector. The resulting reduced density operator is generally mixed due to bra-ket wormholes. An infinite family of ensembles of pure states of the complete theory giving the same effective state is identified. These allow one to equivalently interpret any effective state as being prepared by an ensemble of theories. When computing entropic quantities, bra-ket wormholes give rise to replica wormholes. This causes replica path integrals for the effective theory to not factorize even when the underlying manifold does, as expected from mixing. In contrast, effective theories obtained by derivative expansions have no quantum wormholes and prepare pure states. There exist operators in the algebra of effective theories which can distinguish mixed from pure states, implying a breakdown of non-exact effective theories for sufficiently complex observables.

This framework unifies and provides new insights into much of the phenomena observed in quantum gravity, including the interplay between wormholes and unitarity, the breakdown of bulk effective theory, the factorization puzzle, state ensembles, theory ensembles, quantum error correction, and baby universes. Some interesting lessons are drawn accounting also for characteristic aspects of gravity concerning IR/UV mixing and Kaluza-Klein reductions.
}

\begin{document}

\maketitle

\section{Introduction}

A longstanding problem in quantum gravity, particularly in the context of holography, concerns whether or not certain topologies should be included in a gravitational path integral formulation.
If one treats the gravitational path integral as indiscriminately allowing for all possible manifold topologies consistent with boundary conditions, a puzzle immediately arises: \`a-priori independent boundary theories set up on disjoint spaces seem to get coupled by wormholes in the bulk. In particular, the boundary partition function should factorize across connected components, whereas the bulk geometries seem to break factorization. 

Such a phenomenon would clash with the AdS/CFT correspondence as originally formulated and generally understood \cite{Maldacena:1997re,Witten:1998qj,Gubser:1998bc,Aharony:1999ti}. The puzzle becomes particularly sharp given the existence of wormhole geometries which are classical solutions, and thus saddle points expected to dominate in a controlled semiclassical regime \cite{Maldacena:2004rf,Saad:2018bqo,Harlow:2018tqv,Chandra:2022fwi,VanRiet:2020pcn,Astesiano:2023iql,Hebecker:2018ofv,Arkani-Hamed:2007cpn,Marolf:2021kjc}.\footnote{This discussion concerns spacetime wormholes, and not spatial wormholes such as Einstein-Rosen bridges.}
One may raise concerns about the subtle stability properties of some such solutions, or speculate that perhaps other saddles or non-perturbative contributions could cancel them out. However, lower-dimensional realizations of gravity continue to reaffirm that gravitational wormholes, if allowed to contribute, genuinely introduce statistical correlations between disjoint boundary regions. The various realizations in $3$-dimensional gravity show how the effect of saddle wormholes cannot be reproduced by any single boundary CFT, but only by averages over CFT ensembles or similar randomized constructs \cite{Belin:2020hea,Afkhami-Jeddi:2020ezh,Maloney:2020nni,Cotler:2020ugk,Benjamin:2021wzr,Collier:2021rsn,Chandra:2022bqq,Collier:2023fwi,Collier:2024mgv}. In $2$-dimensional Jackiw-Teitelboim (JT) theories of gravity \cite{Jackiw:1984je,Teitelboim:1983ux,Almheiri:2014cka}, calculations can be taken beyond the semiclassical regime, leading to even more robust results. A full path integral over the moduli space of wormhole geometries of arbitrary topology can be performed in JT and precisely matched by the topological expansion of double-scale matrix integrals \cite{Saad:2019lba,Stanford:2019vob,Johnson:2019eik}. From the perspective of random matrix theory, JT gravity theories genuinely describe the spectral statistics of an ensemble of theories with a random matrix Hamiltonian, where factorization is drastically broken.

Faced with this factorization problem, one may be tempted to just throw away wormholes and somehow restrict the gravitational path integral to not include them. However, there are independent compelling reasons that they should in fact be included.
A prominent one is the recent computations involving evaporating black holes \cite{Almheiri:2019hni,Penington:2019npb,Almheiri:2019psf,Penington:2019kki,Almheiri:2019qdq,Almheiri:2020cfm}, where wormholes are needed to resolve a sharp version of the black hole information paradox \cite{Page:1993df,Page:1993up,Mathur:2009hf,Almheiri:2012rt,Harlow:2014yka,Marolf:2017jkr}.\footnote{The mechanism by which information dynamically escapes the black hole remains largely unknown.} In this particular context, the relevant geometries are replica wormholes whose inclusion in the gravitational path integral is crucial to recover a unitary Page curve for the entropy between a black hole and its radiation. 
Another setting where wormholes are needed is to reproduce the universal late-time behavior of the spectral form factor in chaotic quantum systems \cite{Cotler:2016fpe,Liu:2018hlr,Stanford:2020wkf,Saad:2018bqo,Okuyama:2018gfr,Saad:2019lba,Saad:2019pqd,Winer:2020gdp,Zhou:2023qmk}.
Given these successes, giving up wormhole contributions to the gravitational path integral altogether would seem to be too na\"ive a route.

This paper is at its core motivated by the factorization problem in quantum gravity, and in particular by the quest to understand how the AdS/CFT correspondence could possibly be reconciled with the ensembles that wormholes in effective gravity give rise to. A heuristic viewpoint that resonates across the community is the idea that simple models of gravity, by ignoring microscopic details about their parent, UV-complete theory, should only be able to provide statistical answers. Such statistics would capture the ambiguity in which complete theory gave rise to the effective theory at hand, and reproduce results consistent with the ensemble of all possible such UV completions. While this intuition seems consistent with the observed phenomena, a precise mechanism for how non-factorization and ensembles came to be in the first place is still lacking. The central goal of this paper is to present a very general framework where not only do ensembles and non-factorization naturally arise, but also many other puzzling aspects of quantum gravity can be thoroughly addressed. In particular, as will be seen, features often understood as exclusively quantum gravitational turn out to also arise in field theory, including bra-ket wormholes, replica wormholes, and the breakdown of effective theory with complexity.

\subsection{Overview}
\label{ssec:over}

Our discussion starts by supposing we have a theory which provides a complete description of some system in terms of a path integral. We then consider the construction of a general effective theory for some sector of interest as obtained by path-integrating over any other degrees of freedom of the theory. Note that this broad notion of effective theory is completely independent of renormalization and does not rely on any assumption of energy scale separation.
Importantly, even when the starting point is a local theory, exact functional integration generally leads to an effective theory whose action involves multi-integrals over the underlying space. These non-localities cause the states the path integral of the exact effective theory prepares to generally be mixed due to correlations with the sector that was integrated out.

The non-local correlations in such a theory may be long-range, thus disallowing for a local derivative expansion and possible truncation thereof. However, even when parametrically suppressed, we show that a derivative expansion would lead to results inconsistent with the original theory. 
This becomes manifest in calculations where no matter how short-range, non-localities end up straddling across disjoint spaces. For obvious reasons, we refer to such non-local couplings across connected components of a space as quantum wormholes.

These quantum wormholes are seen to be instrumental in reproducing the fact that states prepared by an exact effective theory are mixed, in which case they take the form of bra-ket wormholes. When it comes to replica calculations they give rise to replica wormholes, which are required for consistency of the framework to reproduce valid results for states with mixing. At a more operational level, quantum wormholes also provide a mechanism for the breakdown of any effective theory where approximations have been made. In particular, no such approximate theory can possibly be able to prepare states capturing the expectation values of sufficiently complex observables in the truly mixed state of the exact effective theory.

We emphasize that there is nothing radical about much of these phenomena, which are in fact well understood in condensed matter theory as applied to open systems. In that context, the Schwinger-Keldysh formalism \cite{Schwinger:1960qe,Keldysh:1964ud,Haehl:2016pec,Haehl:2016uah,Pelliconi:2023ojb} is used to prepare a density operator, integrate out the environment, and obtain a reduced density operator for the system of interest.
Then the evolution of this system is given by a Feynman-Vernon-like influence functional, which describes the non-equilibrium dynamics with the variables intrinsic to the system (cf. an effective theory) \cite{Feynman:1963fq,Lombardo:1995fg,weiss2008quantum}.\footnote{An influence functional is an operator describing Lorentzian time evolution of a reduced density operator, and can be obtained from a path integral by integrating out the environment degrees of freedom. In an operator formulation, making the usual Born-Markov approximations, such evolution of an open quantum system is given by a master equation in Lindblad form
\cite{Lindblad1976OnTG,breuer2002theory,GRABERT1988115,aurell2020operator}. Similar techniques are also used in cosmology \cite{LaFlamme:1990kd,Barvinsky:2023jkl}.} In such a setting, it is rather natural to expect information loss dissipatively to the bath (cf. non-unitarity of the evolution), and for the intrinsic predictions of the system of interest to only be statistical (cf. ensembles averages over the environment) \cite{Caldeira:1981rx}.
The situation in effective theories is entirely analogous and, in our case, simplified by the fact that we will consider state preparation without Lorentzian evolution. As a result, for us the influence functional is nothing but the Euclidean effective path integral, and what our quantum wormholes capture is precisely the mixing of reduced density operators.\footnote{I thank Mukund Rangamani for illuminating discussions on these analogies and useful references.}

Our exposition will be field-theoretic and thus, by holography, allow us to address questions in quantum gravity from different complementary points of view. 
From a boundary perspective, the existence of quantum wormholes naturalizes non-factorization and strong non-localities as a general feature of effective theories. Taking the bulk as holographically defined, in accord with the mantra that the bulk geometrizes boundary quantum correlations, effective gravity wormholes could then be understood as an emergent gravitational description of some of the quantum wormholes of a dual effective theory.\footnote{See e.g. \cite{Betzios:2019rds} for a related discussion of Euclidean wormholes in a holographic setting.} 
In such a setting, the rules for the gravitational path integral should be dictated by the boundary theory, and because there are quantum wormholes in the latter, so there should be geometric wormholes in the former. Interestingly though, as will be seen, not always does effective field theory lead to quantum wormholes, suggesting that perhaps not always should geometric wormholes be allowed.

From a bulk perspective, were one to start from a non-perturbative theory of quantum gravity, it would be desirable to have an intrinsic description of geometric wormholes. Understanding precisely how an effective gravity theory in terms of a gravitational path integral over manifolds with dynamical metrics descends from an exact, UV-complete theory of quantum gravity remains a formidable challenge \cite{Eberhardt:2020bgq,Eberhardt:2021jvj}. 
Nonetheless, quantum wormholes in this case also demystify non-factorization and the emergence of ensembles, and additionally grant alternative plausible interpretations for gravitational wormholes.

A tantalizing possibility is that geometric wormholes are the quantum wormholes that arise upon integrating out whatever quantum gravitational degrees of freedom there may be more fundamental than metrics and gravitons.\footnote{A bottom-up realization of this idea via the renormalization of string effects recently appeared in \cite{Gesteau:2024gzf}.} This seems plausible given some of the distinguishing features quantum gravity exhibits which are usually extraneous to field theory, most noteworthy being diffeomorphism invariance and the phenomenon of IR/UV mixing. Together, if anything, these strongly suggest that effective gravity may exacerbate the strong non-localities that already arise in effective field theory.
Another compelling alternative is to put gravitational wormholes on the same footing as quantum wormholes of quantum fields. Namely, for a field theory defined on gravitating backgrounds of different topology, integrating out gravitational wormholes gives rise to non-localities and quantum wormholes of the same flavor as those obtained by integrating out quantum fields.  Hence it would seem conceivable that geometric wormholes may be equivalently describable as quantum wormholes of a field theory on a background of fixed topology. Without making any strong claims about either of these speculations, we believe the heuristics are sufficient to motivate further explorations.

\subsection{Summary}

Before providing a detailed outline of the contents of this paper, here we summarize some of the main takeaways and where a discussion thereof is presented. The reader is welcome to use this together with the outline that follows as a guide for a targeted reading of this paper that suits their interests.

\paragraph{Takeaways:}

\begin{enumerate}[itemindent=2em,labelsep=.5em,leftmargin=0pt]
    \vspace{-8pt}
    \item An effective theory obtained by exact functional integration from a local interacting theory generally has an action with arbitrarily many multi-integrals over the underlying space. Such a theory provides an exact description of the system of interest, regardless of what environment degrees of freedom were integrated out. In certain cases one can simplify the multi-integrals down to a single integral by performing local derivative expansions. However, even if all terms in the resulting infinite series are kept or even resumed, any derivative expansion of this form already destroys quantum wormholes and gives a strictly different, approximate theory. Upon a possible truncation of higher derivative terms, one simply obtains an even coarser, truncated theory. In order, we refer to these different kinds of effective theories as exact, approximate, and truncated. The effective theories one typically studies in quantum field theory are of the latter type. See \cref{sec:pfs}.
    \item The path integral of an exact effective theory prepares mixed states, even when one considers the full system. At the level of state preparation, this is because integrating out a sector to get an effective theory precisely corresponds to tracing out a density operator over that same sector. From the perspective of the exact effective theory, this is a consequence of bra-ket wormholes corresponding to the the multi-integrals over the space on which the state is prepared. These mixed states capture exactly any observable intrinsic to the effective theory, as reduced density operators do. Approximate and truncated effective theories have no quantum wormholes and thus prepare pure states. No such pure state can possibly reproduce the expectation values of all observables on an exact effective state, indicating a breakdown of such descriptions. See \cref{sec:prep}, particularly \cref{ssec:cmst,ssec:ees,ssec:appeftst}.
    \item In quantum theory, mixed states are classical ensembles of pure states. The partial trace that constructs the reduced density operators of the exact effective theory via a path integral singles out a preferred ensemble of pure states for the system in terms of states of the integrated-out environment. This ensemble defines an orthogonal set of states in the complete theory, but is over-complete and gives rise to null states in the effective theory. The connection of this ensemble to the complete theory allows for an equivalent rewriting of the mixed states of the effective theory as being prepared by an ensemble of theories. The probability density functional that specifies the measure over the ensemble of states fixes the measure over the ensemble of theories. See \cref{sec:prep}, particularly \cref{ssec:ensemble,ssec:theoryensem}.
    \item The amount of mixing of exact effective states can be quantified by entropic quantities obtained from replica calculations. 
    It is still useful to construct a replica manifold, but a direct evaluation of the full path integral of the exact effective theory on this manifold turns out to be na\"ive. The correct prescription requires restricting the action multi-integrals to run over replicas separately to prepare independent states. Nonetheless, bra-ket wormholes end up connecting consecutive replicas and giving rise to replica wormholes, in consistency with mixing. See \cref{fig:triangle} for an illustration. Multi-swap operators can be used to make the discussion of replica symmetry and its breaking more transparent. Under different assumptions of replica symmetry, replica wormhole calculations are seen to be equivalent to powers of partition functions ensemble-averaged over theories. See \cref{sec:reps}.

\begin{figure}[ht]
    \centering
    \includegraphics[width=.5\textwidth]{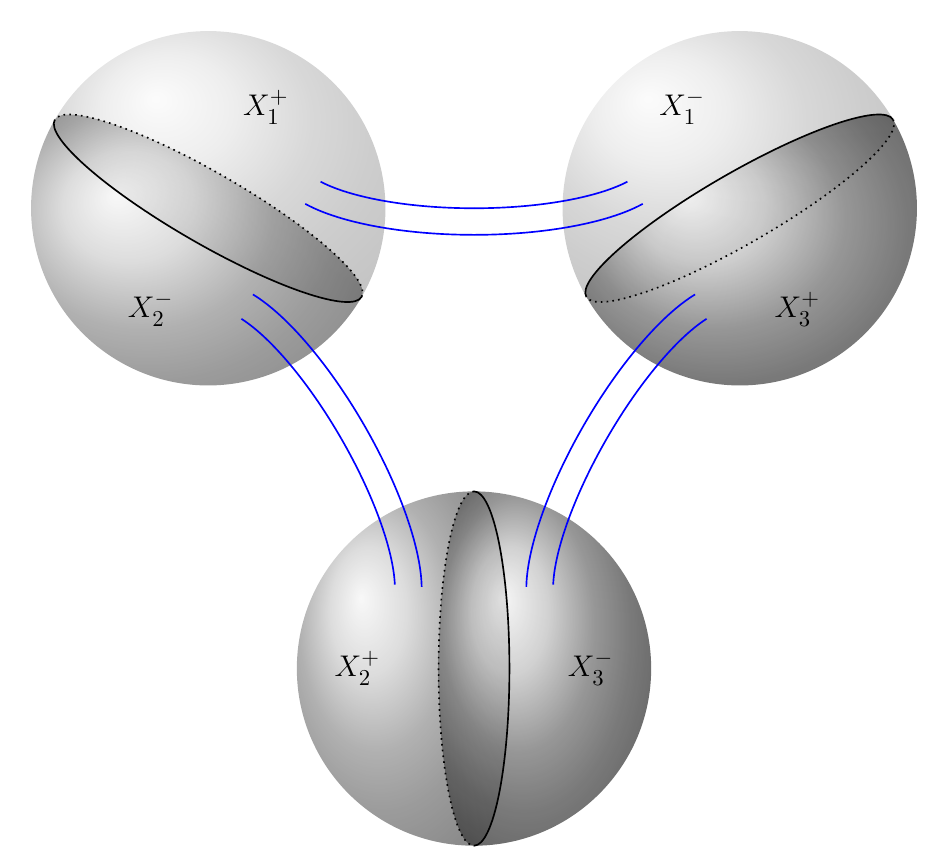}
    \caption{Representation of a $3$-replica calculation. Three independent copies of a density operator are prepared by an exact effective theory path integral on $X_k^+\sqcup X_k^-$ for $k\in\{1,2,3\}$, where boundary conditions on $\partial X_k^+$ and $\partial X_k^-$ are left unspecified. The `bra' and `ket' of the $k^{\text{th}}$ replica are respectively prepared on $X_k^+$ and $X_k^-$, which are coupled by bra-ket wormholes despite being disjoint spaces. Under a trace, boundaries get cyclically identified according to $\partial X_k^+ \sim \partial X_{k+1}^-$ with $k+3\sim k$. As a result, one obtains a nontrivial $3$-replica wormhole. See \cref{sec:reps} for more details.}
    \label{fig:triangle}
\end{figure}

    \item There is no canonical association of operator algebras to subspaces in non-local effective theory. The algebra additively generated by local operators within a subspace is generally a strict subalgebra of the commutant of the additively generated algebra of the complementary subspace. This is a violation of Haag-like duality caused by non-local operators which are only contained in the latter. In an exact effective theory, this violation seems to occur globally due to quantum wormholes. See \cref{sec:algebras}, particularly \cref{ssec:algdu,ssec:qetalg}.
    \item A violation of algebraic duality implies an incompleteness in the description of a state that can be attained by local observables, which cannot distinguish states differing only by expectation values for non-local operators. Relative entropy can be used as an operational measure of distinguishability between such states. When applied to a state prepared by an exact effective theory with respect to one prepared by an approximate or truncated effective theory, the relative entropy is formally infinite. This means that there exist observables which can perfectly distinguish such states, implying a breakdown of non-exact effective theories. For a truncated effective theory, this breakdown can be associated to energy scales that invalidate the neglect of higher derivatives. For an approximate effective theory, what naturally breaks down is the description of complex operators. See \cref{sec:algebras}, particularly \cref{ssec:relent}.
    \item Effective field theory reproduces many features observed in holographic quantum gravity. An exact effective theory is only possibly consistent with the unitary evolution of the parent complete theory if replica wormholes are included in entropy calculations, which parallels the situation encountered in calculations of the Page curve for evaporating black holes. When it comes to partition functions, however, exact effective field theory factorizes. If geometric wormholes are the gravitational avatar of quantum wormholes, this suggests that wormholes should not be included in such calculations. Interestingly, under assumptions of replica symmetry, certain replica calculations precisely yield ensemble averages over partition functions. These allow one to reinterpret bulk calculations of multi-boundary partition functions with wormholes included as replica calculations consistent with a single complete theory on the boundary. Effective theories also exhibit the structure of operator algebra quantum error correction and ensembles of baby universes. See \cref{ssec:effholo}.
    \item The lessons of effective field theory are also applicable to quantum gravity outside the framework of holography. Diffeomorphism invariance and IR/UV mixing suggest that non-localities in effective gravity may be even stronger than in field theory. However, a gravity theory obtained by Kaluza-Klein reduction is more akin to a truncated rather than an exact effective theory. Performing a path integral after such a truncation has catastrophically non-local consequences of the same form as the wormholes one obtains from integrating $\alpha$-parameters. See \cref{ssec:effgrav}.
\end{enumerate}

\paragraph{Outline:}~

In \cref{sec:pfs}, we begin by introducing some basic notation and notions of locality that will be employed throughout. \Cref{ssec:eftnonlo} defines exact effective theories as generally obtained by exact path integration over some sector of a given theory, and addresses how such theories contain non-local interactions which can lead to quantum wormholes. \Cref{ssec:approx} clarifies where such a non-local effective theory stands relative to standard effective field theory treatments where through derivative expansions and truncations one ends up with a local theory. \Cref{ssec:factor} touches for the first time on factorization at the level of partition functions, explaining how quantum wormholes generally do not arise in this case. \Cref{ssec:eg} presents some examples of theories where exact path integration can be performed and illustrates the kind of non-localities that are important for this paper.

In \cref{sec:prep}, we delve into the formalism of state preparation using the path integral, in particular as it applies to the construction of functional density operators. \Cref{ssec:cmst} does so first using the path integral of a local, complete theory. \Cref{ssec:ees} then consider the preparation of states using the path integral of an exact effective theory, and shows how the effective state precisely corresponds to the reduced density operator of the complete state upon partially tracing over the pertinent sector via path integration. \Cref{ssec:appeftst} explains how the generally mixed state that results from this procedure is inequivalent to the states one would obtain were one to perform a derivative expansion of the non-local theory. The inequivalence is due to quantum wormholes: these are the sole sources of their mixing, and any local derivative expansion would destroy them, thereby eliminating any information about the existence of a parent, complete theory. To ask fundamentally fine-grained questions about the theory, this should not be done. \Cref{ssec:ensemble} shows how the path integral singles out a family of ensembles of pure states of the complete theory, all of which realize the same mixed, effective state upon partial tracing. \Cref{ssec:theoryensem} then explains how these mixed states can equivalently be understood as being prepared by an ensemble of theories.

In \cref{sec:reps}, we undertake the exploration of replica path integrals. \Cref{ssec:entrop} motivates entropies as measures of mixing for effective states, and introduces the replica trick to calculate von Neumann entropies. \Cref{ssec:stareps} briefly describes how replica calculations are implemented in a path integral formulation of density operators, and applies it to the simpler case of local, complete theories. \Cref{ssec:nonfact} then proceeds to explain the more subtle case of performing replica path integrals for states prepared by an effective theory. The specific case in which the replica calculation takes place on a space consisting of multiple components is thoroughly explained, as it constitutes a key illustration of how quantum wormholes break factorization even in effective field theory. \Cref{ssec:ensemreps} shows how the von Neumann entropy of the effective states can be written in terms of the classical Shannon entropy of an ensemble of pure states of the parent theory. \Cref{ssec:proj} concludes with alternative representations of the replica path integrals in terms of field projections, which give a more transparent account of how complete and effective theory replicas differ, and what the role of replica symmetry is.

In \cref{sec:algebras}, we approach the observations throughout the paper in terms of the operator algebras of the different theories at play. \Cref{ssec:algdu} considers the association of algebras of operators to subspaces of a theory, and introduces the key notion of duality that allows for a canonical assignment.
\Cref{ssec:qetalg} explores the algebraic structure of effective theories, finding and characterizing the ways in which they violate duality. Quantum wormholes are identified as responsible for a particularly interesting violation of duality on a global scale of the theory. \Cref{ssec:relent} then employs relative entropy as a natural operational measure to quantify the contributions of quantum wormholes to effective states. This provides not only a way of distinguishing between states of an effective theory, but also to distinguish states prepared by different effective theories. We observe that approximate and truncated effective theories cannot possibly capture observables of arbitrary complexity, which provides a mechanism for the breakdown of local effective theory.

In \cref{sec:qg}, we conclude by exploring ways in which the field-theoretic framework of exact effective theory can also be applied to the realm of quantum gravity.

\Cref{ssec:effgrav} approaches this from a holographic viewpoint by applying the lessons of this paper to the boundary theory. This provides an interpretation of the gravitational path integral as the effective gravity realization of a boundary effective theory. The dual to the quantum wormholes in the latter are naturally identified with the geometric wormholes in the former. \Cref{sssec:hawking} illustrates this in the specific setting of black hole evaporation, where the relevant effective boundary theory is the one obtained by integrating out the black hole states. From a purely effective theory perspective, a truncation of quantum wormholes would provide a description inconsistent with unitarity of the parent theory. The bulk results parallel this: replica wormholes are needed for consistency with unitarity, and ignoring them gives information loss. \Cref{sssec:nonrewo} considers the possibility that certain gravitational wormholes in holography may be unrelated to quantum wormholes on the boundary and thus unphysical. \Cref{sssec:tfdtmd} provides an example of an ambiguity in holography between factorizing and non-factorizing quantities which seemingly would be computed by the same gravitational path integral according to standard rules. \Cref{sssec:qec} shows how the algebraic structure of effective theories reproduces all the features of operator algebra quantum error correction in holography. \Cref{sssec:babies} concludes the holographic discussion by identifying the baby universe Hilbert space and the $\alpha$-sectors in an effective theory.

\Cref{ssec:effgrav} finally makes some remarks on how the lessons from the effective theory framework may extrapolate to an effective gravity theory as potentially obtained from within a UV-complete theory of quantum gravity. \Cref{sssec:iruv} argues that the phenomenon of IR/UV mixing may be a mechanism by which non-localities in effective gravity may be even more dramatic than in field theory. Within quantum gravity, the plausible expectation seems to be that geometric wormholes are the effective manifestation of quantum wormholes associated to gravitational degrees of freedom. \Cref{sssec:kkr} briefly reflects on the way manipulations in effective gravity generally involve drastic truncations of the theory rather than integrations over sectors. \Cref{sssec:trunc} concludes with an example of how a na\"ive mode truncation inside a path integral can lead to catastrophically non-local results.

\section{Theory Framework}
\label{sec:pfs}

Let $I$ be the Euclidean action of a non-gravitational theory defined on a Riemannian manifold $M$, and use $\phi$ to collectively denote all quantum fields of the theory.
The partition function of the theory with action $I$ is defined by a Euclidean path integral over the phase space of continuous field configurations of $\phi$, and given by
\begin{equation}
\label{eq:Zpi}
    Z \equiv \int\displaylimits_{M} \mathcal{D}\phi \, e^{-I[\phi]}.
\end{equation}
The subscript $M$ on the path integral indicates that the quantum fields $\phi$ are maps on $M$.
If this theory provides a complete framework for the description of the physics of the system of interest, we refer to it as a \textit{complete theory}.

It will be useful throughout to characterize theories in terms of their local or non-local properties. A theory will be said to be \textit{local} if $I[\phi]$ involves a single integral over $M$, and the highest derivative of any field $\phi$ is of finite order. Otherwise, the theory will be said to be \textit{non-local}. It will be important in what follows to distinguish between different types of non-localities. If $I[\phi]$ involves a single integral over $M$, but includes derivatives of arbitrarily high order for some field, the theory will be qualified to be \textit{weakly non-local}. In contrast, if $I[\phi]$ contains more than one integral over $M$, the theory will be referred to as \textit{strongly non-local}.\footnote{As will be explained in \cref{ssec:approx}, it is sometimes possible to perform a derivative expansion to reduce a strongly non-local theory to a weakly non-local one. Certain non-local contributions are always lost in this process, so in general strong non-locality implies weak non-locality, but the converse is strictly false.} An interesting possibility in a strongly non-local theory is that, if the path integral is evaluated on a space consisting of more than one connected component, the multi-integrals in the action may couple fields on disconnected spaces. Such strongly non-local couplings across distinct connected components of the space will be referred to as \textit{quantum wormholes}.

Hereon we will reserve the symbols $I$ and $Z$ respectively to refer to the action and partition function of a complete theory. Generally, we will be interested in complete theories which are local, so it shall be assumed that the action $I[\phi]$ is local in what follows.

\subsection{Exact Quantum Effective Theory}
\label{ssec:eftnonlo}

If one is only interested in the physics of a specific sector of quantum fields of the theory, a useful procedure is to construct an effective theory for those degrees of freedom. This section makes precise the very general framework of effective theory that is relevant to this paper.\footnote{\label{fn:loaded}The term ``effective'' is a loaded one in physics, and typically triggers a mental association to Wilsonian renormalization. We emphasize here and in multiple other parts of this paper that our construction of effective theories is a lot more general and independent of the considerations of renormalization. In particular, we do not require making any distinction between energy scales or need to assume any form of decoupling.}

Consider a splitting of the quantum fields of a complete theory into two sectors by letting $\phi=(\varphi,\Phi)$. This splitting may be completely arbitrary. Some natural examples correspond to identifying a system of interest and an environment, separating fields into bosonic and fermionic species, or splitting degrees of freedom by energy scales. For instance, these examples are respectively realized when studying open quantum systems, exploring the bosonic sector of a supersymmetric theory, or doing Wilsonian renormalization. Our treatment is completely agnostic as to what distinguishes $\varphi$ and $\Phi$.

An effective theory for the $\IR$ fields can be obtained by integrating out the $\UV$ fields first,\footnote{\label{fn:couple}If the path integral at hand involves insertions of the $\UV$ fields, the resulting quantum effective action will also depend on these nontrivially. From the viewpoint of state preparation (see \cref{sec:prep}), the effective theory for the $\IR$ fields may be different depending on the state being prepared by the path integral over $\UV$ fields.}
\begin{equation}
\label{eq:IEFT}
    e^{-\wtilde{I}[\varphi]} \equiv \int\displaylimits_{M} \mathcal{D}\Phi \, e^{-I[\varphi,\Phi]},
\end{equation}
a path integral which defines the \textit{quantum effective action} $\wtilde{I}$ for $\varphi$. In general, this process gives rise to an action $\wtilde{I}[\varphi]$ which involves infinite series of increasingly many multi-integrals of the $\IR$ fields over $M$.\footnote{This is a general phenomenon which in the example of Wilsonian renormalization follows as an exact functional statement from Polchinski's flow equation \cite{Polchinski:1983gv,Morris:1993qb,Wetterich:1992yh,Rosten:2010vm}.} In other words, an effective theory obtained this way is a strongly non-local theory. Some simple examples of this are provided in \cref{ssec:eg}.
By linearity of the path integral though, functional integration is transitive, so further performing the integral over $\IR$ fields yields back the same partition function as in \cref{eq:Zpi}. Namely, the partition function of the effective theory,
\begin{equation}
\label{eq:Zeft}
    \wtilde{Z} \equiv \int\displaylimits_{M} \mathcal{D}\varphi \, e^{-\wtilde{I}[\varphi]},
\end{equation}
matches $Z=\wtilde{Z}$ in the absence of insertions. Crucially though, $Z$ and $\wtilde{Z}$ are to be understood as mathematically different objects: $Z$ defines a functional integral over $\phi=(\varphi,\Phi)$ fields on $M$, whereas $\wtilde{Z}$ defines a functional integral only over $\varphi$ fields on $M$. In the latter, the path integral over $\Phi$ was already performed in \cref{eq:IEFT} to obtain the functional form of $\wtilde{I}$. The theory that $\wtilde{Z}$ defines only has access to the dynamics of the $\IR$ fields.

\begin{figure}
    \centering
    \includegraphics[width=.7\textwidth]{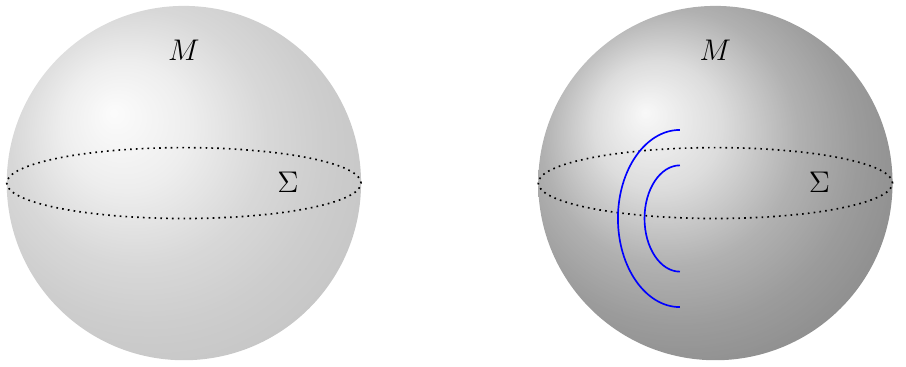}
    \caption{Representation of the path integral computation of partition functions on a spherical space $M$. The equatorial hypersurface $\Sigma$ splits $M$ into two dual halves. On the left, a complete theory with local action $I$ is used to evaluate the partition function $Z$ on $M$ as defined in \cref{eq:Zpi}. On the right, an exact $\QET$ with strongly non-local action $\wtilde{I}$ given by \cref{eq:IEFT} is used to evaluate the partition function $\wtilde{Z}$ defined in \cref{eq:Zeft}. The latter is a strongly non-local theory involving arbitrarily many multi-local couplings. This is depicted in the case of a bilocal interaction with lines straddling across $\Sigma$ between points on the top and bottom hemispheres at arbitrary separation. Numerically $\wtilde{Z}=Z$.}
    \label{fig:partitionfunctions}
\end{figure}

A theory with partition function $\wtilde{Z}$ given by \cref{eq:Zeft} and quantum effective action defined by the path integral in \cref{eq:IEFT} will be referred to as a \textit{quantum effective theory ($\QET$)}.\footnote{In line with \cref{fn:loaded}, the decision to not refer to these objects by the initialism `EFT' usually used for effective field theory has been deliberate. An EFT obtained by renormalization is just an example of a $\QET$.} In particular, if \cref{eq:IEFT} is evaluated by exact functional integration without any approximations, the resulting theory will be said to be an \textit{exact $\QET$}. Tilded symbols like $\wtilde{I}$ and $\wtilde{Z}$ will always be used to refer to exact $\QET$ objects. The path integrals on $M$ that give the partition function $Z$ for a complete theory and $\wtilde{Z}$ for an exact $\QET$ are illustrated in \cref{fig:partitionfunctions}.

This choice of terminology is easily motivated.
The general quantum field theoretic concept of an effective action refers to the generating functional of one-particle irreducible correlation functions. Explicitly, introducing a classical current $J$ sourcing a quantum field $\phi$, one first defines the generating functional of connected correlation functions $W[J]$ by
\begin{equation}
\label{eq:WJ}
    e^{-W[J]} \equiv \int_M \mathcal{D}\phi\, e^{- I[\phi] - \int J\phi}.
\end{equation}
\ignore{
The effective action is then defined as the Legendre transform of $W[J]$, which reads
\begin{equation}
    \Gamma[\bar{\phi}] \equiv W[J_{\bar{\phi}}] - \int_M J_{\bar{\phi}} \, {\bar{\phi}}, \qquad {\bar{\phi}} \equiv \frac{\delta W[J]}{\delta J},
\end{equation}
where the last expression defines the conjugate ${\bar{\phi}}$ to $J$, often called the classical field, and defines $J_{\bar{\phi}}$ implicitly in terms of it. In terms of the effective action, the sourced partition function of the theory is
\begin{equation}
    Z[J] = \int_M \mathcal{D}\phi\, e^{- I[\phi] - \int J\phi} = e^{-\Gamma[\bar{\phi}] - \int J \, {\bar{\phi}}},
\end{equation}
where no path integral remains to be done on the right-most expression.}
The effective action is then the Legendre transform of $W[J]$, in terms of which one obtains an explicit form for the sourced partition functional of the theory with no path integrals left to be done.
The way we define a $\QET$ through \cref{eq:IEFT} is entirely analogous, with a couple of important qualitative differences. In general the action in \cref{eq:IEFT} takes the form
\begin{equation}
\label{eq:iii}
    I[\varphi, \Phi] = I_\varphi[\varphi] + I_\Phi[\Phi] + I_{\smalltext{int}}[\varphi,\Phi],
\end{equation}
with independent free actions $I_\varphi$ and $I_\Phi$ for each sector, and an interaction term $I_{\smalltext{int}}$ that couples them. Comparing with \cref{eq:WJ}, we see that only $\Phi$ is integrated out in our case, with $\varphi$ left untouched. Hence we have $\IR$ fields playing the role of currents, which source the $\UV$ fields through the $I_{\smalltext{int}}$ term. This is why we call the action $\wtilde{I}[\varphi]$ effective.\footnote{Note however that this source-type term is generally nontrivial, so the resulting effective action no longer has any of the convexity properties of a Legendre transform.} Additionally, our current-like objects $\varphi$ are not classical, but actual quantum fields that we still want to treat dynamically. This is why we call the action $\wtilde{I}[\varphi]$ quantum.
Finally, while in quantum field theory the effective action may often be obtained perturbatively or approximated to leading order by the classical action, note that its definition is given non-perturbatively by functional integration. This is why we refer to the resulting $\QET$ that $\wtilde{I}[\varphi]$ defines as exact.

\subsection{Approximations and Truncations}
\label{ssec:approx}

Although in quantum field theory one most often works with some effective theory in the above sense, rarely does one encounter a non-local action, let alone a strongly non-local one. This is because for most purposes an exact $\QET$ is an overkill, and truncations of it often suffice for describing the physics of interest to the desired degree of accuracy, but not always. Any such truncation gives rise to a coarser theory which may give good approximations for certain observables but is bound to fail completely for others. Here we briefly review this procedure in the context of renormalization, where a breakdown of truncated theories can easily be seen to be associated to energy scales. However, later in \cref{sec:prep,sec:algebras} we will see that truncated theories also fail at preparing states in concrete ways, which implies that such theories will also fail to reproduce the expectation values of sufficiently complex observables.

As explained below \cref{eq:IEFT}, even in Wilsonian renormalization the process of integrating out some heavy $\UV$ fields generates a strongly non-local action for light $\IR$ fields with infinitely many terms involving arbitrary multi-integrals over $M$. All of these non-localities account for the effects the now virtual $\UV$ fields have in mediating interactions among $\IR$ fields (cf. \cref{fig:feynman}). However, provided the $\UV$ fields have a smallest positive mass $m_{\Phi}$, such interactions will decay exponentially in distance scales of order $m_{\Phi}^{-1}$ and, additionally, higher integrals will exhibit a polynomial suppression in powers of $m_{\Phi}^{-1}$. This decoupling phenomenon guarantees that the physics of $\IR$ fields behave approximately as local at energy scales $\Lambda \ll m_{\Phi}$. 

At a mathematical level, this can be seen as follows. Under certain assumptions, small energies $\Lambda \ll m_{\Phi}$ allow one to expand all integrands in the strongly non-local $\QET$ action about any choice of reference point in $M$. Except for the reference point, all other points now appear outside $\IR$ fields, and their corresponding integrals over $M$ can be explicitly performed. This way one obtains an action which contains infinitely many derivatives of the $\IR$ fields coming from the expansion about a reference point, but involves a single integral of that reference point over $M$. In other words, the resulting $\QET$ action is only weakly non-local.
Now, this expansion is under control provided the energy scale of $\IR$ field derivatives is small compared to the mass scale of the $\UV$ fields, i.e., so long as one probes the $\QET$ at energies $\Lambda\ll m_{\Phi}$, as noted above. Because higher derivative terms are suppressed, at small enough $\Lambda$ energies one may recover an apparently local theory by truncating the expansion at some sufficiently high order. Hence, working at suitably small energies, one can often reduce a strongly non-local theory down to a weakly non-local one, and eventually to a local one by expanding and truncating. This procedure is exemplified in \cref{ssec:eg}.
Of course, while the original theory is exact, the others are only progressively coarser approximations. In particular, these approximate theories will deviate significantly from the exact theory or even break down when probed at energies $\Lambda$ comparable the mass scale $m_{\Phi}$ of the $\UV$ fields that were integrated out.

More abstractly, it may also be possible to apply these approximations to an exact $\QET$ obtained outside the framework of renormalization whenever one is only interested in some limit of the theory. The terminology that follows will be used in this broader sense. 
Let us first emphasize that, starting from an exact $\QET$ that is strongly non-local, any local derivative expansion already leads to an inequivalent theory.
This is simply because such an expansion destroys quantum wormholes, whether or not the infinite series of derivatives is kept or even resummed after.\footnote{\label{fn:resum}Doing a resummation before performing any integrals would give back a strongly non-local theory, but one where there no longer are quantum wormholes. If the local expansion is not resummed before path integrating, the results cannot possibly reproduce those of the exact QET, even if one keeps all infinitely many terms. This is because convergence of a Taylor series to a given function everywhere requires that function to be analytic everywhere; however, path integrals generally run over functions which are not even differentiable.} Indeed, if convergent, such a series can at most reproduce the action only within a connected component of the space, and not across connected components. The immediate goal of performing a derivative expansion is always to be able to evaluate all but one of the multi-integrals in the action. The resulting theory thus has infinitely many derivative terms but a single integral in its action, and is therefore only weakly non-local. A weakly non-local theory obtained this way from an exact $\QET$ without neglecting any terms will be referred to as an \textit{approximate $\QET$}.
If one then truncates the infinite series to some finite derivative order, the result is an even coarser but now local theory. A local theory obtained this way from an approximate $\QET$ will be referred to as a \textit{truncated $\QET$}. The states and operator algebras of all these theories are related in interesting and operationally distinguishable ways, as analysed in \cref{sec:prep,sec:algebras}.

We conclude this section by setting some expectations as to which basic properties these theories may or may not possess, which will help in anticipating some findings later on. Suppose our starting point is a complete theory which is local, unitary, and causal, and assume its fields $\varphi$ and $\Phi$ interact nontrivially. As emphasized in \cref{ssec:over}, an intuitive way to think about the resulting exact $\QET$ is as a theory describing an open system where the environment has been integrated out. This suggests that the sector the exact $\QET$ describes will evolve non-unitarily as only part of the complete, interacting theory, an expectation which is borne out due to quantum wormholes. Strong non-localities would also seem to make acausal physics plausible, since in Lorentzian signature spacelike separated regions would be non-locally coupled. However, microcausality for operators within disjoint domains of dependence will still hold since any non-local interactions are mediated by the integrated-out fields, which behave causally in the complete theory.
It is crucial to bear in mind that the way an exact $\QET$ may be non-local and non-unitary will always be in consistency with how the complete theory genuinely behaves as a local and unitary theory. Put differently, any apparently pathological behavior of an exact $\QET$ will be reproducing precisely what is required to restore locality, unitarity, and causality of the complete theory.

A logical converse naturally suggests that any approximation or truncation of an exact $\QET$ may fail in recovering this behavior of the complete theory. For instance, by destroying quantum wormholes, one may expect that an approximate $\QET$ no longer behaves like an open system and thus describes a decoupled system which should obey unitarily. Such an evolution in which the system is artificially decoupled from the environment cannot possibly be consistent with the actual coupled, unitary evolution of the complete theory. In a similar vein, the way a truncated $\QET$ describes $\varphi$ in terms of local dynamics cannot possibly be reconciled with the actual local dynamics of $\varphi$ coupled to $\Phi$ in the complete theory. As we know though, nothing prevents us from performing these approximations and working with local effective theories which, provided they satisfy certain effective theory bounds, are perfectly unitary and causal. 
The lesson starting from an exact $\QET$ is that restoring these properties via approximations is incompatible with describing a complete theory with those same properties. Indeed, from the perspective of the complete theory, the predictions of an exact $\QET$ for the $\varphi$ are perfectly valid, whereas those of a truncated $\QET$ would appear non-local and non-unitary.

\subsection{Factorization}
\label{ssec:factor}

This section addresses the behavior of the partition function of our different theories, and anticipates how subtle the appearance of quantum wormholes happens to actually be.
Indeed, as it turns out, quantum wormholes will only make their first explicit appearance when preparing states in \cref{sec:prep}.

In general, the manifold $M$ on which we define a theory may be disjoint. Let $\{M_k\}$ denote the connected components of $M$, i.e., the maximal connected subsets which partition it. By definition, these are disjoint, nonempty, and their union is the whole space. If the theory is local, the action $I[\phi]$ involves a single integral which by additivity will decompose into a sum of integrals over the different connected components. Schematically, the action reads
\begin{equation}
\label{eq:Isplit}
    \int\displaylimits_{M} (\,\cdot\,) = \sum_{k} \int\displaylimits_{M_k} (\,\cdot\,).
\end{equation}
If $\phi_k$ is the restriction of $\phi$ to $M_k$, then clearly the integral over $M_k$ only depends on $\phi_k$. By extending $\phi_k$ trivially to all $M$ such that it continues to only be supported on $M_k$, one can write $\phi = \sum_{k} \phi_k$. This way, the path integral in \cref{eq:Zpi} can be easily seen to factorize as\footnote{For $I[\phi_k]$, it makes no difference whether the action integral is taken to run over $M$ or just $M_k$, so continuing to use the same notation for the action is unambiguous.}
\begin{equation}
\label{eq:Zfact}
    Z = \prod_{k} Z_k, \qquad Z_k \equiv \int\displaylimits_{M_k} \mathcal{D}\phi_k \, e^{-I[\phi_k]}.
\end{equation}
In fact, if the theory is only weakly non-local, then \cref{eq:Isplit} still holds and the path integral similarly factorizes. Indeed, higher derivatives lead to non-localities within path-connected parts of the manifold, but never to interactions across distinct connected components.
Strongly non-local theories have the potential to behave rather differently.
In particular, a path integral in such a theory would fail to factorize if there are quantum wormholes across connected components of the underlying space. While one could certainly make up a strongly non-local theory that behaves this way, an exact $\QET$ turns out to not do this at the level of partition functions.

\begin{figure}
    \centering
    \includegraphics[width=.7\textwidth]{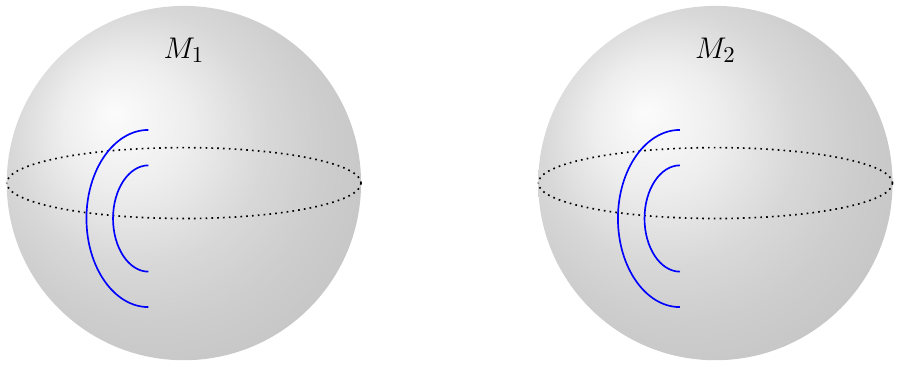}
    \caption{Representation of the partition function $\wtilde{Z}$ of an exact $\QET$ defined on a space $M$ consisting of two distinct connected components, $M_1$ and $M_2$. The partition function factorizes as in \cref{eq:Zeftfact} since exact integration starting from a complete theory does not generate quantum wormholes. Nonetheless, strong non-localities do arise within each connected component of the space.}
    \label{fig:qetfactorization}
\end{figure}

Consider again calculating the partition function, now for an exact $\QET$. This is a specific type of strongly non-local theory for which the starting point would be a complete theory defined on $M$. If this complete theory is local, the discussion that led to \cref{eq:Zfact} still applies. Hence, one may start integrating out $\UV$ fields from this factorized expression, with each path integral involving fields $\phi_k$ supported only on $M_k$. For each $Z_k$, there will clearly appear multi-integrals in the $\QET$ action $\wtilde{I}[\varphi_k]$, but these will only involve the $M_k$ component of $M$. The non-localities are thus only generated within each connected component, but never across them. In other words, there are no quantum wormholes in the partition function of an exact $\QET$, which still factorizes as
\begin{equation}
\label{eq:Zeftfact}
    \wtilde{Z} = \prod_{k} \wtilde{Z}_k, \qquad \wtilde{Z}_k \equiv \int\displaylimits_{M_k} \mathcal{D}\varphi_k \, e^{-\wtilde{I}[\varphi_k]},
\end{equation}
where $\wtilde{I}$ is obtained as in \cref{eq:IEFT} by performing the path integral over $\Phi_k$ fields supported on just $M_k$. The structure of $\wtilde{Z}$ in \cref{eq:Zeftfact} is illustrated in \cref{fig:qetfactorization}.
This situation shall be contrasted with the discussion that follows in \cref{sec:prep,sec:reps}, where it will be shown how other $\QET$ constructions do give rise to, and require, quantum wormholes.

Let us finish by pointing out that since fields are continuous maps on $M$, the constant function $\Phi(x)=\Phi_0$ is a perfectly valid configuration. However, if the path integral for $\Phi$ is restricted to just run over constant functions, what one gets is a dynamical cosmological constant taking the same value on every connected component of $M$. This is a very non-local object which would actually break factorization not only in \cref{eq:Zeftfact}, but also in \cref{eq:Zfact}. A global gauge symmetry would have similar consequences. This also illustrates an important point: in a theory where $M$ itself is dynamically determined (cf. a gravitational path integral), non-factorization upon path integration would seem rather natural.

\subsection{Examples}
\label{ssec:eg}

Here we cover some simple examples of the type of effective theories discussed in previous sections, which will serve as an illustration for the more general results of this paper.

Consider a Euclidean action of the form in \cref{eq:iii} involving two coupled scalars on a $d$-dimensional space $M$ with free parts
\begin{equation}
    I_\varphi[\varphi] = \frac{1}{2} \int\displaylimits_M d^dx \left[ (\nabla\varphi(x))^2 + m_\varphi^2 \varphi(x)^2 \right], \quad     I_\Phi[\Phi] = \frac{1}{2} \int\displaylimits_M d^dx \left[ (\nabla\Phi(x))^2 + m_\Phi^2 \Phi(x)^2 \right],
\end{equation}
where the masses $m_\varphi,m_\Phi\geq0$ are left arbitrary. For clarity, if $M$ is a curved space we leave any metric determinants implicit in the integral measures and elsewhere.

The simplest possible interaction term one may consider is linear in $\Phi$ and of the form
\begin{equation}
\label{eq:simint}
    I_{\smalltext{int}}[\varphi,\Phi] = \lambda \int\displaylimits_M d^dx \, f_\varphi(x) \, \Phi(x) ,
\end{equation}
for some arbitrary function $f_\varphi$ that depends only on $\varphi$.\footnote{This may be as simple as e.g. a quadratic term $f_\varphi = \frac{1}{2} \varphi^2$.} For this interaction term the theory is well-defined so long as $m_\Phi>0$. By virtue of being linear in $\Phi$, \cref{eq:simint} admits the standard quantum field theory treatment of a source term.
Integrating by parts with appropriate boundary conditions and completing the square to perform the functional Gaussian integral over $\Phi$ leads to
\begin{equation}
\label{eq:nonloeg}
    \int\displaylimits_M \mathcal{D}\Phi \, e^{ - I[\varphi,\Phi]} \propto \exp( -I_\varphi[\varphi] + \frac{\lambda^2}{2} \int\displaylimits_{M^2} d^dx \, d^dy \, f_\varphi(x) \, G(x,y) \, f_\varphi(y) ),
\end{equation}
where $G$ denotes the free propagator for $\Phi$, i.e., the Green's function on $M$ defined by
\begin{equation}
\label{eq:Ggreen}
    (-\nabla_x^2 + m_\Phi^2) \, G(x,y) = \delta(x-y).
\end{equation}
From \cref{eq:nonloeg} we can thus quote the final exact $\QET$ action as
\begin{equation}
\label{eq:fstqet}
    \wtilde{I}[\varphi] = I_\varphi[\varphi] - \frac{\lambda^2}{2} \int\displaylimits_{M^2} d^dx \, d^dy \, f_\varphi(x) \, G(x,y) \, f_\varphi(y),
\end{equation}
which is our first example of a strongly non-local theory arising as an exact $\QET$ by integrating out $\Phi$. In this case the strong non-locality takes the form of just a double integral or bilocal interaction involving two copies of the original space $M$.

The construction of an approximate $\QET$ from \cref{eq:fstqet} requires carefully checking that certain manipulations of the series expansion and integrals are allowed for a given $M$ and corresponding $G$. Assuming this is the case for now and choosing $x$ as the reference point, the desired local derivative expansion may be written
\begin{equation}
    f_\varphi(y) = \sum_{k=0}^\infty \frac{f_\varphi^{(k)}(x)}{k!} (y-x)^k. 
\end{equation}
Using this in \cref{eq:fstqet} allows one to express the action as
\begin{equation}
\label{eq:fstqetapp}
    \wtilde{I}[\varphi] = I_\varphi[\varphi] - \frac{\lambda^2}{2} \int\displaylimits_{M} d^dx \sum_{k=0}^\infty \frac{g_k(x)}{m_{\Phi}^{2+k}} f_\varphi(x) f_\varphi^{(k)}(x),
\end{equation}
where the $g_k$ functions are the effective couplings induced by the multi-integrals,
\begin{equation}
\label{eq:effco}
    g_k(x) \equiv \frac{m_{\Phi}^{2+k}}{k!} \int\displaylimits_{M} d^dy \, G(x,y) (y-x)^k,
\end{equation}
and the factors of $m_{\Phi}$ have been introduced to make $g_k$ dimensionless.
Now \cref{eq:fstqetapp} takes the form of the weakly non-local action of an approximate $\QET$. In this action, higher orders in the derivative expansion are suppressed by powers of $m_\Phi$. Hence this expansion is under control so long as the higher derivatives $f_\varphi^{(k)}$ are all small in units of $m_{\Phi}^{-1}$. If this is the case, a truncation of the series to a finite order would be justified, thereby further reducing the theory to a truncated $\QET$ where locality is recovered.

For concreteness, suppose $M$ is just a circle of radius $R$. Then the Green's function that obeys \cref{eq:Ggreen} can be written
\begin{equation}
\label{eq:propaaa}
    G(x,y) = \frac{1}{2 m_\Phi} \frac{\cosh \left( m_\Phi (\pi R -| x-y| )\right)}{\sinh \left( m_\Phi \pi R \right)} = \frac{1}{2 m_{\Phi}} \sum_{k\in\mathbb{Z}} e^{- m_{\Phi} | x - y + 2\pi R \, k|},
\end{equation}
where on the circle $|x-y|< \pi R$. Notice that this propagator exhibits the familiar exponential decay with distance characteristic of massive fields.\footnote{For instance, in flat space, at long distances satisfying $|x-y|\gg m_\Phi^{-1}$ the propagator would behave as $G(x,y) \sim {|x-y|^{-d/2}}e^{-m_\Phi|x-y|}$. For $m_\Phi = 0$ it would be everywhere just the corresponding power law.} Since $G$ only depends on $|x-y|$, the couplings in \cref{eq:effco} are constants. These coupling constants vanish for odd $k$ by symmetry. For even $k$ they can be compactly written as finite series,
\begin{equation}
    g_k = 1- \frac{1}{\sinh \left(m_\Phi \pi R\right)} \sum_{l=1}^{k/2} \frac{\left(m_\Phi \pi R \right)^{2 l-1}}{(2 l-1)!}.
\end{equation}
As $k\to\infty$ the finite series above becomes precisely the Taylor expansion of $\sinh \left(m_\Phi \pi R\right)$, implying that these dimensionless coupling constants go to zero at higher orders. In particular, at large $k$ one finds the asymptotics
\begin{equation}
\label{eq:gkasymp}
    g_k \sim \frac{1}{k!} \frac{(m_\Phi \pi R)^k}{\sinh (m_\Phi \pi R)}.
\end{equation}
Having introduced a new scale into the problem, we see by comparing \cref{eq:gkasymp} to \cref{eq:fstqet} that the dimensionful parameter that is now important for the derivative expansion is $R$. Namely, for the expansion to be under control it must be the case that higher derivatives $f_\varphi^{(k)}$ are small in units of $R$. Since $R$ is the length scale of the system, it becomes manifest that this local expansion only makes sense if the physics is indeed dominated by local interactions.

The example above only led to a bilocal coupling in the exact $\QET$. Consider a slightly less trivial interaction term by making $\Phi$ quadratic \cite{skinnernotes},
\begin{equation}
    I_{\smalltext{int}}[\varphi,\Phi] = \frac{\lambda}{2} \int\displaylimits_M d^dx \, f_\varphi(x)\, \Phi(x)^2.
\end{equation}
Integrating by parts, the resulting path integral over $\Phi$ gives rise to a functional determinant,
\begin{equation}
    \int\displaylimits_M \mathcal{D}\Phi \, e^{ - I[\varphi,\Phi]} \propto e^{-I_\varphi[\varphi]} \det(- \nabla^2 + m_\Phi^2 + \lambda f_\varphi)^{-1/2},
\end{equation}
which up to constants gives the exact $\QET$ action formally as\footnote{We are implicitly assuming here that operators are trace class.}
\begin{equation}
\label{eq:intvp}
    \wtilde{I}[\varphi] = I_\varphi[\varphi] + \frac{1}{2} \tr\log(- \nabla^2 + m_\Phi^2 + \lambda f_\varphi).
\end{equation}
Under traces this gives the useful operator identity
\begin{equation}
\label{eq:opid}
    \tr\log(- \nabla^2 + m_\Phi^2 + \lambda f_\varphi) = \tr\log(- \nabla^2 + m_\Phi^2) + \tr\log( 1 + \lambda \, G \, f_\varphi ).
\end{equation}
The first term on the right-hand side is independent of $\varphi$ and thus gives an uninteresting contribution to the quantum effective action. The $\varphi$-dependent term admits the expansion
\begin{equation}
    \tr\log( 1 + \lambda \, G \, f_\varphi ) = -\sum_{n=1}^\infty \frac{(-\lambda)^n}{n} \tr \left( G \, f_\varphi \right)^n,
\end{equation}
where one can now explicitly write out the cyclic trace terms as
\begin{equation}
    \tr \left( G \, f_\varphi \right)^n = \int\displaylimits_{M^n} d^dx_1 \cdots d^dx_n \, G(x_n,x_1) \, f_\varphi(x_1) \, G(x_1,x_2) \, f_\varphi(x_{2}) \cdots G(x_{n-1},x_n) \, f_\varphi(x_n).
\end{equation}
Altogether, one finally arrives at the following action for the exact $\QET$:
\begin{equation}
\label{eq:nonono}
    \wtilde{I}[\varphi] = I_\varphi[\varphi] - \frac{1}{2} \sum_{n=1}^\infty \frac{(-\lambda)^n}{n} \int\displaylimits_{M^n} d^dx_1 \cdots d^dx_n \, \prod_{k=1}^n G(x_{k-1},x_{k}) \,  f_\varphi(x_k),
\end{equation}
with $x_{0}\equiv x_n$. We thus see that already a simple Gaussian functional integral gives rise to a strongly non-local action involving infinitely many multi-integrals over arbitrarily many copies of the original space $M$. 
As for the previous example, one could similarly consider performing a derivative expansion of \cref{eq:nonono} to simplify the theory in some regime. We leave this as an exercise for the reader.

\begin{figure}[ht]
    \centering
    \includegraphics[width=.9\textwidth]{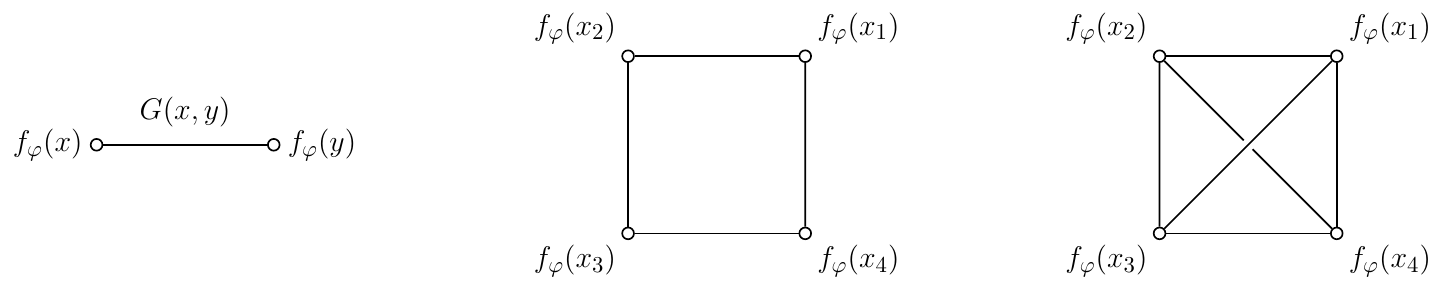}
    \caption{Feynman diagrams for the construction of the strongly non-local action of an exact $\QET$. We consider a complete theory with interaction terms of the form $f_\varphi \Phi^k$ for $k\in\mathbb{Z}^+$ and where $\Phi$ is integrated out. The multi-integrals in the action come from integrating all points in these diagrams over $M$.
    On the left, the only diagram that contributes for $k=1$, and which gives rise to the bilocal action in \cref{eq:nonloeg}. In the middle, the diagram which contributes the $n=4$ term in \cref{eq:nonono} for the theory with a $k=2$ interaction term. Similar loops with all possible numbers $n$ of vertices give all terms in the series in \cref{eq:nonono}. On the right, an example of a diagram which would also contribute a $4$-local interaction among $f_\varphi$ factors if the complete theory contained a $k=3$ interaction.}
    \label{fig:feynman}
\end{figure}

Let us conclude by emphasizing that the construction of such strongly non-local actions characteristic of any exact $\QET$ in fact admits a standard perturbative treatment \`a la Feynman. This is because the local derivative expansion that destroys strong non-locality has nothing to do with the standard expansions at small couplings of perturbation theory. Importantly, this means that strongly non-local actions can actually be constructed using standard Feynman diagrams for arbitrarily complicated interacting theories at weak coupling. In other words, the framework that is relevant to this paper is as broad as perturbative quantum field theory techniques apply, and not at all limited to simple cases where exact path integration is directly doable as in the examples above.
See \cref{fig:feynman} for a representation of the Feynman diagrams that arise in the construction of an exact $\QET$.

\section{State Preparation}
\label{sec:prep}

This section thoroughly studies the states that the path integrals of different theories naturally prepare, their general properties, and how they are related across theories.

\subsection{Complete States}
\label{ssec:cmst}
Let $X$ be a manifold with boundary $\partial X$. Imposing boundary conditions $\phi_0$ for the quantum fields on $\partial X$, the path integral on $X$ computes the wavefunctional
\begin{equation}
\label{eq:wvfnal}
    \Psi[\phi_0] \equiv \langle \phi_0 | \Psi \rangle = \frac{1}{\sqrt{Z}} \int\displaylimits_{X}^{\evalat{\phi}{\partial X}=\phi_0} \mathcal{D}\phi \, e^{-I[\phi]},
\end{equation}
where the normalization is given by \cref{eq:Zpi} with $M$ constructed by gluing two copies of $X$ along $\partial X$ (cf. \cref{fig:partitionfunctions} with $\partial X \sim \Sigma$).
Wavefunctionals corresponding to different states can be prepared by inserting distinct combinations of local operators in the path integral. The insertions that specify the state will be left implicit to avoid cluttering the notation. 
The wavefunctional in \cref{eq:wvfnal} is a representation of the state $\ket{\Psi}$ in the space of field configurations $\phi_0$ on $\partial X$.
By leaving the boundary conditions on $\partial X$ in \cref{eq:wvfnal} unspecified, one obtains the state $\ket{\Psi}$ itself as a representation-independent functional operator,
\begin{equation}
\label{eq:pure}
	\ket{\Psi} = \frac{1}{\sqrt{Z}} \int\displaylimits_{X} \mathcal{D}\phi \, e^{-I[\phi]}.
\end{equation}
The adjoint state $\bra{\Psi}$ can be obtained by conjugating the fields in the path integral and evaluating it on a CRT-conjugate copy $X^*$ of $X$. The Hilbert space $\mathcal{H}$ of the theory consists of the span of all quantum states $\ket{\Psi}$ prepared this way, endowed with the inner product that the path integral naturally defines,
\begin{equation}
\label{eq:normpr}
    \langle \Psi' | \Psi \rangle \equiv \int\displaylimits_{\partial X \sim\partial X^*} \mathcal{D} \phi_0 \, \langle \Psi' | \phi_0 \rangle \langle \phi_0 | \Psi \rangle.
\end{equation}
That states are appropriately unit-normalized can be easily seen by transitivity of the path integral. Namely, \cref{eq:normpr} involves independent path integrals over $X$ and $X^*$ with identified boundary conditions along $\partial X \sim \partial X^*$ which are also integrated over. This is equivalent to just doing the path integral on $M$ without cuts, which is by definition $Z$.\footnote{With appropriate insertions corresponding to the state being prepared, if any.} The following formal completeness relation can be read off from \cref{eq:normpr}:
\begin{equation}
\label{eq:completeness}
    \int\displaylimits_{\partial X} \mathcal{D} \phi_0 \ket{\phi_0} \bra{\phi_0} = \mathds{I}_{\partial X},
\end{equation}
which provides a resolution of the identity on $\partial X$. This relation itself can be taken as the definition of the inner product on $\mathcal{H}$. Furthermore, it follows from \cref{eq:normpr} that these configuration states obey the orthogonality condition\footnote{
To see this, note that by definition \cref{eq:wvfnal} can also be written as
$$
\langle \phi_0 | \Psi \rangle = \frac{1}{\sqrt{Z}} \int\displaylimits_{X} \mathcal{D}\phi \, \delta(\evalat{\phi}{\partial X}-\phi_0) \, e^{-I[\phi]}.
$$
Inserting \cref{eq:completeness} on the left-hand side and comparing to the right-hand side leads to \cref{eq:ortho}. In canonical quantization, this orthogonality relation is simply declared. In the path integral formulation, it is a consequence of the functional formalism.}
\begin{equation}
\label{eq:ortho}
    \langle \phi_0' | \phi_0 \rangle = \delta(\phi_0' - \phi_0).  
\end{equation}
Together with the completeness relation in \cref{eq:completeness}, this implies that the $\ket{\phi_0}$ states form an orthonormal basis for the Hilbert space $\mathcal{H}$ of states of the theory on $\partial X$.

\begin{figure}
    \centering
    \includegraphics[width=.3\textwidth]{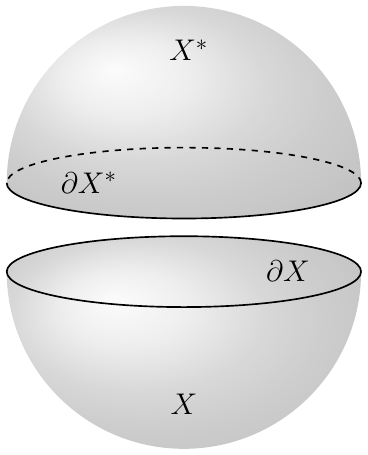}
    \caption{Representation of the path integral preparation of the density operator in \cref{eq:rhopure} by a complete theory. Because the theory is local, the path integral factorizes between the two dual spaces $X$ and $X^*$. The path integral on $X$ prepares $\Ket{\Psi}$ on $\partial X$, and the CRT-conjugate path integral on $X^*$ prepares $\Bra{\Psi}$ on $\partial X^*$. 
    The identification $\partial X\sim \partial X^*$ defines the space $M\cong X \cup X^*$ on which the path integral computes the trace of the density operator.}
    \label{fig:densityoperator}
\end{figure}

This formalism also allows one to directly prepare a density operator of the pure state by combining the preparation of a state and its dual into a single path integral,
\begin{equation}
\label{eq:rhopure}
	\ket{\Psi}\bra{\Psi} = \frac{1}{Z} \int\displaylimits_{X \sqcup X^*} \mathcal{D}\phi \, e^{-I[\phi]}.
\end{equation}
This is depicted in \cref{fig:densityoperator}. Starting from the partition function definition of the path integral as in \cref{fig:partitionfunctions}, the space $X \sqcup X^*$ can also be thought of as obtained by cutting open $M\cong X \cup X^*$ along $\Sigma\sim\partial X \sim \partial X^*$. Matrix elements of this operator are obtained by independently fixing boundary conditions for the quantum fields on $\partial X$ and $\partial X^*$. Correspondingly, the trace is computed by identifying these configurations between both boundaries and path integrating over them as in \cref{eq:normpr}.

\begin{figure}
    \centering
    \hspace{10pt}
    \includegraphics[height=.3\textwidth]{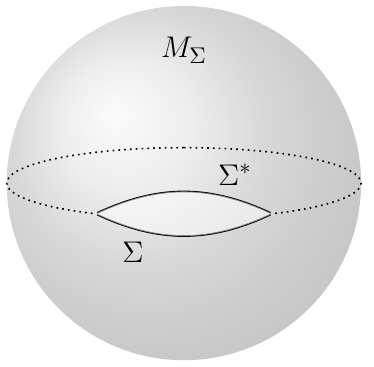}
    \hfill
    \includegraphics[height=.3\textwidth]{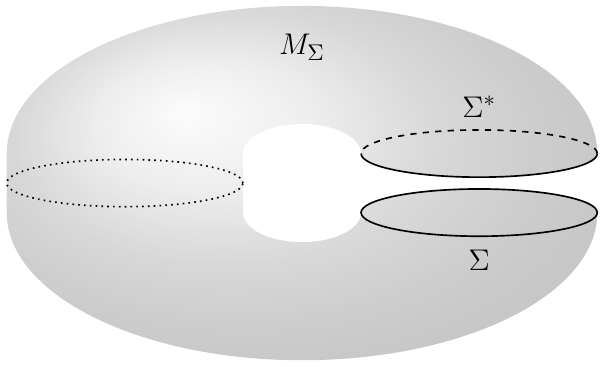}
    \hspace{10pt}
    ~
    \caption{Representation of the path integral on spaces $M_\Sigma$ obtained by cutting open $M$ along a non-splitting hypersurface $\Sigma$. The non-splitting $\Sigma$ on the left has $\partial \Sigma \neq \varnothing$, whereas on the right it is a cycle with $\partial \Sigma = \varnothing$. By leaving unspecified the boundary conditions for field configurations on the two sides of the cut $\Sigma$ and $\Sigma^*$, the path integral prepares a general density operator on $\Sigma$. In field theory, such a state is generally mixed.}
    \label{fig:nonsplitting}
\end{figure}

More generally, one may want to construct a state on some subspace $\Sigma\subseteq\partial X$. If $\Sigma = \partial X$ as above, it splits $M\cong X \cup X^*$ into two dual halves, so we refer to it as a \textit{splitting} hypersurface (cf. $\Sigma$ in \cref{fig:partitionfunctions}).\footnote{If the manifold were Lorentzian, we would refer to such a hypersurface as a complete Cauchy surface.}
If instead it is a proper subspace, $\Sigma \subset \partial X$, then cutting $M$ open along $\Sigma$ does not split it into two dual halves; rather, $X$ and $X^*$ remain glued along the portion of their boundaries outside $\Sigma$. This is illustrated in \cref{fig:nonsplitting}. In general, the space obtained by cutting open $M$ along $\Sigma$ will be denoted by $M_\Sigma$. Performing a path integral on $M_\Sigma$ can be easily seen to be equivalent to doing it on $X\sqcup X^*$, identifying the fields along $\Sigma \smallsetminus \partial X$ and $\Sigma \smallsetminus \partial X^*$, and integrating over configurations therein. This is nothing but the path integral implementation of a partial trace over configurations outside $\Sigma$, which is precisely what is needed to obtain the state of quantum fields on only $\Sigma$. Hence the path integral on $M_\Sigma$ prepares the state on $\Sigma$ as a reduced functional density operator which we write as\footnote{The operator algebra of a quantum field theory on splitting hypersurfaces $\Sigma$ is generally type-I von Neumann, which guarantees that pure states and traces like those given in \cref{eq:rhopure,eq:completeness} are indeed well-defined. For a non-splitting $\Sigma$ states will be mixed, but so long as $\Sigma$ itself has no boundaries, the algebra will be type II and admit a trace. Otherwise, if $\Sigma$ has boundaries the algebra is generally type III, which does not admit pure states, density operators, or even a trace. The reader is welcome to assume $\Sigma$ has no boundaries for all purposes of this paper, whether it is splitting or not.}
\begin{equation}
\label{eq:rhoexact}
    \rho = \frac{1}{Z} \int\displaylimits_{M_\Sigma} \mathcal{D}\phi \, e^{-I[\phi]}.
\end{equation}

In field theory this is generally a mixed state unless $\Sigma$ splits $M$.
Since we used the path integral of the complete theory defined in \cref{eq:Zpi} to prepare the state in \cref{eq:rhoexact}, we regard such a $\rho$ as a \textit{complete state}. Computing the trace of $\rho$ involves a path integral that now just runs over field configurations on $\Sigma\subset\partial X$, and a subspace completeness relation follows:
\begin{equation}
    \int\displaylimits_{\Sigma} \mathcal{D} \phi_0 \ket{\phi_0}\bra{\phi_0} = \mathds{I}_{{\Sigma}}.
\end{equation}
The orthogonality condition from \cref{eq:ortho} similarly holds on $\Sigma$. For future reference, we also quote the relations this would induce under the splitting of $\phi$ into $\IR$ and $\UV$ sectors,
\begin{equation}
\label{eq:inneriruv}
    \mathds{I}_{{\Sigma}} = \mathds{I}^{{\varphi}}_{{\Sigma}} \otimes \mathds{I}^{{\Phi}}_{{\Sigma}}, \qquad \int\displaylimits_{\Sigma} \mathcal{D} \varphi_0 \ket{\varphi_0}\bra{\varphi_0} = \mathds{I}^{{\varphi}}_{{\Sigma}}, \quad \int\displaylimits_{\Sigma} \mathcal{D} \Phi_0 \ket{\Phi_0}\bra{\Phi_0} = \mathds{I}^{{\Phi}}_{{\Sigma}}.
\end{equation}
where the Hilbert space of $\IR$ fields $\mathcal{H}_{{\varphi}}$ is spanned by $\ket{\varphi_0}$ states, and the Hilbert space of $\UV$ fields $\mathcal{H}_{{\Phi}}$ is spanned by $\ket{\Phi_0}$ states. The same argument that led to \cref{eq:ortho} for the complete states of the theory applies to the $\IR$ and $\UV$ states using the inner products from \cref{eq:inneriruv},
\begin{equation}
\label{eq:orthoiruv}
    \langle \varphi_0' | \varphi_0 \rangle = \delta(\varphi_0' - \varphi_0), \qquad \langle \Phi_0' | \Phi_0 \rangle = \delta(\Phi_0' - \Phi_0).
\end{equation}

\subsection{Exact Effective States}
\label{ssec:ees}

If interested only in the physics of $\IR$ fields, however, one may want to consider the preparation of states using just the $\QET$ whose action is given by \cref{eq:IEFT}. The state analogous to \cref{eq:rhoexact} that the $\QET$ prepares on $\Sigma$ can be obtained similarly by cutting open the path integral in \cref{eq:Zeft} along the same hypersurface $\Sigma$,\footnote{As emphasized in \cref{fn:couple}, if the $\QET$ action $\wtilde{I}$ depends on $\UV$ field insertions, so will this $\QET$ state.}
\begin{equation}
\label{eq:rhoeft}
    \wtilde{\rho} = \frac{1}{\wtilde{Z}} \int\displaylimits_{M_\Sigma} \mathcal{D}\varphi\, e^{-\wtilde{I}[\varphi]}.
\end{equation}
Since this state preparation involves the path integral of the exact $\QET$ defined in \cref{eq:Zeft}, we regard such a $\wtilde{\rho}$ as an \textit{exact $\QET$ state}.
Importantly, the matrix elements of $\wtilde{\rho}$ involve boundary conditions for the $\IR$ fields only; the $\UV$ fields were already integrated out in computing the quantum effective action $\wtilde{I}$. States of this kind define the Hilbert space of the exact $\QET$, and their inner products are taken by the path integral for $\mathds{I}^{{\varphi}}_{{\Sigma}}$ in \cref{eq:inneriruv}.

\begin{figure}
    \centering
    \includegraphics[width=.3\textwidth]{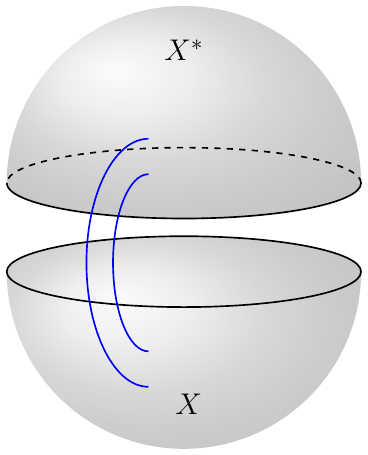}
    \caption{Representation of the path integral preparation of the density operator in \cref{eq:rhoeft} by an exact $\QET$. Here $\Sigma$ has been taken to be a splitting hypersurface giving $M_\Sigma \cong X \sqcup X^*$. Because the theory is strongly non-local, the path integral does not factorize, but involves quantum wormholes coupling the two dual spaces $X$ and $X^*$. As a result, what would be the `bra' and `ket' of the state are in fact coupled, meaning that the resulting state is generally mixed despite $\Sigma$ being splitting.}
    \label{fig:statewormholes}
\end{figure}

There is a very important lesson to be drawn from \cref{eq:rhoeft} about state preparation in strongly non-local theories, particularly when they descend from some complete theory as an exact $\QET$. For simplicity, here let $M$ be connected and suppose $\Sigma$ is not splitting, such that after cutting $M$ open, the space $M_\Sigma$ is still connected. All integrals inside the $\QET$ action $\wtilde{I}$ in \cref{eq:rhoeft} then clearly run over all of $M_\Sigma$. By continuously extending $\Sigma$ into a splitting hypersurface, one eventually reaches the disjoint limit in which $M_\Sigma \cong X \sqcup X^*$, for $X$ and $X^*$ the two dual half-spaces that make up $M$. Expecting continuity of the path integral in this limiting procedure, all integrals inside the $\QET$ action $\wtilde{I}$ will thus eventually have to run over both $X$ and $X^*$. Schematically, for a double integral, this limit corresponds to
\begin{equation}
\label{eq:nonfactor}
    \int\displaylimits_{M_{\Sigma}}\int\displaylimits_{M_{\Sigma}} (\,\cdot\,) = \int\displaylimits_{X}\int\displaylimits_{X} (\,\cdot\,) + \int\displaylimits_{X^*}\int\displaylimits_{X^*} (\,\cdot\,) + 2\int\displaylimits_{X}\int\displaylimits_{X^*} (\,\cdot\,).
\end{equation}
The appearance of the last term is remarkable: it couples the two disconnected spaces $X$ and $X^*$. This is our first encounter of quantum wormholes, which we illustrate in \cref{fig:statewormholes}.
This cross-term in \cref{{eq:nonfactor}} may seem surprising, but as will be seen turns out to be required by consistency of the framework.
The basic reason it must be there is that the functional integral over $\UV$ fields
that defined the exact $\QET$ action $\wtilde{I}$ involved no cuts on the space $M$. As a result, the strong non-localities already present in the theory at the level of $\wtilde{I}$ are insensitive to the $\Sigma$ cut. This in particular means that even if $\Sigma$ is splitting, the two dual halves $X$ and $X^*$ of the resulting disjoint space $M_\Sigma$ will remain coupled.

Correlations between the two dual spaces $X$ and $X^*$ in the preparation of a state are general indicators of mixing. One should thus suspect that quantum wormholes may cause exact $\QET$ density operators to be mixed.
The linearity of the path integrals involved in \cref{eq:rhoeft} allow for a suggestive rewriting that confirms this intuition:
\begin{equation}
\label{eq:tildeout}
    \wtilde{\rho} = \frac{1}{\wtilde{Z}} \int\displaylimits_{M} \mathcal{D}\Phi \int\displaylimits_{M_\Sigma} \mathcal{D}\varphi \, e^{-I[\varphi,\Phi]},
\end{equation}
where we have simply used the definition of $\wtilde{I}$ from \cref{eq:IEFT}.
In particular, one can easily see that, altogether, \cref{eq:tildeout} is the result of splitting the path integral in \cref{eq:rhoexact} into $\IR$ and $\UV$ fields, and computing the one over the latter with $\Phi$ identified across the cut. From this point of view, $\wtilde{\rho}$ is the reduced density operator obtained from a state $\rho$ by performing a partial trace over $\UV$ fields. Hence \cref{eq:rhoeft} can be interpreted as\footnote{In the context of Wilsonian renormalization in momentum space, this observation was also made in \cite{Balasubramanian:2011wt}, where an explicit perturbative calculation of the entanglement entropy of the low-energy reduced density operator was performed for various scalar theories.}
\begin{equation}
\label{eq:ffF}
    \wtilde{\rho} = \Tr_{{\Phi}} \rho,
\end{equation}
where the trace over the $\UV$ sector is defined by the path integral for $\mathds{I}^{{\Phi}}_{{\Sigma}}$ in \cref{eq:inneriruv}.
The structure and consequences of this trace will be the subject of \cref{ssec:ensemble}.
For now, \cref{eq:ffF} already demystifies why we should expect quantum wormholes to generally cause exact $\QET$ states to be mixed even when $\Sigma$ is splitting.
The identification of $\wtilde{\rho}$ as a reduced density operator for the $\IR$ fields will offer useful insights into the behavior of state replicas in \cref{sec:reps} and its operational distinguishability from states prepared by non-exact $\QET$ path integrals in \cref{sec:algebras}.

\subsection{Approximate Effective States}
\label{ssec:appeftst}

A reasonable question at this point is whether the inclusion of quantum wormholes in the preparation of a state is optional.\footnote{The quantum gravity reader may find this reminiscent of the analogous question of whether or not to include gravitational bra-ket wormholes in the preparation of the quantum state of the universe \cite{Hartle:1983ai,Hawking:1983hj,Page:1986vw,Anous:2020lka,Chen:2020tes}.} Here we address what the consequences are of choosing not to include quantum wormholes.

Suppose one took an exact $\QET$ and decided to treat it as if it were the full theory of a system, i.e., as if it were a (strongly non-local) complete theory. Then one could consider constructing two completely separate pure states as in \cref{eq:pure}. By dualizing one of them and then putting them together into a functional density operator as in \cref{eq:rhopure}, one would obtain a pure state which we will call $\rho_{\times}$. In the path integral that prepares $\rho_{\times}$, the strongly non-local action would not involve any multi-integrals coupling $X$ and $X^*$. For the double integral example in \cref{eq:nonfactor}, the last cross-term would be absent. 
Following this recipe, one is essentially preparing a pure state by truncating away quantum wormholes.

One may thus wonder how the pure state $\rho_{\times}$ is related to the mixed state $\wtilde{\rho}$ of the actual exact $\QET$, both prepared by seemingly the same theory. In particular, how did $\rho_{\times}$ come to be a pure state of the $\IR$ fields despite their generic correlations with $\UV$ fields in the complete theory? The answer is that in changing the rules of the path integral, one is in fact defining a strictly different theory. The resulting theory is precisely the outcome of erasing quantum wormholes from the exact $\QET$. As explained in \cref{ssec:approx} (cf. \cref{fn:resum}), this is basically what happens if one performs a local derivative expansion and then decides to resum it back. Put differently, this new theory that prepares pure states like $\rho_{\times}$ is secretly no more than an approximate $\QET$ behaving like a weakly non-local theory with no quantum wormholes.\footnote{As a weakly non-local theory in disguise, the path integral inner product of this new theory would behave rather unconventionally. Namely, upon sewing $X\sqcup X^*$ by integrating over boundary conditions of fields identified across $\Sigma$, the multi-integrals would continue to run only over either $X$ or $X^*$ independently. This would be inequivalent to performing a full strongly non-local path integral over $X\cup X^*$ without cuts.}

Let us dissect further the preparation of these states to ascertain how much information is being lost by the approximate $\QET$. A crucial realization is that strong non-localities are very important even when $\Sigma$ is not splitting. To see this, suppose one decided to start by reducing an exact $\QET$ to an approximate $\QET$ in which the non-localities are only caused by derivatives of the fields of arbitrarily high order. As emphasized above, for a splitting $\Sigma$ there would be no way of making sense of cross terms like those in \cref{eq:nonfactor}. However, more generally, even if a state is being prepared on a non-splitting $\Sigma$, such higher derivatives would be unable to couple fields right across the cut. In the exact $\QET$ these are genuinely coupled: indeed, from the perspective of $\UV$ fields, points right across the cut are in a local neighborhood of each other. Once again, this is because the exact $\QET$ is obtained by integrating out the $\UV$ fields without cuts, so the non-localities these induce persist across $\Sigma$ as if the manifold had not been cut open (because for the $\UV$ fields it was not).
All this information is completely lost once one performs a derivative expansion of the strongly non-local theory, and then cuts open the path integral.

In intuitive terms, the exact $\QET$ still remembers the $\Phi$ sector, whereas the approximate $\QET$ has forgotten much of it. This is because in the preparation of $\wtilde{\rho}$ in \cref{eq:rhoeft}, the path integral is still treating the complete theory with action $\wtilde{I}$ as the full theory, and the states being prepared know they are missing information about the $\UV$ fields. This is how the exact $\QET$ path integral genuinely works, and mixed states like $\wtilde{\rho}$ are the states the exact $\QET$ path integral naturally prepares.\footnote{We remain agnostic as to whether there are pure states in an exact $\QET$. For splitting hypersurfaces, the algebra of operators of an exact $\QET$ may be of type II, in contrast with the type I algebra of a complete theory (cf. a complete theory on a non-splitting hypersurface with no boundary, as on the right of \cref{fig:nonsplitting}).}
In contrast, in preparing $\rho_{\times}$, the path integral is instead treating the exact $\QET$ with action $\wtilde{I}$ as the full theory. In so doing, the only degrees of freedom this path integral accounts for are those associated to $\IR$ fields, and the $\UV$ fields from the old complete theory no longer have any imprint on the states the path integral prepares. As a result the state $\rho_{\times}$ of $\IR$ fields on a splitting $\Sigma$ hypersurface is a perfectly valid pure state of this approximate $\QET$.

\ignore{
A word on what distinguishes the Hilbert spaces between the theories is clarifying. In the exact $\QET$, the strongly non-local interactions appearing in the action as multi-integrals over the full space

. In the path integral these capture global non-localities arising from integrating out $\UV$ fields. These interactions are solely responsible for the mixing of states like $\wtilde{\rho}$ when prepared on splitting hypersurfaces, where the state would have been pure had it been prepared by the complete theory. In other words, the exact $\QET$ prepares states in the Hilbert space of the complete theory, and then partially traces over the $\UV$ sector, as explicit from \cref{eq:ffF}. In contrast, the state $\rho_{\times}$ is the result of first truncating away the sector of the exact $\QET$ which involves strongly non-local couplings across connected components coming from the integration of $\UV$ fields. In doing so, the approximate $\QET$ actually prepares states in a Hilbert subspace, obtained by truncating away such non-local interactions first. The couplings that were responsible for the mixing of $\wtilde{\rho}$ in the exact $\QET$ are no longer observable in this reduced Hilbert space. The state $\rho_{\times}$ that the approximate $\QET$ prepares on this reduced Hilbert space is, as a result, a pure. These statements are made more precise in terms of the algebras of operators of the different theories in \cref{sec:algebras}.
}

\subsection{Ensembles of States}
\label{ssec:ensemble}

In trying to learn about complete theories, our primary interest is in understanding exact $\QET$ states $\wtilde{\rho}$ and how the exact $\QET$ path integrals that prepare them behave. The goal of this section is to first obtain a representation of $\wtilde{\rho}$ as a classical mixture of pure states of the exact $\QET$, and then use these to identify ensembles of pure states of the complete theory, all of which yield the same $\wtilde{\rho}$ upon partial traces. This exercise will crystallize the sense in which exact $\QET$ states correspond to no single complete state of the theory, but to ensembles thereof.

Recall that we are denoting the Hilbert space of $\varphi$ fields of the exact $\QET$ by $\mathcal{H}_{{\varphi}}$, and that of $\Phi$ fields of the complete theory by $\mathcal{H}_\Phi$.
As usual, a mixed state on a given Hilbert space admits expansions into pure states on the same Hilbert space. These states may or may not be orthogonal, and in general this expansion is highly non-unique. Suppose $\wtilde{\rho}_{{*}}[\alpha]$ is an orthogonal basis of pure $\QET$ states for $\mathcal{H}_{{\varphi}}$ on $\Sigma$ labeled by some function $\alpha$. By definition these must obey the orthogonality and completeness relations
\begin{equation}
\label{eq:orthoeft}
    \Tr \left(\wtilde{\rho}_{{*}}[\alpha]\wtilde{\rho}_{{*}}[\alpha']\right) = \delta(\alpha- \alpha'), \qquad \int \mathcal{D}\alpha \, \wtilde{\rho}_{{*}}[\alpha] = \mathds{I}_{\Sigma}^{{\varphi}},
\end{equation}
for some appropriate formal measure over $\alpha$. On the left, the trace is over  $\mathcal{H}_{{\varphi}}$, and is defined by the path integral over $\IR$ fields in \cref{eq:inneriruv}. The desired expansion of a mixed, exact $\QET$ state $\wtilde{\rho}$ in terms of the $\wtilde{\rho}_{{*}}$ states then takes the form
\begin{equation}
\label{eq:mixpurex}
    \wtilde{\rho} = \int \mathcal{D}\alpha \, P_*[\alpha]  \, \wtilde{\rho}_{{*}}[\alpha],
\end{equation}
where the probability density ${P}_*$ can be obtained by the orthogonality property in \cref{eq:orthoeft},
\begin{equation}
\label{eq:alphatr}
	{P}_*[\alpha] = \Tr\left( \wtilde{\rho}_{{*}}[\alpha] \, \wtilde{\rho} \right).
\end{equation}
Put differently, $\wtilde{\rho}$ can be interpreted as an average over an ensemble of $\wtilde{\rho}_{{*}}[\alpha]$ states with probability density function ${P}_*[\alpha]$. Of course, this density depends on the choice of spanning $\wtilde{\rho}_{{*}}$ states, which here has been arbitrary. However, as it turns out, a canonical choice of pure states spanning $\wtilde{\rho}$ descends rather naturally from the state of the complete theory.

Such a canonical choice can be constructed by exploiting the reduced density operator form of $\wtilde{\rho}$ in \cref{eq:ffF}. Making \cref{eq:tildeout} even more explicit, the functional form of the partial trace over $\UV$ fields reads
\begin{equation}
\label{eq:expantilde}
    \wtilde{\rho} = \int\displaylimits_{\Sigma} \mathcal{D}\Phi_0 \, \bra{\Phi_0} \rho \ket{\Phi_0}, \qquad \bra{\Phi_0} \rho \ket{\Phi_0} = \frac{1}{\wtilde{Z}} \, \int\displaylimits_{M_\Sigma} \mathcal{D}\varphi\, \int\displaylimits_{M}^{\evalat{\Phi}{\Sigma} = \Phi_0} \mathcal{D}\Phi \, e^{-I[\varphi,\Phi]}.
\end{equation}
In obtaining the second expression, we have simply noticed that the projection $\bra{\Phi_0}\rho\ket{\Phi_0}$ corresponds to a constrained preparation of the state $\rho$ as in \cref{eq:rhoexact} with the $\UV$ fields clamped onto the configuration $\Phi_0$ on $\Sigma$. Appropriately normalized, each such projection of $\rho$ naturally defines a state on the $\IR$ sector:
\begin{equation}
\label{eq:irstates}
    \wtilde{\rho}[\Phi_0] \equiv \frac{\wtilde{Z}}{Z^*_{\Phi_0}} \bra{\Phi_0} \rho \ket{\Phi_0} = \frac{1}{Z^*_{\Phi_0}} \int\displaylimits_{M_\Sigma} \mathcal{D}\varphi\, \int\displaylimits_{M}^{\evalat{\Phi}{\Sigma} = \Phi_0} \mathcal{D}\Phi \, e^{-I[\varphi,\Phi]},
\end{equation}
where the normalization factor can be easily seen to be
\begin{equation}
	\label{eq:zphi0}
    Z^*_{\Phi_0} \equiv \int\displaylimits_{M} \mathcal{D}\varphi\, \int\displaylimits_{M}^{\evalat{\Phi}{\Sigma} = \Phi_0} \mathcal{D}\Phi \, e^{-I[\varphi,\Phi]}.
\end{equation}
The states defined by \cref{eq:irstates} are states on $\mathcal{H}_{{\varphi}}$ for the $\IR$ fields, and will henceforth be referred to as \textit{projected states}.

We would like to identify these projected states as the canonical choice for an expansion like \cref{eq:mixpurex} of an exact $\QET$ state. In particular, we are interested in the case in which the state is mixed solely because the theory that prepares it is an exact $\QET$, and not the complete theory of the system. In other words, we want to understand the expansion of $\wtilde{\rho}$ on hypersurfaces $\Sigma$ which are splitting, and on which the complete state $\rho$ would be pure. For the projected states to be such a canonical choice, one just needs to check that they indeed are a set of pure states spanning $\wtilde{\rho}$. In fact, it will also be possible to think of them as an orthogonal basis in a precise sense from the perspective of the $\UV$ sector.

That they span $\wtilde{\rho}$ trivially follows from their definition in \cref{eq:irstates} applied to \cref{eq:expantilde}. 
Regarding purity, note that by construction these states are obtained by projections of the $\UV$ sector. In general, if $\rho$ is a pure state of the complete theory, so will be any projection, which includes these projected states. Although this argument suffices, understanding how this happens from the path integral preparation in \cref{eq:irstates} is informative. If the projection were not there, this would precisely correspond to the preparation of an exact $\QET$ state as in \cref{eq:rhoeft}, which generally is mixed due to correlations between $\IR$ and $\UV$ fields. What the projection does is destroy correlations between $\IR$ and $\UV$ fields from the get go. More precisely, for a splitting $\Sigma$, the projection on $\Sigma$ causes the path integral over $\UV$ fields to factorize into separate path integrals over each half-space of $M$. Hence, if this path integral is performed first, no quantum wormholes are generated between the two dual spaces into which $M$ is being cut open. The result is thus the preparation of two state vectors dual to each other.\footnote{The argument for purity from the perspective of the complete theory is trivial and uninformative: \cref{eq:irstates} is obviously pure as it corresponds to preparing a complete state as in \cref{eq:rhopure}, but with partially specified boundary conditions on the sector of $\UV$ fields.} Explicitly, following the notation in \cref{eq:pure} and below, if $\rho=\ket{\Psi}\bra{\Psi}$, projected states as functional vectors in $\mathcal{H}_{{\varphi}}$ read
\begin{equation}
\label{eq:pureirs}
    \ket{\psi}_{\Phi_0} \equiv \frac{\langle \Phi_0 | \Psi \rangle}{\sqrt{Z^*_{\Phi_0}}} = \frac{1}{\sqrt{Z^*_{\Phi_0}}} \int\displaylimits_{X} \mathcal{D}\varphi\, \int\displaylimits_{X}^{\evalat{\Phi}{\partial X} = \Phi_0} \mathcal{D}\Phi \, e^{-I[\varphi,\Phi]},
\end{equation}
which manifestly demonstrates their purity. It thus follows that the states $\wtilde{\rho}[\Phi_0]$ defined in \cref{eq:irstates} form the desired canonical spanning set of pure states for the exact $\QET$ state in \cref{eq:expantilde}. Explicitly, the decomposition of the mixed state $\wtilde{\rho}$ in \cref{eq:mixpurex} takes the form
\begin{equation}
	\label{eq:decomprho}
	\wtilde{\rho} = \int\displaylimits_{\Sigma} \mathcal{D}\Phi_0 \, {{{P}}}[\Phi_0] \, \wtilde{\rho}[\Phi_0], \qquad {{{P}}}[\Phi_0] \equiv \frac{Z^*_{\Phi_0}}{\wtilde{Z}}.
\end{equation}
where $Z^*_{\Phi_0}$ was defined in \cref{eq:zphi0}. 
This realizes the exact $\QET$ state $\wtilde{\rho}$ as an ensemble average over $\wtilde{\rho}[\Phi_0]$ states with probability density ${{{P}}}[\Phi_0]$, and motivates introducing the following notation for expectation values in this ensemble:
\begin{equation}
\label{eq:expval}
	\langle \,\cdot\,\rangle_{{{{P}}}} \equiv \int\displaylimits_{\Sigma} \mathcal{D}\Phi_0 \, {{{P}}}[\Phi_0] \, (\,\cdot\,),
\end{equation}
which note is indeed correctly normalized to give $\langle 1 \rangle_{{{{P}}}} = 1$. Using this, \cref{eq:decomprho} reads
\begin{equation}
\label{eq:ensemrho}
    \wtilde{\rho} = \left\langle \wtilde{\rho}[\Phi_0] \right\rangle_{{{{P}}}}.
\end{equation}
As quoted, \cref{eq:ensemrho} holds in fact for states on any $\Sigma$ hypersurface. If $\Sigma$ is splitting, then \cref{eq:ensemrho} indeed  is a decomposition of the mixed exact $\QET$ state into pure projected states. If instead $\Sigma$ is not splitting, then this is a decomposition into mixed states, but whose mixing is solely caused by spatial correlations of quantum fields (both $\IR$ and $\UV$) between $\Sigma$ and its complementary subsystem. This way, for general $\Sigma$, \cref{eq:decomprho} succeeds in decomposing $\wtilde{\rho}$ into a family of states with no correlations between $\IR$ and $\UV$ fields. 

While \cref{eq:decomprho} takes the same form as \cref{eq:mixpurex}, the projected states do not enjoy an orthogonality property like \cref{eq:orthoeft} on $\mathcal{H}_{{\varphi}}$, and thus their ${{{P}}}$ density measure cannot be extracted as in \cref{eq:alphatr}. In general the failure of orthogonality only allows for the bound $\Tr (\wtilde{\rho}[\Phi_0]\wtilde{\rho}) \geq {{{P}}}[\Phi_0]$.
The measure ${{{P}}}[\Phi_0]$ is more naturally associated to $\UV$ field configurations $\Phi_0$, which do obey the orthogonality condition in \cref{eq:orthoiruv}. Our projected states on $\mathcal{H}_{{\varphi}}$ can be lifted back to states acting on the full Hilbert space of the complete theory $\mathcal{H}$ in arbitrary ways, such as
\begin{equation}
\label{eq:liftrhoxi}
	\rho_\xi[\Phi_0] \equiv \wtilde{\rho}[\Phi_0]\otimes \ket{\xi_{\Phi_0}}\bra{\xi_{\Phi_0}}, \qquad \wtilde{\rho}[\Phi_0] = \Tr_{{\Phi}} \rho_\xi[\Phi_0],
\end{equation}
for any pure state $\ket{\xi_{\Phi_0}}$ in $\mathcal{H}_{{\Phi}}$.
Since \cref{eq:liftrho} holds for any family of states $\ket{\xi_{\Phi_0}}$ labeled by $\Phi_0$, we indeed see that this provides an infinite family of ensembles of pure states of the complete theory, all of which can be combined to reproduce the exact same quantum state $\wtilde{\rho}$ on $\mathcal{H}_{{\varphi}}$. There is though an obvious canonical choice corresponding to $\xi_{\Phi_0}=\Phi_0$ and which we denote by
\begin{equation}
\label{eq:liftrho}
	\Omega[\Phi_0] \equiv \wtilde{\rho}[\Phi_0]\otimes \ket{{\Phi_0}}\bra{{\Phi_0}}.
\end{equation}
This lifting does not invert the projection of $\rho$ in \cref{eq:expantilde} (i.e., $\Omega\neq\rho$ in general); rather, it identifies a canonical ensemble of states of the complete theory with orthogonal $\UV$ field configurations which precisely reproduces the exact $\QET$ state $\wtilde{\rho}$ in its $\IR$ sector.
Using \cref{eq:expval}, this ensemble of states we are referring to can be written
\begin{equation}
\label{eq:omegaensem}
    \Omega \equiv \left\langle \Omega[\Phi_0] \right\rangle_{{{{P}}}}.
\end{equation}
Equipped with this notation, we reproduce the statistics of the $\IR$ sector that the exact $\QET$ state captures as coming from a classical ensemble via (cf. \cref{eq:expantilde})
\begin{equation}
\label{eq:classomega}
    \wtilde{\rho} = \Tr_{{\Phi}} \Omega.
\end{equation}
In addition, because the $\Omega[\Phi_0]$ states are orthogonal on $\mathcal{H}_{{\Phi}}$, the ${{{P}}}$ density measure can now be extracted from the ensemble by projection onto the $\UV$ sector,
\begin{equation}
\label{eq:alphasec}
	{{{P}}}[\Phi_0] = \Tr \left( \ket{\Phi_0}\bra{\Phi_0} \, \Omega \right),
\end{equation}
where we are using that projected states are properly normalized to $\Tr \wtilde{\rho}[\Phi_0] = 1$.

Instead of recovering orthogonality by lifting projected states to act on $\mathcal{H}$, one may consider constructing an orthogonal set out of them. For a given $\wtilde{\rho}$, the projected states $\wtilde{\rho}[\Phi_0]$ may only act on a subspace of $\mathcal{H}_\varphi$. However, more generally, the set of all $\wtilde{\rho}[\Phi_0]$ for all possible exact $\QET$ states $\wtilde{\rho}$ is guaranteed to act on all of $\mathcal{H}_\varphi$. Here we will focus on constructing an orthogonal set only using projected states $\wtilde{\rho}[\Phi_0]$ for a given $\wtilde{\rho}$. We would like to diagonalize the subspace $\wtilde{\mathcal{H}} \subseteq \mathcal{H}_\varphi$ on which these act. By construction, in general $\dim \wtilde{\mathcal{H}}$ is upper bounded by the smallest of $\dim \mathcal{H}_\varphi$ and $\dim \mathcal{H}_\Phi$. If $\dim \wtilde{\mathcal{H}} < \dim \mathcal{H}_\Phi$, the projected states will be guaranteed to form an over-complete basis for $\wtilde{\mathcal{H}}$. In other words, there will be combinations of them which give null states,
\begin{equation}
\label{eq:nulls}
    \int\displaylimits_{\Sigma} \mathcal{D}\Phi_0 \, \nu[\Phi_0] \, \wtilde{\rho}[\Phi_0] = 0,
\end{equation}
for certain choices of functionals $\nu$. The space of null states corresponds to the kernel of the map $\Omega[\Phi_0]\to\wtilde{\rho}[\Phi_0]$, whose dimension will be $\dim \mathcal{H}_\Phi-\dim \wtilde{\mathcal{H}}$ or zero, whichever is greater.
Combinations of projected states can also be used to form an orthonormal basis for $\wtilde{\mathcal{H}}$. To make contact with the notation in \cref{eq:orthoeft,eq:mixpurex,eq:alphatr}, we write these as
\begin{equation}
\label{eq:chista}
    \wtilde{\rho}_{*}[\alpha] \equiv \int\displaylimits_{\Sigma} \mathcal{D}\Phi_0 \, \alpha[\Phi_0] \, \wtilde{\rho}[\Phi_0].
\end{equation}
In other words, we are identifying $\alpha$ as the functionals giving combinations of the projected states obeying the orthogonality relation in \cref{eq:orthoeft}. The states in \cref{eq:chista} for a fixed $\wtilde{\rho}$ may only obey a completeness relation on $\wtilde{\mathcal{H}}$, but that is all we need. As combinations of projected states, the $\wtilde{\rho}_{*}[\alpha]$ states can be thought of as coming from a superposition of states of $\Phi$ fields with different configurations on $\Sigma$. Because they form an orthonormal basis of operators on $\wtilde{\mathcal{H}}$, they can always be identified as pure states on that subspace. This notion of purity corresponds to purity on the exact $\QET$ Hilbert space $\mathcal{H}_\varphi$ when $\dim\mathcal{H}_\Phi \geq \dim\mathcal{H}_\varphi$ and $\wtilde{\mathcal{H}}\cong \mathcal{H}_\varphi$, but is also guaranteed by purity of the projected states if $\Sigma$ is splitting even if $\wtilde{\mathcal{H}}\subset\mathcal{H}_\varphi$.
The expansion for $\wtilde{\rho}$ with density measure ${P}_*[\alpha]$ over $\wtilde{\rho}_{*}[\alpha]$ states can be related to the expansion with density measure ${{{P}}}[\Phi_0]$ over $\wtilde{\rho}[\Phi_0]$ states using \cref{eq:alphatr}, obtaining

\begin{equation}
\label{eq:palpha}
	{P}_*[\alpha] = \int\displaylimits_{\Sigma} \mathcal{D}\Phi_0 \, \int\displaylimits_{\Sigma} \mathcal{D}\Phi_0' \, \alpha[\Phi_0] \, {{{P}}}[\Phi_0'] \, \Tr\left( \wtilde{\rho}[\Phi_0] \, \wtilde{\rho}[\Phi_0']\right).
\end{equation}

Altogether, the discussion in this section allows for a precise identification of the exact $\QET$ state $\wtilde{\rho}$ as descending from an ensemble average over orthogonal states $\Omega[\Phi_0]$ of the complete theory with probability density functional ${{{P}}}[\Phi_0]$, or as an ensemble average over orthogonal states $\wtilde{\rho}_{*}[\alpha]$ with probability density functional ${{{P}}}_*[\alpha]$. 
In the former interpretation, we are taking the viewpoint that there is a unique complete theory, and that the ensemble captures the failure of the $\QET$ to distinguish $\UV$ field configurations. In the latter, we are simply using the complete theory to select a natural over-complete basis for the exact $\QET$ states, and identifying sets of null states and spanning states.
These constructs are the field-theoretic analogue to $\alpha$-states in the wormhole literature, as discussed in \cref{ssec:effholo}.

\subsection{Ensembles of Theories}
\label{ssec:theoryensem}

An interesting alternative perspective arises if one decides to begin by declaring a family of theories functionally parameterized by $\Phi_0$, with actions defined by
\begin{equation}
\label{eq:uvcompi}
    \frac{1}{\wtilde{Z}} e^{-I_{\Phi_0}[\varphi]} \equiv \frac{1}{Z^*_{\Phi_0}}\int\displaylimits_{M}^{\evalat{\Phi}{\Sigma} = \Phi_0} \mathcal{D}\Phi \, e^{-I[\varphi,\Phi]},
\end{equation}
where $\wtilde{Z}$ and $Z^*_{\Phi_0}$ appear as normalization constants for fixed $\Phi_0$.
Because of their obvious relation to the projected states defined in \cref{eq:irstates}, these theories will be referred to as \textit{projected theories}.
In terms of this family of theories, the exact $\QET$ state is prepared by
\begin{equation}
\label{eq:theoryensem}
    \wtilde{\rho} = \frac{1}{\wtilde{Z}} \int\displaylimits_{M_{\Sigma}} \mathcal{D}\varphi\, 
    \left\langle e^{-I_{\Phi_0}[\varphi]} \right\rangle_{{{{P}}}}.
\end{equation}
This expression allows one to interpret $\wtilde{\rho}$ as being prepared by an ensemble of projected theories, rather than by any single theory. Correspondingly, the $\QET$ partition function takes the form of an ensemble average over partition functions of these theories,
\begin{equation}
\label{eq:theoryzsn}
    \wtilde{Z} = \left\langle Z_{\Phi_0} \right\rangle_{{{{P}}}}, \qquad Z_{\Phi_0} \equiv \int\displaylimits_{M} \mathcal{D}\varphi \, e^{-I_{\Phi_0}[\varphi]}.
\end{equation}
The objects $Z_{\Phi_0}$ in \cref{eq:theoryzsn} and $Z^*_{\Phi_0}$ in \cref{eq:uvcompi} should not be confused.
In particular, $Z^*_{\Phi_0}$ is just a normalization constant that is evaluated and set once and for all when the projected theory action $I_{\Phi_0}[\varphi]$ is defined. In contrast, the partition function $Z_{\Phi_0}$ defines a functional integral over $\varphi$ fields on $M$. In fact, in the absence of insertions, note that numerically $Z_{\Phi_0}$ evaluates to $\wtilde{Z}$, not $Z^*_{\Phi_0}$. With insertions, however, the path integral that $Z_{\Phi_0}$ defines gives results which depend on $\Phi_0$ nontrivially, whereas that of $\wtilde{Z}$ of course does not.

A projected theory with partition function $Z_{\Phi_0}$ may be regarded as a complete theory in its own right. It is not an exact $\QET$: the operation being performed on the $\UV$ fields in \cref{eq:uvcompi} is a projection onto a specific field configuration state $\ket{\Phi_0}$ on $\Sigma$, not a partial trace over states on $\Sigma$.
This theory is strongly non-local because of the preparation of the state of the $\Phi$ fields, but it has no quantum wormholes across $\Sigma$ due to the projection onto $\Phi_0$. Hence these projected theories naturally prepare pure states on splitting hypersurfaces, which are no other than the projected states in \cref{eq:irstates}.

In terms of state preparation, these projected theories are complete theories for the $\varphi$ fields in a superselection sector of the original theory where the configuration state of the $\Phi$ fields on $\Sigma$ is given by $\Phi_0$. This state is superselected in the sense that no matter what operators one decides to insert into the path integral, the theory stays within the sector in which $\evalat{\Phi}{\Sigma} = \Phi_0$. This notion of superselection is entirely analogous to that of $\alpha$-states and baby universes in quantum gravity, as we elaborate on in \cref{sssec:babies}.
From the perspective of the original complete Euclidean theory on $M$, these projected theories are the result of postselecting on the state of $\Phi$ fields on the $\Sigma$ hypersurface. Indeed, the path integral is not preparing a state that happens to freely evolve deterministically to the configuration $\Phi_0$ for the $\Phi$ fields on $\Sigma$; rather, this configuration is enforced by postselection.

Optionally, one could consider performing a derivative expansion of the strong non-localities of a projected theory and a truncation down to a local theory. The ensemble average analogous to \cref{eq:theoryzsn} over such truncated projected theories would give rise to an averaged truncated $\QET$. At the level of the truncated $\QET$ action, the coupling constants would thus acquire a statistical meaning as coming from an ensemble of theories with different coupling constants. Every theory in the ensemble comes from a different state of the $\Phi$ fields of the original complete theory, and thus with an associated probability of being measured by an observer. This literally implies an indeterminacy in the couplings an observer would measure in trying to describe the complete theory in terms of a truncated $\QET$. More strikingly, in general the states of the $\Phi$ fields will have no symmetries and be position dependent. Correspondingly, one should expect a poor description of the physics of the complete theory if the couplings are assumed to be constant across space.
Observations of this flavor have also been made in the context of effective theories in quantum gravity \cite{Hawking:1982dj,Giddings:1988cx,Coleman:1988cy,Hawking:1991vs,Balasubramanian:2020lux,Giddings:2020yes}. In string theory phenomenology it is well understood that low-energy couplings are really expectation values of dynamical background fields. Here we see these as general $\QET$ features, with no need to appeal to gravity or string theory.

In conclusion, we find that the theory ensemble interpretation of the exact $\QET$ that \cref{eq:theoryzsn} provides and the previous state ensemble interpretation of the exact $\QET$ states from \cref{eq:classomega} are two sides of the same coin.\footnote{See e.g. \cite{Pollack:2020gfa,Freivogel:2021ivu} for a quantum gravity realization of this idea.} Summarizing: the exact $\QET$ is an ensemble of projected theories, the exact $\QET$ states are ensembles of projected states, and the connection between these two facts is that projected states are precisely the states that the path integral of the projected theory prepares.

\section{Replica Path Integrals}
\label{sec:reps}

A sharp distinguishing feature among the theories in \cref{sec:prep} is whether and why the states they prepare are pure or mixed. This section is devoted to quantifying this in terms of entropic measures which are computable via path integrals, and thus further elucidate how exact $\QET$ path integrals work.

\subsection{Entropic Measures}
\label{ssec:entrop}

From \cref{eq:ffF}, we see that the $\QET$ state $\wtilde{\rho}$ is the reduced density operator obtained by tracing out the $\UV$ fields from $\rho$. From a quantum mechanical perspective, one would like to know how much information about the $\IR$ fields in the complete state $\rho$ is still retained in $\wtilde{\rho}$ intrinsically within $\IR$ fields. This depends on the amount of correlations between the $\IR$ and $\UV$ fields, particularly their shared entanglement.
Information theoretic measures of such correlations typically involve the calculation of entropies.

Given a system with density operator $\sigma$, a quantity of interest is its von Neumann entropy,
\begin{equation}
\label{eq:vNS}
    S(\sigma) \equiv - \Tr \sigma \log \sigma.
\end{equation}
Because $S$ is a nonlinear functional of $\sigma$, standard path integral techniques cannot be straightforwardly applied to compute the trace above. Instead, a fruitful strategy is to employ the following mathematical identity:
\begin{equation}
    \log x = \lim_{n\to0} \frac{x^n-1}{n} = \lim_{n\to0} \partial_n x^n.
\end{equation}
Applying this to \cref{eq:vNS} and using $\Tr\sigma=1$ leads to
\begin{equation}
\label{eq:reptrick}
    S(\sigma) = - \lim_{n\to1} \frac{\Tr \sigma^{n}-1}{n-1} = - \lim_{n\to1} \partial_n \Tr \sigma^{n}.
\end{equation}
This is the starting point for a path integral procedure known as the replica trick for the von Neumann entropy. This mathematical trick consists of calculating traces of powers of the density operator, and analytically continuing them near $n=1$. At values $n\neq1$, these objects are of interest themselves and compute R\'enyi entropies,
\begin{equation}
    S_n(\sigma) \equiv - \frac{1}{n-1} \log \Tr \sigma^n,
\end{equation}
which can be easily checked to yield the limit $\lim_{n\to1} S_n(\sigma) = S(\sigma)$.\footnote{Note, in particular, that one can write $\lim_{n\to1} \partial_n \Tr \sigma^{n} = \lim_{n\to1} \partial_n \log \Tr \sigma^{n}$.} Crucially, for positive integers $n\in\mathbb{Z}^+$, an appropriate path integral construct does allow one to compute these R\'enyi entropies. This amounts to calculating traces of powers of density operators, $\Tr \sigma^n$, to which we refer as \textit{state replicas}.

\subsection{Replica Manifolds}
\label{ssec:stareps}

Consider a state $\rho$ prepared on $\Sigma\subset M$ by the local path integral of a complete theory on $M_\Sigma$. It will be convenient here to refer to the two sides of the cut of $M_\Sigma$ by $\Sigma^{\pm}$.
Recall that the path integral calculates $\Tr \rho$ by identifying fields across the $\Sigma$ cut and integrating over them. As explained in \cref{sec:prep}, this is equivalent to performing the path integral on $M_\Sigma$ after identifying $\Sigma^{+}\sim\Sigma^{-}$, which gives back the manifold $M$.

Consider now $\Tr \rho^2$, which requires two copies of the manifold, ${M_\Sigma}_1$ and ${M_\Sigma}_2$, to build each replica of the state inside the trace. At the level of matrix elements it is clear what the path integral should do. For the product $\rho^2$, we would like to identify the field configurations at $\Sigma_1^+$ and $\Sigma_2^-$, and integrate over them. To then take the trace $\Tr\rho^2$, we additionally identify the fields at $\Sigma_2^+$ and $\Sigma_1^-$, and integrate over them. Again, this field identifications can also be implemented by performing analogous identifications at the level of the spaces themselves, specifically $\Sigma_1^+ \sim \Sigma_2^-$ and $\Sigma_2^+ \sim \Sigma_1^-$. The resulting space consists of two copies of $M_\Sigma$ which are appropriately identified along their respective cuts. Because the theory is local, doing the path integral on each copy separately with the desired identification of the fields along cuts is equivalent to just doing the path integral on the space consisting of the two copies of $M_\Sigma$ that are themselves identified along their cuts.

The more general case of calculating $\Tr \rho^n$ for $n\in\mathbb{Z}^+$ replicas of the state $\rho$ easily follows. The desired periodic identification between field configurations from cut to cut across replicas under the trace can be equivalently implemented by the construction of an appropriate \textit{replica manifold} $M_{\Sigma}^{(n)}$. This space is obtained by taking $n$ copies of $M_\Sigma$, and cyclically gluing them along their $\Sigma$ cuts, such that the $\Sigma_k^+$ side of the $k^{\text{th}}$ replica is identified with the $\Sigma_{k+1}^-$ side of the $(k+1)^{\text{th}}$ replica, with $k+n \sim k$. The resulting replica manifold $M_\Sigma^{(n)}$ can be easily described: it is an $n$-sheeted covering of $M_\Sigma$ branched over $\Sigma$. 

It will clearly be important in this section to keep track of which manifold a path integral is to be evaluated on. We will make so explicit as follows: if $Z$ is the partition function of the theory that prepares the state $\rho$, then we will use $Z[X]$ to denote the path integral of this theory performed on $X$.
The upshot of the discussion above is thus the following path integral prescription for computing state replicas in local theories:\footnote{A na\"ive application of this formula to a strongly non-local theory would be wrong.}
\begin{equation}
\label{eq:reptrace}
    \Tr \rho^n = \frac{Z[M_\Sigma^{(n)}]}{{Z[M]}^n}.
\end{equation}
The denominator is a normalizing factor that makes $\Tr \rho^n=1$ for $\rho$ a pure state. To see this, recall that for a field theory a path integral may prepare a pure state on $\Sigma$ only if it is a splitting hypersurface.
When this is the case, the components in which $M$ is split by $\Sigma$ can be cyclically re-glued into copies of $M$ itself. In particular, if ${M_\Sigma}_k \cong X_k^- \sqcup X_k^+$, the half-space $X_k^+$ gets glued to the half-space $X_{k+1}^-$ of the next replica, thus giving a space isomorphic to $M$.
Altogether, the replica manifold reassembles itself into $M_\Sigma^{(n)}\cong M^n$, i.e., $M_\Sigma^{(n)}$ is isomorphic to $n$ disjoint copies of $M$ with no cuts. If the theory is local, then from the factorization property in \cref{eq:Zfact} it would follow that $Z[M^n]=Z[M]^n$, thus yielding $\Tr \rho^n=1$, as desired.

For later comparison, it will be useful to write out \cref{eq:reptrace} explicitly. Using the definition of the path integral for the complete theory in \cref{eq:Zpi}, the replica calculation for such a local theory reads
\begin{equation}
\label{eq:trrho}
    \Tr \rho^n = \frac{1}{Z^n} \int\displaylimits_{M_{\Sigma}^{(n)}} \mathcal{D}\phi \, e^{-I[\phi]} = \frac{1}{Z^n} \int\displaylimits_{M_{\Sigma}^{(n)}} \mathcal{D}\varphi \int\displaylimits_{M_{\Sigma}^{(n)}} \mathcal{D}\Phi \, e^{-I[\varphi,\Phi]},
\end{equation}
where in the second equality we are simply showcasing the splitting into $\IR$ and $\UV$ fields. In either expression, locality of the theory means the action takes the form of a single integral over $M_{\Sigma}^{(n)}$. Hence if $M_{\Sigma}^{(n)}\cong M^n$, the action just decomposes as the right-hand side of \cref{eq:Isplit} with $M_k$ labeling the $k^{\text{th}}$ replica of $M$.
Correspondingly, the path integral in \cref{eq:trrho} factorizes into a product of $n$ identical integrals over $M$.

A potential pitfall at this point is to conclude that, just because $M_{\Sigma}^{(n)}$ looks like $n$ disjoint copies of $M$ for $\Sigma$ splitting, for a general path integral the replica calculation on $M_{\Sigma}^{(n)}$ will always factorize. In fact, as explained in the next section, the replica path integral for an exact $\QET$ on $M_{\Sigma}^{(n)}$ does not factorize. This is to be contrasted with the behavior of an exact $\QET$ on $M^n$, which as in \cref{eq:Zeftfact} would indeed give a factorizing answer. The crucial point is that, although isomorphic, the spaces $M_{\Sigma}^{(n)}$ and $M^n$ are obtained very differently.

\subsection{Replica Wormholes}
\label{ssec:nonfact}

Before jumping into the technical details, let us first emphasize why the factorization of a general replica path integral for $M_\Sigma^{(n)} \cong M^n$ would be an inconsistency of the framework itself. 
Consider the path integral that prepares the exact $\QET$ state $\wtilde{\rho}$ on $\Sigma$. In \cref{eq:ffF}, $\wtilde{\rho}$ was identified as a reduced density operator obtained from $\rho$ by tracing out $\UV$ fields. The state $\wtilde{\rho}$ will be mixed so long as $\rho$ has entanglement between $\IR$ and $\UV$ fields, whether or not $\Sigma$ is splitting. In other words, how mixed $\wtilde{\rho}$ is depends on two phenomena: entanglement between $\IR$ fields inside and outside $\Sigma$ (if $\Sigma$ is not splitting), and entanglement between $\IR$ fields in $\Sigma$ and $\UV$ fields everywhere. If only the latter happens, the replica manifold will be isomorphic to $M^n$, as argued above, but the entanglement between $\IR$ and $\UV$ fields will still cause $\wtilde{\rho}$ to be mixed. A general fact about a mixed density operator is that $\Tr \wtilde{\rho}^{\,n}$ is a strictly decreasing function of $n$ so, in particular, $\Tr \wtilde{\rho}^{\,n} < 1$ for all $n>1$.\footnote{This follows from monotonicity of the $p$-norm, which also leads to monotonicity of the R\'enyi entropies.} Hence, in a replica calculation like \cref{eq:reptrace}, the $\QET$ path integral with partition function $\wtilde{Z}$ that prepares $\wtilde{\rho}$ cannot possibly factorize.
How this non-factorization occurs at the level of path integrals is explained next.

As emphasized throughout \cref{ssec:stareps}, locality played an important role in deriving \cref{eq:reptrace} as a path integral prescription for calculating state replicas. However, suppose the path integral under consideration is that of an exact $\QET$ with a strongly non-local action. Na\"ively, given \cref{eq:reptrace}, to calculate state replicas for the exact $\QET$ state $\wtilde{\rho}$, one could proceed to evaluate $\wtilde{Z}[M_\Sigma^{(n)}]$. If $\Sigma$ is splitting, by the isomorphism $M_\Sigma^{(n)} \cong M^n$ one would be tempted to declare a factorizing answer as in \cref{eq:Zeftfact}, which once again would be wrong. But even worse, if $\Sigma$ is not splitting, this would correspond to evaluating the full strongly non-local path integral on all of $M_\Sigma^{(n)}$, meaning that all multi-integrals in the action would run over the full replica manifold $M_\Sigma^{(n)}$. This would be very different from what the replica calculation is supposed to do.

The reason is simple: evaluating the strongly non-local theory on the full replica manifold would lead to an action with multi-integrals coupling all replicas ${M_\Sigma}_k$ that make up $M_\Sigma^{(n)}$. In particular, there would be multi-integrals mixing ${M_\Sigma}_k$ and ${M_\Sigma}_{k'}$ for all $k'\neq k$. This should not happen, since the preparation of identical copies of the state should involve completely independent path integrals. Namely, the full path integral on ${M_\Sigma}_k$ that prepares the $k^{\text{th}}$ copy of $\rho$ involves strong non-localities with multi-integrals in the action over ${M_\Sigma}_k$ only. Field configurations are then to be identified along $\Sigma$ cuts, but never does this step introduce strong non-localities taking the form of multi-integrals over distinct replicas.

From these observations, we learn that in the replica calculation for the $\wtilde{Z}$ path integral, the multi-integrals in the action should not run over the full replica manifold, but only over each different replica independently. We capture this behavior of the action for a replica calculations on $M_\Sigma^{(n)}$ with the following notation:
\begin{equation}
\label{eq:icirc}
    \wtilde{I}_\circ \equiv \sum_{k=1}^n \wtilde{I}_k,
\end{equation}
where $\wtilde{I}_k$ is the full exact $\QET$ action evaluated on the $k^{\text{th}}$ replica ${M_\Sigma}_k$. Correspondingly, the desired path integral to be performed on $M_\Sigma^{(n)}$ is
\begin{equation}
\label{eq:zcirc}
    \wtilde{Z}_{\circ}[M_\Sigma^{(n)}] \equiv \int\displaylimits_{M_\Sigma^{(n)}} \mathcal{D}\varphi \, e^{-\wtilde{I}_\circ[\varphi]},
\end{equation}
where even though the action decomposes as in \cref{eq:icirc}, the path integral still is over continuous field configurations $\varphi$ on $M_\Sigma^{(n)}$. The appropriate generalization of \cref{eq:reptrace} for the calculation of state replicas is thus
\begin{equation}
\label{eq:trcirc}
    \Tr \wtilde{\rho}^{\,n} = \frac{\wtilde{Z}_{\circ}[M_\Sigma^{(n)}]}{\wtilde{Z}[M]^n}.
\end{equation}
This prescription now holds for the path integral of any theory, be it a complete theory, an exact $\QET$, an approximate $\QET$, or a truncated $\QET$.

\begin{figure}
    \centering
    \includegraphics[width=.5\textwidth]{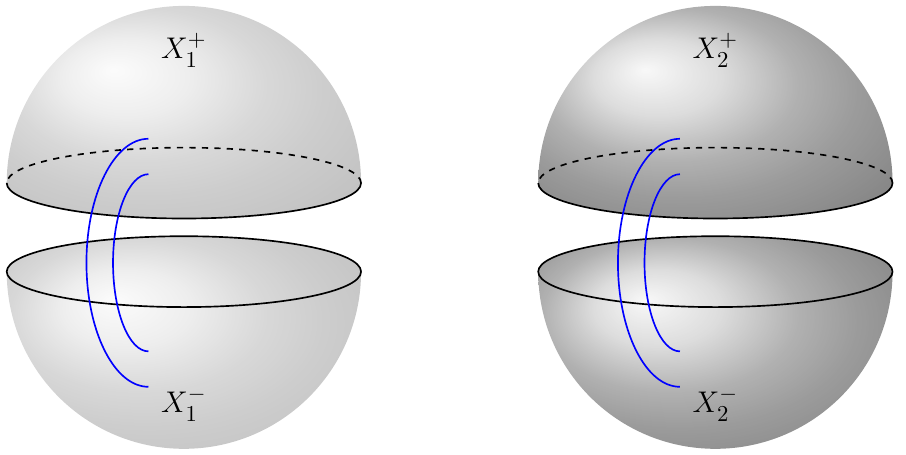}\\\vspace{.2cm}
    \includegraphics[width=.5\textwidth]{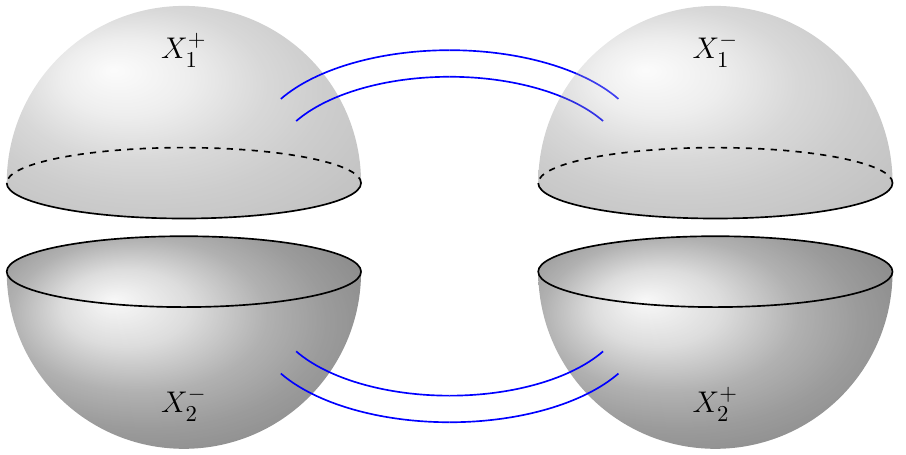}\\\vspace{.2cm}
    \includegraphics[width=.5\textwidth]{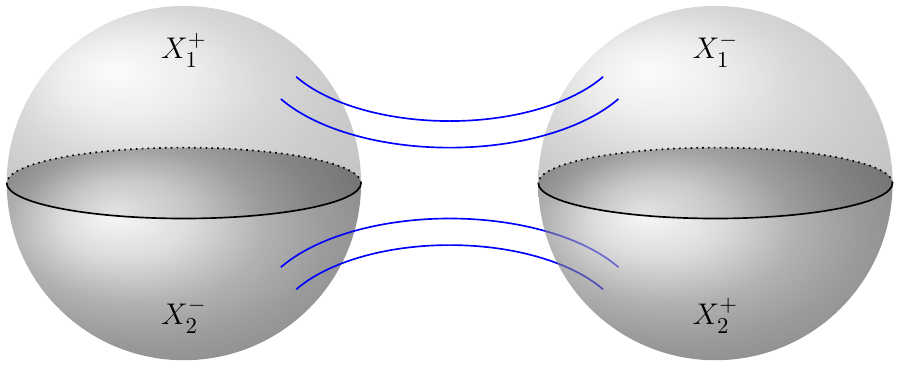}
    \caption{Step-wise representation of the state replica calculation in \cref{eq:trcirc} for an exact $\QET$. For simplicity, the figures show the $n=2$ case that computes the purity, and consider a state prepared on a splitting hypersurface.
    In the first figure, the two replicas of the state are separately prepared by independent path integrals on distinct copies of $M_{\partial X} \cong X^- \sqcup X^+$ (cf. replicating \cref{fig:statewormholes}). The second figure is just a convenient redrawing of the first one anticipating the cyclic identifications that the trace implements, keeping track of which spaces are coupled by quantum wormholes. In the last figure, the trace is performed by identifying $\partial X_1^+ \sim \partial X_2^-$ and $\partial X_2^+ \sim \partial X_1^-$, i.e., by acting with the `bra' of the first replica onto the `ket' of the second one and by cyclicity also acting with the `bra' of the latter onto the `ket' of the former. The resulting replica manifold, $M_{\partial X}^{(2)}$, though isomorphic to $M^2$, swaps halves and is given by \cref{eq:repm}, not \cref{eq:nonrepm}. This is how replica wormholes arise.}
    \label{fig:replicas}
\end{figure}

In light of the decomposition of the action in \cref{eq:icirc}, one may worry that the path integral $\wtilde{Z}_{\circ}$ would end up factorizing for $M_\Sigma^{(n)}\cong M^n$, contradicting the consistency arguments at the beginning of this section. In fact, factorization does not occur, and the mechanism for this is as follows.
Assume $\Sigma$ is splitting, since this is the sharp case where factorization would imply a qualitative failure of the replica calculation. The $k^{\text{th}}$ replica of the manifold $M_\Sigma$ on which $\wtilde{\rho}$ is prepared then splits as
\begin{equation}
    M_{\partial X_k} \cong X_k^-\sqcup X_k^+,
\end{equation}
where $X_k^\pm$ are the two dual halves that make up $M$ when glued along their respective boundaries $\partial X_k^\pm$. As explained in \cref{ssec:stareps} and particularly below \cref{eq:reptrace}, these are then glued back into the replica manifold $M_{\partial X}^{(n)}$ in a cyclic fashion. Namely, the cut identifications are of the form $\partial X_k^+ \sim \partial X_{k+1}^-$, thereby joining the $X_k^+$ half of the $k^{\text{th}}$ replica to the $X_{k+1}^-$ half of the next one. As a result, the reassembling of half-spaces into $n$ disjoint copies of $M$ involves dual halves from consecutive replicas giving spaces $X_k^+ \cup X_{k+1}^- \cong M$, with $k+n \sim k$. Though isomorphic, the replica labels matter, since these keep track of which specific spaces the action integrals run over. The replica manifold for a splitting hypersurface one obtains is precisely given by:
\begin{equation}
\label{eq:repm}
    M_{{\partial X}}^{(n)} = \bigsqcup_{k=1}^n M_{k\mid k+1}, \qquad M_{k\mid k+1} \equiv 
    X_k^+ \cup X_{k+1}^-.
\end{equation}
An illustration of this structure can be seen in \cref{fig:replicas}.
In contrast, the usual notation $M^n$ refers to a standard disjoint union of the $n$ original replicas,
\begin{equation}
\label{eq:nonrepm}
    M^n = \bigsqcup_{k=1}^n M_{k}, \qquad M_{k} \equiv X_{k}^+ \cup X_k^-.
\end{equation}

While obviously isomorphic $M_{{\partial X}}^{(n)} \cong M^n$ as claimed throughout, these two spaces are different and explain why the path integral in \cref{eq:zcirc} indeed does not factorize. Written out using the notation in \cref{eq:icirc}, we have
\begin{equation}
\label{eq:ztfactno}
    \wtilde{Z}_{\circ}[M_{\partial X}^{(n)}]~~ = \int\displaylimits_{\bigsqcup_{k=1}^n M_{k\mid k+1}} \mathcal{D}\varphi \, \prod_{k=1}^n e^{- \wtilde{I}_k[\varphi]},
\end{equation}
where recall $\wtilde{I}_k$ is the action with integrals only on $M_k=X_{k}^- \cup X_k^+$. Hence we see that although the path integral is over spaces $M_{k\mid k+1}$ which are disjoint for distinct $k$, the multi-integrals in the action $\wtilde{I}_k$ which run over $M_k$ will couple the $X_k^-$ half of $M_{k-1\mid k}$ to the $X_k^+$ half of $M_{k\mid k+1}$. Because the disjoint spaces the path integral runs over are genuinely coupled by quantum wormholes, as illustrated in \cref{fig:replicas}, it is clear that the replica path integral does not factorize. In addition, because mixing causes $\Tr\wtilde{\rho}^{\,n}\leq1$, for general $n\in\mathbb{Z}^+$ one expects 
\begin{equation}
\label{eq:factineq}
    \wtilde{Z}_{\circ}[M_{\partial X}^{(n)}] \leq \wtilde{Z}[M]^n,
\end{equation}
with equality if and only if $n=1$ or the state $\wtilde{\rho}$ prepared by the path integral on $M_{\partial X}$ is pure. Moreover, the monotonicity property of $\Tr\wtilde{\rho}^{\,n}$ discussed at the end of \cref{ssec:stareps} implies that for all $n\in\mathbb{Z}^+$,
\begin{equation}
\label{eq:mono}
    \wtilde{Z}_{\circ}[M_{\partial X}^{(n+1)}] \leq \wtilde{Z}_{\circ}[M_{\partial X}^{(n)}] \, \wtilde{Z}[M],
\end{equation}
which leads to \cref{eq:factineq} as a special case. Of course, this phenomenon relies on the fact that $\wtilde{I}_k$ is a strongly non-local action that can sustain quantum wormholes. In contrast, even a non-local theory that is only weakly so would break into separate integrals over $X_k^-$ and $X_k^+$, thereby restoring factorization. Hence, once again we see that quantum wormholes are crucial for the exact $\QET$ replica calculation to give a sensible answer.

\subsection{Ensembles of Replicas}
\label{ssec:ensemreps}

The quantum wormholes giving non-factorizing replicas came from the path integral over $\UV$ fields.
It will be illustrative to restore these fields by going back to the complete theory in order to understand precisely how the non-factorization in the exact $\QET$ occurs.
We can do this by expressing the $\QET$ action in \cref{eq:ztfactno} back in terms of the path integral that defines it using \cref{eq:IEFT}. This leads to
\begin{equation}
\label{eq:ztfactUV}
    \wtilde{Z}_{\circ}[M_{\partial X}^{(n)}] = \int\displaylimits_{\bigsqcup_{k=1}^n M_{k\mid k+1}} \mathcal{D}\varphi \int\displaylimits_{\bigsqcup_{k=1}^n M_k} \mathcal{D}\Phi \, e^{- I[\varphi,\Phi]},
\end{equation}
where we have used locality of the complete theory to combine the path integrals that define the $\QET$ on each $M_k$ into a single path integral over $M^n$. From \cref{eq:ztfactUV} we see that the path integrals over $\Phi$ and $\varphi$ are perfectly out of phase. This is what causes the strong non-localities arising from the path integral over the former to lead to quantum wormholes for the latter.

Going back to the general case in which we have a hypersurface $\Sigma$ which is not necessarily splitting, \cref{eq:ztfactno,eq:ztfactUV} lead to the following explicit form for \cref{eq:trcirc}:
\begin{equation}
\label{eq:eftreps}
     \Tr\wtilde{\rho}^{\,n} = \frac{1}{\wtilde{Z}^n} \int\displaylimits_{M_\Sigma^{(n)}} \mathcal{D}\varphi \, \prod_{k=1}^n e^{- \wtilde{I}_k[\varphi]} = \frac{1}{\wtilde{Z}^n} \int\displaylimits_{M_\Sigma^{(n)}} \mathcal{D}\varphi \int\displaylimits_{M^n} \mathcal{D}\Phi \, e^{- I[\varphi,\Phi]}.
\end{equation}
The difference between \cref{eq:trrho,eq:eftreps} is here manifest: in the latter, the $\UV$ fields are only allowed to propagate on disjoint copies of $M$ instead of on the replica manifold $M_\Sigma^{(n)}$. In other words, when calculating $\Tr \wtilde{\rho}^{\,n}$, the path integral first traces out the $\UV$ fields to define a $\QET$ on each replica of $M$, and then the calculation on the replica manifold $M_\Sigma^{(n)}$ is done for the only dynamical fields remaining, which are the $\IR$ fields. In contrast, when calculating $\Tr\rho^n$, the full theory was path-integrated on $M_\Sigma^{(n)}$, allowing both $\IR$ and $\UV$ fields to propagate across the replica manifold.

The lack of factorization of $\Tr\wtilde{\rho}^{\,n}$ also admits an interpretation in terms of the averages over ensembles discussed in \cref{ssec:ensemble,ssec:theoryensem}. In particular, consider expanding $\wtilde{\rho}$ in terms of the $\wtilde{\rho}_*[\alpha]$ states from \cref{eq:chista}. Using orthogonality, the state replicas for $\wtilde{\rho}$ can be easily seen to reduce to
\begin{equation}
\label{eq:ensempure}
    \Tr\wtilde{\rho}^{\,n} = \int \mathcal{D}\alpha \, {{{P}}}_*[\alpha]^n \, \Tr \wtilde{\rho}_{{*}}[\alpha]^n,
\end{equation}
where the trace term is unity if $\wtilde{\rho}_{{*}}[\alpha]$ is pure. It is straightforward to apply the replica trick in \cref{eq:reptrick} to \cref{eq:ensempure}, obtaining
\begin{equation}
\label{eq:alphass}
    S(\wtilde{\rho}) = H[{{{P}}}_*] + \left\langle S(\wtilde{\rho}_*[\alpha]) \right\rangle_{{{{P}}}_*},
\end{equation}
where we have defined the Shannon entropy $H$ of a classical probability distribution,
\begin{equation}
    H[P_*] \equiv - \int \mathcal{D}\alpha \, {{{P}}}_*[\alpha] \, \log {{{P}}}_*[\alpha].
\end{equation}
This is of course what one would obtain by simply applying \cref{eq:vNS} to \cref{eq:ensemrho}. This result says how for the state $\wtilde{\rho}$, as a classical mixture of quantum states, the von Neumann entropy decomposes into the Shannon entropy $H[P_*]$ of the classical ensemble, plus the average von Neumann entropy of the $\wtilde{\rho}_*[\alpha]$ states across the ensemble.
The former accounts for purely classical uncertainty in the state $\wtilde{\rho}$ due to ignorance of the state of $\Phi$ fields, whereas the latter captures genuinely quantum correlations intrinsic to the $\varphi$ fields within each member of the ensemble.
Recall in particular that for a sufficiently large Hilbert space $\mathcal{H}_\Phi$, the $\wtilde{\rho}_*[\alpha]$ may be pure even for a non-splitting $\Sigma$.
Hence we expect the classical term in \cref{eq:alphass} to be extensive in $\mathcal{H}_\Phi$, and the quantum term to only be nontrivial if the $\wtilde{\rho}_*[\alpha]$ states fail to act with unit rank on the Hilbert space $\mathcal{H}_\varphi$ of the exact $\QET$.

We can also express the state replicas for $\wtilde{\rho}$ in terms of traces of the spanning family of projected states defined in \cref{eq:irstates}. However, this family is not orthogonal on $\mathcal{H}_{{\varphi}}$ and consists of pure states only if $\Sigma$ is splitting. The path integrals can still be pulled out by linearity, but the resulting expression does not simplify as \cref{eq:ensempure}. We can write it as
\begin{equation}
\label{eq:ugly}
    \Tr \wtilde{\rho}^{\,n} = \left\langle \Tr \left( \prod_{k=1}^n \wtilde{\rho}[{\Phi_0}_k] \right) \right\rangle_{{{{P}}}^n}.
\end{equation}
The notation for the ensemble average $\langle \,\cdot\, \rangle_{{{{P}}}^n}$ in \cref{eq:ugly} instructs us to introduce a copy of the measure ${{{P}}}[{\Phi_0}_{k}]$ for every replica, i.e.,
\begin{equation}
\label{eq:multimess}
    \langle \,\cdot\,\rangle_{{{{P}}}^n} \equiv \int\displaylimits_{\Sigma^n} \prod_{k=1}^n \left( \mathcal{D}{\Phi_0} \, {{{P}}}[{\Phi_0}_k]  \right) \, (\,\cdot\,),
\end{equation}
which is correctly normalized to $\langle 1\rangle_{{{{P}}}^n} = 1$ for any $n$. The von Neumann entropy of $\wtilde{\rho}$ one gets now reads
\begin{equation}
\label{eq:projents}
    S(\wtilde{\rho}) = H[{{{P}}}] + \left\langle S(\wtilde{\rho}[\Phi_0]) \right\rangle_{{{{P}}}},
\end{equation}
which has a structure similar to \cref{eq:alphass}. Of course both equations give the same total entropy for $\wtilde{\rho}$. Here, however, what each term corresponds to is qualitatively different.
The classical term in \cref{eq:projents} accounts for correlations of the $\UV$ fields on $\Sigma$ with every other field, whereas the quantum term quantifies spatial correlations between $\varphi$ fields on $\Sigma$ and any other field outside of $\Sigma$. Hence, if $\Sigma$ is splitting, the former captures correlations between $\IR$ and $\UV$ fields, and the latter vanishes.
In such a case it is clear that $H[{{{P}}}]$ will scale with the interaction term in \cref{eq:iii}. In particular, were one to introduce a dimensionless parameter $\kappa$ in front of $I_{\smalltext{int}}$, one would find that $H[{{{P}}}] \propto \kappa$, with a vanishing entropy as $\kappa\to0$. In other words, the ensemble trivializes as the interactions between $\IR$ and $\UV$ fields are turned off.

One may expect that, even if the projected states $\wtilde{\rho}[\Phi_0]$ are not trace-orthogonal, the trace on the right-hand side of \cref{eq:ugly} may still be dominated by the diagonal term where all states are equal across replicas. This is plausible if there are not many null states among projected states like in \cref{eq:nulls}, and also motivated by the fact that such configurations respect the replica symmetry that the path integral manifestly satisfies. With this expectation, one may write
\begin{equation}
\label{eq:trome}
    \Tr \wtilde{\rho}^{\,n}
    = \int\displaylimits_{\Sigma} \mathcal{D}{\Phi_0} \, {{{P}}}[{\Phi_0}]^n \, \Tr \wtilde{\rho}[\Phi_0]^n +  O(\text{RSB}_{\evalat{\Phi}{\Sigma}}) \geq \left\langle \Tr \wtilde{\rho}[\Phi_0]^n \right\rangle_{{{{P}}}^n},
\end{equation}
where the corrections correspond to replica-symmetry breaking (RSB) configurations of $\Phi$ across the different $\Sigma$ cuts, and are guaranteed to be nonnegative.
On the right-most expression, note that we are using the notation introduced in \cref{eq:trome}, but with the understanding that, because every replica has the same configuration ${\Phi_0}_k=\Phi_0$, there is just a single integral over $\Phi_0$ and every copy of ${{{P}}}$ is evaluated on this $\Phi_0$.
For $n>1$, the trace over projected states in \cref{eq:trome} gives unity if and only if $\Sigma$ is splitting, and for $n=1$ \cref{eq:trome} becomes a trivial equality. Using \cref{eq:trcirc,eq:decomprho}, the analogous expression to \cref{eq:trome} that holds at the level of the exact $\QET$ partition function for a splitting $\Sigma$ is
\begin{equation}
\label{eq:tromewa}
    \wtilde{Z}_\circ[M_\Sigma^{(n)}] = \left\langle Z_{\Phi_0}[M]^n \right\rangle_{{{{P}}}^n} +  O(\text{RSB}_{\evalat{\Phi}{\Sigma}}) \geq \left\langle Z_{\Phi_0}[M]^n \right\rangle_{{{{P}}}^n},
\end{equation}
where we are using the partition functions of projected theories from \cref{eq:theoryzsn}, and the fact that $\wtilde{\rho}[\Phi_0]$ is pure for $\Sigma$ splitting. This says that the replica calculation for the exact $\QET$ partition function on disconnected replicas does not factorize, but is approximated and lower bounded by an ensemble average over partition functions of theories which do factorize across replicas. In \cref{sssec:tfdtmd} we relate \cref{eq:tromewa} to the factorization problem in holography.

In fact, the right-hand side of \cref{eq:trome} does provide an exact expression for the replicas of a different object: the classical ensemble $\Omega$ of pure states of the complete theory defined in \cref{eq:omegaensem}. This is because the lifted states $\Omega[\Phi_0]$ from \cref{eq:liftrho}, which act on the full Hilbert space $\mathcal{H}$ of the complete theory, are actually orthogonal on the $\UV$ sector. As a result, the replica calculation for this ensemble yields
\begin{equation}
\label{eq:ensemore}
    \Tr \Omega^n =  \left\langle \Tr \wtilde{\rho}[\Phi_0]^n \right\rangle_{{{{P}}}^n},
\end{equation}
an equation that holds exactly due to the orthogonality property in \cref{eq:orthoiruv}. The von Neumann entropy of the $\Omega$ still nicely reduces to that of $\wtilde{\rho}$, as expected by construction, 
\begin{equation}
\label{eq:somega}
    S(\Omega) = S(\wtilde{\rho}).
\end{equation}
These expressions provide precise relations between the exact $\QET$ state $\wtilde{\rho}$, and the statistics of the corresponding canonical ensemble $\Omega$ of states of the complete theory.

\subsection{Multi-Swap Operators}
\label{ssec:proj}

In this section, we obtain alternative representations of the path integrals for state replicas by performing field space projections rather than identifications on the underlying manifolds. The projection operators we obtain can be understood as multi-swap operators. These operators realize an alternative way of thinking of state replicas as measurements performed on states prepared by different copies of a theory, rather than as identical copies of a state prepared by a single theory.\footnote{In quantum theory this distinction is operationally meaningful while mathematically innocuous. In quantum gravity, these two interpretations of state replicas can lead to different results \cite{Marolf:2020rpm,Marolf:2021ghr}.}
This exercise will also make the differences and similarities between the replica calculations in a complete theory and in an exact $\QET$ more transparent.

The path integrals that compute state replicas started life as preparations of states on independent copies of the cut-open space $M_{\Sigma}$. In previous sections we implemented field identifications across the $\Sigma_k$ cuts of the different replicas by doing the path integral on a space constructed by gluing copies of $M_{\Sigma}$ appropriately. This gave rise to the replica manifold $M_\Sigma^{(n)}$, and also to the disjoint union $M^n$. One can reproduce these identifications in terms of projection operators in field space instead. In particular, here we will keep the path integrals just running over the disjoint union $M_{\Sigma}^n$ of $n$ independent copies of $M_\Sigma$. Because these path integrals are completely independent of each other, one may think of them as $n$ copies of the same theory, each independently preparing a single copy of the pertinent state on a different $M_{\Sigma_k}$. In the cyclic trace, these copies then get related by a projection operator which can be interpreted as performing a measurement across all replicas (cf. a multi-swap test).

Comparing \cref{eq:trrho,eq:eftreps}, we see that at the level of the path integral for the $\varphi$ fields these two calculations are identical. What actually distinguishes the replica path integrals for complete and exact $\QET$ states is fully captured by what happens to the $\Phi$ fields.
The focus here will thus be on understanding the differences between $\Tr \rho^n$ and $\Tr \wtilde{\rho}^{\,n}$ in terms of projection operators on the $\Phi$ fields.

\begin{figure}
    \centering
    \hspace{10pt}
    \includegraphics[width=0.4\textwidth]{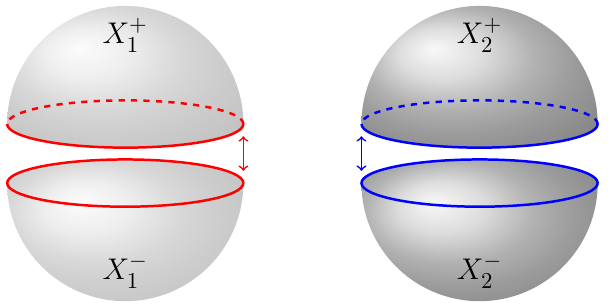}
    \hfill
    \includegraphics[width=0.4\textwidth]{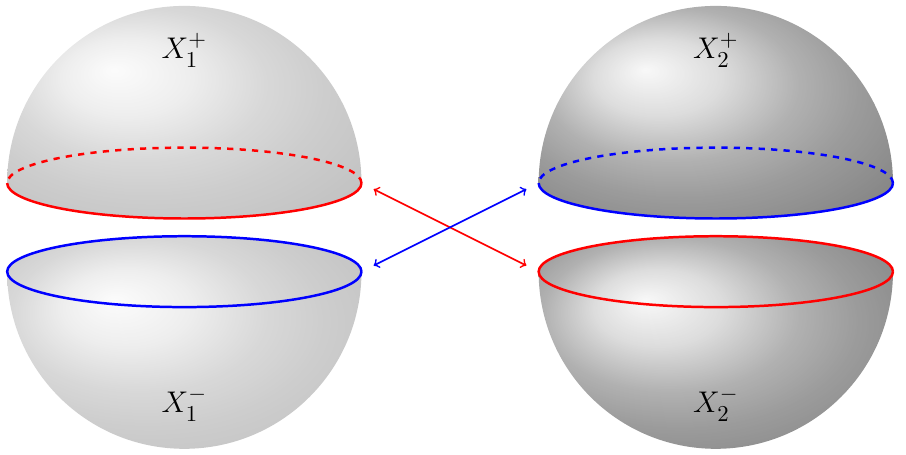}
    \hspace{10pt}
    ~
    \caption{Illustration of the projection operators implementing field identifications in replica path integrals for $n=2$. On the left, the operator $\mathbf{P}_{\Pi_n}$ defined in \cref{eq:piproj} performs identifications of field configurations between the boundaries of dual half-spaces within each replica. On the right, the operator $\mathbf{P}_{\Xi_n}$ defined in \cref{{eq:repproj}} performs these identifications between dual half-spaces of consecutive replicas. For $n=2$, this corresponds to a standard swap operator. The combined projection operator $\mathbf{P}_{\mathbb{Z}_n}$ defined in \cref{eq:allproj} would implement all identifications depicted above at once.} 
    \label{fig:operators}
\end{figure}

For $\Tr \wtilde{\rho}^{\,n}$, the space $M^n$ corresponds to simply gluing back each $M_{\Sigma_k}$ into $M_k$ along its own $\Sigma_k$ cut. The identifications this implements on $\Phi$ can be captured by
\begin{equation}
\label{eq:piproj}
    \mathbf{P}_{\Pi_n}[\Phi] \equiv \prod_{k=1}^n \delta\left( \evalat{\Phi}{\Sigma_k^-} - \evalat{\Phi}{\Sigma_k^+} \right).
\end{equation}
Explicitly, what this means is that the $\UV$ path integral in \cref{eq:eftreps} can equivalently be written as the following path integral on $M_\Sigma^n$:
\begin{equation}
\label{eq:projs}
    \int\displaylimits_{M^n} \mathcal{D}\Phi \, (\,\cdot\,) = \int\displaylimits_{M_\Sigma^n} \mathcal{D}\Phi \, \mathbf{P}_{\Pi_n}[\Phi] \, (\,\cdot\,).
\end{equation}
On the other hand, for $\Tr \rho^n$, the field identifications that the replica manifold $M_\Sigma^{(n)}$ enforces can be implemented by
\begin{equation}
\label{eq:repproj}
    \mathbf{P}_{\Xi_n}[\Phi] \equiv \prod_{k=1}^n \delta\left( \evalat{\Phi}{\Sigma_k^+} - \evalat{\Phi}{\Sigma_{k+1}^-} \right),
\end{equation}
where $k+n\sim k$. The $\UV$ path integral in \cref{eq:trrho} can thus be reproduced on $M_\Sigma^n$ by
\begin{equation}
\label{eq:pirepproj}
    \int\displaylimits_{M_{\Sigma}^{(n)}} \mathcal{D}\Phi \, (\,\cdot\,) = \int\displaylimits_{M_\Sigma^n} \mathcal{D}\Phi \, \mathbf{P}_{\Xi_n}[\Phi] \, (\,\cdot\,).
\end{equation}
The operators defined in \cref{eq:repproj,eq:piproj} are depicted in \cref{fig:operators}. Notice that for both \cref{eq:trrho,eq:eftreps} the projector in action for the $\varphi$ fields is $\mathbf{P}_{\Xi_n}[\varphi]$, since these fields are dynamical in both the complete and the exact $\QET$.

By inserting the projector $\mathbf{P}_{\Pi_n}$ into \cref{eq:pirepproj} and the projector $\mathbf{P}_{\Xi_n}$ into \cref{eq:projs} one gets matching right-hand sides, which leads to the following identity for the left-hand sides:
\begin{equation}
\label{eq:projrel}
    \int\displaylimits_{M_{\Sigma}^{(n)}} \mathcal{D}\Phi \, \mathbf{P}_{\Pi_n}[\Phi] \, (\,\cdot\,)  = \int\displaylimits_{M^n} \mathcal{D}\Phi \, \mathbf{P}_{\Xi_n}[\Phi]\, (\,\cdot\,).
\end{equation}
This path integral provides a precise relation between \cref{eq:trrho,eq:eftreps}. We can use this to better understand what makes the complete theory and the exact $\QET$ differ in calculations of state replicas. From \cref{eq:projrel}, we learn that in order to relate the path integrals on $M_{\Sigma}^{(n)}$ and $M^n$ one has to project out configurations from both. Namely, there are field configurations on the replica manifold $M_{\Sigma}^{(n)}$ which need to be projected out by $\mathbf{P}_{\Pi_n}$, and also field configurations on the disjoint union $M^n$ which need to be projected out by $\mathbf{P}_{\Xi_n}$. Put differently, the path integrals on $M_{\Sigma}^{(n)}$ and $M^n$ respectively receive contributions which are generally not accounted for by the other.

However, suppose the path integral over $\UV$ fields admits a saddle point approximation. In both cases, the functional extremization problem over $\Phi$ takes place on $M_\Sigma^n$. The difference between extremizing fields on $M_{\Sigma}^{(n)}$ and $M^n$ is which saddles are allowed to contribute, which depends on boundary conditions. If the extrema on $M_\Sigma^n$ happen to automatically be consistent with the identifications that both $\mathbf{P}_{\Xi_n}$ and $\mathbf{P}_{\Pi_n}$ impose, then the saddle-point approximations to the path integrals on $M_\Sigma^{(n)}$ and $M^n$ will naturally agree.
Hence at the level of saddle points, the state replicas for $\wtilde{\rho}$ and $\rho$ agree if the dominant saddles are not projected out by the joint action of $\mathbf{P}_{\Xi_n}$ and $\mathbf{P}_{\Pi_n}$. Using $\Sigma_k$ to indistinctively refer to both $\Sigma_k^{\pm}$ cut sides, this combined projection can be expressed as
\begin{equation}
\label{eq:allproj}
    \mathbf{P}_{\mathbb{Z}_n}[\Phi] \equiv \mathbf{P}_{\Xi_n}[\Phi] \, \mathbf{P}_{\Pi_n}[\Phi] =  \int\displaylimits_{\Sigma} \mathcal{D}\Phi_0 \prod_{k=1}^n \delta\left( \evalat{\Phi}{\Sigma_k} - \Phi_0\right),
\end{equation}
i.e., their joint action collapses the configurations on all $\Sigma_k$ cuts to be identical. Because $\mathbf{P}_{\mathbb{Z}_n}$ enforces the $\mathbb{Z}_n$ replica symmetry at the level of configurations on the $\Sigma_k$ cuts, we refer to it as the \textit{replica symmetry operator}.\footnote{This operator was already secretly in action when going from \cref{eq:ugly} to \cref{eq:trome}.}
In the path integrals that compute state replicas, even if off-shell, there are always field configurations which break replica symmetry, and are thus in the kernel of $\mathbf{P}_{\mathbb{Z}_n}$. All such configurations of the $\Phi$ fields will make replica calculations for the complete theory and the exact $\QET$ automatically differ.
The path integral over $\UV$ fields with the operator in \cref{eq:allproj} inserted yields
\begin{equation}
    \int\displaylimits_{M_\Sigma^n} \mathcal{D}\Phi \, \mathbf{P}_{\mathbb{Z}_n}[\Phi] \, e^{-I[\varphi,\Phi]} = \int\displaylimits_{\Sigma} \mathcal{D}{\Phi_0} \, \prod_{k=1}^n \evalat{{{{P}}}[{\Phi_0}_k] \, e^{- I_{{\Phi_0}_k}[\varphi]}}{{\Phi_0}_k=\Phi_0},
\end{equation}
where the ${{{P}}}$ measures were introduced in \cref{eq:decomprho}, and $I_{{\Phi_0}_k}$ is the action on the $k^{\text{th}}$ replica of a projected theory, as defined in \cref{eq:uvcompi}. By additionally acting with the path integral over $\varphi$ on $M_\Sigma^{(n)}$ to prepare the corresponding state replica we obtain
\begin{equation}
    \label{eq:exprepo}
    \int\displaylimits_{M_\Sigma^{(n)}} \mathcal{D}\varphi \int\displaylimits_{M_\Sigma^n} \mathcal{D}\Phi \, \mathbf{P}_{\mathbb{Z}_n}[\Phi] \, e^{-I[\varphi,\Phi]}
     =  \frac{1}{\wtilde{Z}^n}
     \left\langle \Tr \wtilde{\rho}[\Phi_0]^n \right\rangle_{{{{P}}}^n}
     = \frac{1}{\wtilde{Z}^n} \Tr \Omega^n,
\end{equation}
where we have used the notation from \cref{eq:multimess}, and in the last equality we have recognized the replica result for the $\Omega$ ensemble in \cref{eq:ensemore}. 
One may interpret \cref{eq:exprepo} as giving the expectation value of the replica symmetry operator $\mathbf{P}_{\mathbb{Z}_n}[\Phi]$ in the complete theory.
Hence contributions to $\Tr\rho^n$ and $\Tr\wtilde{\rho}^{\,n}$ which are common to both path integrals are given precisely by $\Tr\Omega^n$. Conversely, it is contributions to the $\Phi$ path integral in the kernel of $\mathbf{P}_{\mathbb{Z}_n}[\Phi]$ which make the state replicas for complete and exact $\QET$ states differ, as claimed above.

The focus in this section up to this point has been on understanding the difference between the complete theory and the exact $\QET$ when it comes to replica calculations, which as discussed at the beginning is fully captured by the difference in how the $\Phi$ fields are treated.
There is however also an important observation to be made about replica symmetry and its breaking at the level of the $\varphi$ path integrals. 
Using the operator $\mathbf{P}_{\Xi_n}$ from \cref{eq:repproj}, we can write the replica path integral in \cref{eq:zcirc} as
\begin{equation}
\label{eq:repopp}
    \wtilde{Z}_{\circ}[M_\Sigma^{(n)}] = \int\displaylimits_{M_\Sigma^{n}} \mathcal{D}\varphi \, \mathbf{P}_{\Xi_n}[\varphi] \, e^{-\wtilde{I}_\circ[\varphi]}.
\end{equation}
Consider now constructing a resolution of the identity by using the operator in \cref{eq:piproj} and its negation,
\begin{equation}
    \mathds{I} \equiv \mathbf{P}_{\Pi_n} + \not{\!\mathbf{P}}_{\Pi_n},
\end{equation}
which can in fact be taken as a definition of $\not{\!\mathbf{P}}_{\Pi_n}$. Combined with $\mathbf{P}_{\Xi_n}$, using \cref{eq:allproj} gives
\begin{equation}
    \mathbf{P}_{\Xi_n} \equiv \mathbf{P}_{\mathbb{Z}_n} + \mathbf{P}_{\Xi_n} \not{\!\mathbf{P}}_{\Pi_n}.
\end{equation}
This is simply saying that the multi-swap operator $\mathbf{P}_{\Xi_n}$ which implements the replica manifold identifications in \cref{eq:repopp} can be decomposed into a projection onto replica-symmetric configurations captured by $\mathbf{P}_{\mathbb{Z}_n}$, and RSB ones captured by the other term. 
Note that here we are talking about replica symmetry at the level of field configurations on $\Sigma$ only, so the path integral is free to run over all possible configurations away from the cut in both terms. Since $\mathbb{Z}_n$ is a symmetry of the path integral, it is often reasonable to assume that RSB contributions are subleading, so we may write \cref{eq:repopp} as
\begin{equation}
\label{eq:rsbsmall}
    \wtilde{Z}_{\circ}[M_\Sigma^{(n)}] = \int\displaylimits_{M_\Sigma^{n}} \mathcal{D}\varphi \, \mathbf{P}_{\mathbb{Z}_n}[\varphi] \, e^{-\wtilde{I}_\circ[\varphi]} +  O(\text{RSB}_{\evalat{\varphi}{\Sigma}}).
\end{equation}
Using \cref{eq:allproj}, for the replica-symmetric path integral we find the expression
\begin{equation}
\label{eq:donona}
    \int\displaylimits_{M_\Sigma^{n}} \mathcal{D}\varphi \, \mathbf{P}_{\mathbb{Z}_n}[\varphi]  \, e^{-\wtilde{I}_\circ[\varphi]} = \int\displaylimits_{\Sigma} \mathcal{D}\varphi_0 \, \wtilde{{{{P}}}}[\varphi_0]^n \, \wtilde{Z}_{\varphi_0}^n,
\end{equation}
where we have defined a partition function $\wtilde{Z}_{\varphi_0}$ and a normalized distribution $\wtilde{{{{P}}}}$ by
\begin{equation}
\label{eq:zvphi0}
    \wtilde{Z}_{\varphi_0} \equiv \frac{\wtilde{Z}}{\wtilde{Z}^*_{\varphi_0}} \int\displaylimits_{M}^{\evalat{\varphi}{\Sigma} = \varphi_0} \mathcal{D}\varphi \, e^{- \wtilde{I}[\varphi]}, \qquad \wtilde{{{{P}}}}[\varphi_0] \equiv \frac{\wtilde{Z}^*_{\varphi_0}}{\wtilde{Z}},
\end{equation}
and the constant $\wtilde{Z}^*_{\varphi_0}$ is given by
\begin{equation}
\label{eq:starphi}
    \wtilde{Z}^*_{\varphi_0} \equiv \int\displaylimits_{M}^{\evalat{\varphi}{\Sigma} = \varphi_0} \mathcal{D}\varphi \, e^{- \wtilde{I}[\varphi]}.
\end{equation}
The normalizations here have been picked so as to mimic the definition of the projected theories $Z_{\Phi_0}$ in \cref{eq:theoryzsn}.\footnote{The reader might wonder what motivates the convoluted definitions in \cref{eq:zvphi0,eq:starphi}, since plugging these into \cref{eq:donona} just gives $(\wtilde{Z}^*_{\varphi_0})^n$ for the integrand. The point is that partition functions are only meaningful up to normalization, and we are using this freedom to extract a normalized measure $\wtilde{{{{P}}}}$. This measure may seem arbitrary at this level, but what fixes it uniquely is that it is the measure that appears when relating observables computed by $\wtilde{Z}$ and $\wtilde{Z}_{\varphi_0}$, for which normalizations drop out. In other words, it is the measure $\wtilde{{{{P}}}}$ that picks out a preferred normalization for the partition functions with fixed $\varphi_0$ configurations.} Once again, as explained below \cref{eq:theoryzsn} for those, here $\wtilde{Z}^*_{\varphi_0}$ is also just a constant to be evaluated as in \cref{eq:starphi}, whereas $\wtilde{Z}_{\varphi_0}$ is to be treated as a partition function which defines a path integral for the $\varphi$ fields on $M$.
Hence we obtain the following form for \cref{eq:rsbsmall} (cf. \cref{eq:tromewa} for the $\Phi$ fields):
\begin{equation}
\label{eq:rsbphi}
    \wtilde{Z}_{\circ}[M_\Sigma^{(n)}] = \left\langle \wtilde{Z}_{\varphi_0}[M]^n \right\rangle_{\wtilde{{{{P}}}}^n} +  O(\text{RSB}_{\evalat{\varphi}{\Sigma}}),
\end{equation}
where we are using the notation from \cref{eq:multimess}, but for the new measure $\wtilde{{{{P}}}}$ and for configurations of $\varphi_0$ of $\varphi$.
Applied to \cref{eq:trcirc}, this says that up to RSB effects the state replicas for $\wtilde{\rho}$ are given by an ensemble average over $\wtilde{Z}_{\varphi_0}$ partition functions. This is used in \cref{sssec:tfdtmd} to offer a different perspective on the factorization problem in holography.

To conclude, let us quote the identities which allow one to interpret the operators introduced in this section as performing measurements on states prepared by different copies of a theory. The Hilbert space of $n$ copies of the exact $\QET$ on each $M_k$ replica is a tensor product $\mathcal{H}_\varphi^{\otimes^n}$, i.e., a multi-theory Hilbert space where each factor corresponds to one exact $\QET$. The states prepared by these theories similarly separate as $\wtilde{\rho}^{\otimes^n}$. Then if $\Tr_n$ denotes the trace on the multi-theory Hilbert space, one has the multi-swap test identity
\begin{equation}
\label{eq:expswapp}
    \Tr_n \left( \mathbf{P}_{\Xi_n}^{{\varphi}} \wtilde{\rho}^{\otimes^n} \right) = \Tr \wtilde{\rho}^{\,n},
\end{equation}
where the operator $\mathbf{P}_{\Xi_n}^{{\varphi}}$ is being used here to multi-swap the $\varphi$ field states. From the point of view of \cref{eq:expswapp}, the state replicas for $\wtilde{\rho}$ can be understood as expectation values of the multi-swap operator $\mathbf{P}_{\Xi_n}^{{\varphi}}$.

\section{Operator Algebras}
\label{sec:algebras}

The algebraic picture that follows helps clarify the qualitative difference among the various types of theories considered throughout. Namely, what distinguishes the algebra of operators of a complete theory (local), an exact $\QET$ (strongly non-local), an approximate $\QET$ (weakly non-local), and a truncated $\QET$ (local). For this algebraic detour we follow \cite{Casini:2019kex,Casini:2020rgj,Casini:2021zgr,Arveson1976-wp}.

\subsection{Algebraic Duality}
\label{ssec:algdu}

Let us begin by introducing some basic notions that we will need to analyse the algebraic structure of these theories. Since we are interested in understanding the properties of very different theories, no assumptions will be made about relations that do not simply follow axiomatically.
We denote algebras by $\mathcal{A}$. The set of operators that commute with a subalgebra $\mathcal{A}$ is denoted by $\mathcal{A}'$ and called the commutant of $\mathcal{A}$. For a quantum theory on some space $M$, let $S$ be a connected splitting hypersurface $S\subset M$. To any subspace $\Sigma\subseteq S$, consider associating an algebra of local operators supported on it, and write it as $\mathcal{A}(\Sigma)$. Doing this for the complementary subspace $\Sigma'\equiv S\smallsetminus \Sigma$ one obtains the algebra $\mathcal{A}(\Sigma')$. In general, for such locally generated algebras the following causality relation holds:
\begin{equation}
\label{eq:causality}
    \mathcal{A}(\Sigma) \subseteq (\mathcal{A}(\Sigma'))'.
\end{equation}
Indeed, for a Lorentzian theory where $S$ is spacelike, this is the statement of microcausality.
The algebra associated to $\Sigma$ is said to satisfy duality (or Haag duality for ball regions in vacuum \cite{horuzhy2012,Haag_1996}) if
\begin{equation}
\label{eq:haag}
    \mathcal{A}(\Sigma) = (\mathcal{A}(\Sigma'))'.
\end{equation}
This relation does not always holds, as will be important for us.
In particular, duality can be violated by making $\Sigma$ disjoint or non-simply connected \cite{Casini:2020rgj}.
It will be simpler to investigate the former case in which $\Sigma$ consists of more than one connected component. With this purpose, consider two subspaces $\Sigma_1,\Sigma_2\subseteq S$. If one is contained in the other, their algebras obey isotony,
\begin{equation}
    \mathcal{A}(\Sigma_1) \subseteq \mathcal{A}(\Sigma_2), \qquad \Sigma_1 \subseteq \Sigma_2.
\end{equation}
If instead they are disjoint, let $\Sigma \equiv \Sigma_1 \cup \Sigma_2 \subseteq S$. The smallest algebra one can possibly associate to $\Sigma$ is the one generated additively by the local operators in the algebras $\mathcal{A}(\Sigma_1)$ and $\mathcal{A}(\Sigma_2)$. This defines the minimal additively generated algebra 
\begin{equation}
    \mathcal{A}_{\smalltext{min}}(\Sigma) \equiv \mathcal{A}(\Sigma_1) \vee \mathcal{A}(\Sigma_2),
\end{equation}
where $\mathcal{A}_1 \vee \mathcal{A}_2 \equiv (\mathcal{A}_1 \cup \mathcal{A}_2)''$. If $\mathcal{A}$ is a von Neumann algebra, then $\mathcal{A}''=\mathcal{A}$ by the bicommutant theorem \cite{Arveson1976-wp}.
The notation $\mathcal{A}(\Sigma)$ used up to this point implicitly refers to $\mathcal{A}_{\smalltext{min}}(\Sigma)$; hereon, only the latter, explicit notation will be used when referring to locally generated algebras of $\Sigma$ subspaces.
This is important because there generally is no canonical algebra one may associate to a subspace.
In particular, there is also a largest possible algebra one may associated to $\Sigma$ in a way that is still consistent with the the causality relation in \cref{eq:causality}. This corresponds to taking the set of all operators which commute with those additively generated in the complementary subspace. More explicitly, this maximal algebra is given by the commutant of the additively generated algebra of the complement, 
\begin{equation}
\label{eq:maxadef}
    \mathcal{A}_{\smalltext{max}}(\Sigma) \equiv \left(\mathcal{A}_{\smalltext{min}}(\Sigma')\right){'}.
\end{equation}
The causality relation in \cref{eq:causality} becomes the general statement that
\begin{equation}
    \mathcal{A}_{\smalltext{min}}(\Sigma) \subseteq \mathcal{A}_{\smalltext{max}}(\Sigma).
\end{equation}
These definitions allow one to make concrete statements about the algebraic structure of a theory without having to commit to any choice of association of algebras to subspaces. This in particular means we can extend the notion of duality in \cref{eq:haag} to a general statement about the theory. The algebra of the theory is said to satisfy duality if for any subspace $\Sigma$, minimal and maximal algebras agree, i.e.,
\begin{equation}
    \mathcal{A}_{\smalltext{min}}(\Sigma) = \mathcal{A}_{\smalltext{max}}(\Sigma), \qquad \forall\Sigma\subseteq S.
\end{equation}
In \cite{Casini:2019kex,Casini:2020rgj,Casini:2021zgr}, theories are characterized as complete if this notion of duality holds, and non-complete otherwise. The complete, local theories studied throughout this paper are assumed to be complete in this sense.
Because violations of duality for effective theories will be important for us, let us briefly comment on both their cause and consequences. 

A violation of duality takes the form of the proper inclusion $\mathcal{A}_{\smalltext{min}}(\Sigma) \subset \mathcal{A}_{\smalltext{max}}(\Sigma)$ for some subspace $\Sigma$. In words, this says that there are operators in the maximal algebra one could associate to $\Sigma$ which cannot be locally constructed within $\Sigma$. Put differently, the mismatch between $\mathcal{A}_{\smalltext{min}}$
and $\mathcal{A}_{\smalltext{max}}$ is solely caused by the existence of non-local operators in the latter which are absent in the former. In the context of this paper, this is obviously why duality violations are relevant. The algebraic consequences of the failure of duality are noteworthy too. Assuming von Neumann’s bicommutant theorem applies, one can easily see that if $\Sigma$ violates duality, so will its complement:
\begin{equation}
    \mathcal{A}_{\smalltext{max}}(\Sigma') = (\mathcal{A}_{\smalltext{min}}(\Sigma))' \supset (\mathcal{A}_{\smalltext{max}}(\Sigma))'=\mathcal{A}_{\smalltext{min}}(\Sigma').
\end{equation}
This in particular says that $\mathcal{A}_{\smalltext{max}}(\Sigma')$ contains operators outside the commutant of $(\mathcal{A}_{\smalltext{max}}(\Sigma))'$. In other words, there are operators in $\mathcal{A}_{\smalltext{max}}(\Sigma')$ which do not commute with some operators in $\mathcal{A}_{\smalltext{max}}(\Sigma)$, and vice versa. As a result, it follows that the expectation values of such operators cannot be locally determined, but depend non-locally on each other. As understood in \cite{Casini:2021zgr}, these are related by generalized symmetries. In our context, it is the explicit existence of non-local couplings which will relate those expectation values.

\subsection{Effective Theory Algebras}
\label{ssec:qetalg}

Equipped with this machinery, it is now possible to make precise distinctions between the structural properties of the algebras of operators in our theories of interest. In each theory, we would like to characterize the kind of subspaces on which duality is violated, if any. On one end we have some complete, local theory, which as mentioned above is taken to respect duality. On the opposite end we have a truncated $\QET$, which by virtue of being again local may also obey duality, although this is not necessarily always true. The intermediate cases of an exact $\QET$ and an approximate $\QET$ do violate duality in general and are thus more interesting. Given that non-localities are strictly stronger in an exact than in an approximate $\QET$, so do we expect violations of duality to be strictly stronger in the former than in the latter. Let us introduce the minimal structure necessary to qualitatively distinguish these two on the same setting.

\begin{figure}
    \centering
    \hspace{10pt}
    \includegraphics[height=2.2cm]{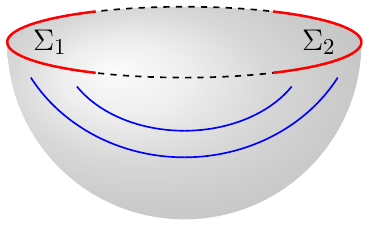}
    \hfill
    \includegraphics[height=2.2cm]{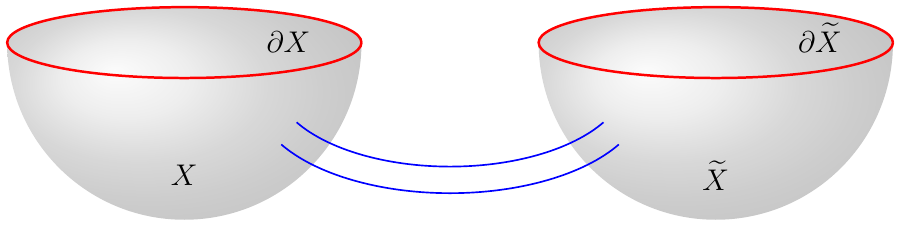}
    \hspace{10pt}
    ~
    \caption{Illustration of the settings consider for understanding the violations of duality in $\QET$ operator algebras. On the left, the maximal algebra for $\Sigma=\Sigma_1 \cup \Sigma_2$ includes non-local operators straddling between $\Sigma_1$ and $\Sigma_2$ which are not in the minimal algebra. This gives the characteristic violation of duality in an approximate $\QET$ due to weak non-locality. On the right, the maximal algebra for the full space $\partial X$ includes non-local operators which cannot be locally generated within the theory on $\partial X$. These are associated to quantum wormholes which are more naturally understood as connecting the two half-spaces which prepare a canonical purification of the state on $\partial X$. These globally non-local operators give the characteristic violation of duality in an exact $\QET$ due to strong non-locality and, in particular, the existence of quantum wormholes.}
    \label{fig:violations}
\end{figure}

As we saw in \cref{ssec:factor}, partition functions factorize across connected components of a space even for an exact $\QET$. Hence let our theory of interest be defined on a connected space $M\cong X\cup X^*$, where $X$ and $X^*$ are the usual CRT-dual half-spaces. The path integrals in \cref{sec:prep} allowed us to construct a Hilbert space for a given theory by preparing states and defining an inner product on $\partial X$. We would like to understand the algebras of operators acting on this Hilbert space and, in particular, their structure at the level of subspaces $\Sigma \subseteq \partial X$. As it turns out, it suffices to consider a connected boundary $\partial X$, and to make $\Sigma$ consist of just two connected components, $\Sigma_1$ and $\Sigma_2$. This setting is illustrated on the left in \cref{fig:violations}.

Consider first an approximate $\QET$, which recall is a weakly non-local theory. These theories are characterized by actions with arbitrarily high derivative orders, which give rise to non-local interactions within connected components. By virtue of being connected in $X$, the subspaces $\Sigma_1$ and $\Sigma_2$ will correspondingly be non-locally coupled. Hence we expect there to be non-local operators across these two subspaces with nontrivial expectation values. This would imply a violation of duality on $\Sigma$. This type of violation is characteristic of an approximate $\QET$, but would of course also happen for an exact $\QET$. A failure of duality that occurs for an exact $\QET$ but not for an approximate one must take place at a more global level.

An exact $\QET$ is a strongly non-local theory, the characteristic non-localities of which are associated to quantum wormholes. Recall that quantum wormholes correspond to non-local couplings across connected components, which are in particular lost in an approximate $\QET$. Remarkably, this suggests that we should expect the violation of duality characteristic of an exact $\QET$ to occur even on $\partial X$, which is a full connected component. This is surprising because $\partial X$ is the full space where the path integral defines the Hilbert space of the theory. In particular, $\partial X$ has an empty complement, $\partial X ' = \varnothing$, and the empty set has no local operators, so those in its commutant which define $\mathcal{A}_{\smalltext{max}}(\partial X)$ must genuinely give the full algebra of the theory. Hence we see that a violation of duality on $\partial X$ would imply that the full algebra of the theory includes operators which cannot be locally generated on the full $\partial X$ space.

A more concrete way of understanding how a global violation of duality on $\partial X$ can possibly happen is to remember that $X$ is just one half of $M\cong X\cup X^*$, the space where the exact $\QET$ is originally defined. Once the path integral is cut open, quantum wormholes straddle between $X$ and $X^*$, and the density operators prepared on $\partial X$ are thus mixed. But if one considers CRT-conjugating $X^*$ back to another copy $\wtilde{X}$ of $X$, then what the corresponding path integral would be preparing are canonical purifications of those density operators. This is illustrated in \cref{fig:violations}.
From this viewpoint, $\partial\wtilde{X}$ can be thought of as the complementary subspace to $\partial X$, a more familiar setting when discussing algebras associated to subspaces. Then quantum wormholes are non-local interactions between $\partial X$ and $\partial \wtilde{X}$, which would give rise to nontrivial expectation values between non-local operators associated to $\partial X \sqcup \partial \wtilde{X}$. This identification allows for a more explicit realization of the maximal algebra of the exact $\QET$ on $\partial X$ as
\begin{equation}
\label{eq:twocopies}
    \mathcal{A}_{\smalltext{max}}(\partial X) = ( \mathcal{A}_{\smalltext{min}}(\partial \wtilde{X}))' 
\end{equation}
The point of view of canonical purifications used above also serves to demystify the global violation of duality of an exact $\QET$. Namely, similarly to how for an approximate $\QET$ the maximal algebra of a subspace $\Sigma$ is given by the commutant of the local algebra of its purifying subspace $\Sigma'$, for an exact $\QET$ so is the maximal algebra of the theory given by the commutant of the local algebra of a purifying theory on $\partial \wtilde{X}$.

Intuitively, an approximate $\QET$ reveals its non-completeness when spatial regions are traced out, but is otherwise globally complete in the sense that duality holds on splitting hypersurfaces. In contrast, an exact $\QET$ reveals its non-completeness already at the global level of splitting hypersurfaces. This is because an exact $\QET$ still remembers that it came from a complete theory by tracing out some sector. Because this global type of non-completeness corresponds to a violation of duality at the level of the full theory, one may say that the non-local operators in $\mathcal{A}_{\smalltext{max}}$ which are not in $\mathcal{A}_{\smalltext{min}}$ are actually not part of the theory at all. However, we already saw quantum wormholes in action in \cref{sec:reps} giving nontrivial correlations across replicas which would result in nontrivial expectation values for non-local operators across replicas.
The crux to reconcile these observations is to embrace the idea that the global violation of duality of an exact $\QET$ is in a specific sense not associated to non-local operators on a single copy of the theory, but to non-local operators on multiple copies of it. This perspective was made precise in \cref{ssec:proj} by interpreting state replicas as multi-swap measurements. As usual, this emphasizes how global the properties that replica calculations probe are.

Let us conclude with an interesting observation about the global algebras of our theories. For the approximate $\QET$, the absence of quantum wormholes guarantees duality on splitting hypersurfaces. For the exact $\QET$, it is solely quantum wormholes which cause a violation of duality. But these strong non-localities associated to quantum wormholes are precisely what distinguish an exact $\QET$ from an approximate $\QET$. This is particularly manifest if one decides to treat an exact $\QET$ as if it were a complete theory, which as explained in \cref{ssec:approx,ssec:appeftst} is secretly equivalent to operating with an approximate $\QET$ (cf. a resummed derivative expansion). This leads to the expectation that the locally generated algebras of both theories should agree on splitting hypersurfaces. Using duality of the approximate $\QET$, this means
\begin{equation}
\label{eq:exapp}
    \mathcal{A}_{\smalltext{max}}(\partial X)^{\smalltext{approx}} \cong \mathcal{A}_{\smalltext{min}}(\partial X)^{\smalltext{exact}},
\end{equation}
i.e., the full operator algebra of the approximate $\QET$ should be isomorphic to the full locally generated operator algebra of the exact $\QET$. One may also generalize this to a relation involving the maximal algebra of the approximate $\QET$ for an arbitrary subspace $\Sigma$. In the exact $\QET$, this would be reproduced by an intermediate algebra consisting of all operators in $\mathcal{A}_{\smalltext{max}}(\Sigma)$ which are not in the commutant of $\mathcal{A}_{\smalltext{min}}(\partial X)$.

\subsection{Breakdown of Effective Theory}
\label{ssec:relent}

In a complete theory, minimal and maximal algebras for any subspace agree. This way, duality grants a canonical association of an operator algebra to any subspace, and also guarantees that this canonical algebra is always locally generated.
In a non-complete theory, the failure of duality makes such an association of algebras to subspaces ambiguous. If one associates $\mathcal{A}_{\smalltext{min}}(\Sigma)$ to $\Sigma$, non-local operators will be missed. If instead one associates $\mathcal{A}_{\smalltext{max}}(\Sigma)$, non-local operators will be accounted for, but the algebras of complementary regions will contain operators which do not commute. An operational way of distinguishing the physics these two algebras capture when acting on states is thus desirable. When addressing our different theories, this will provide a diagnostic of which physics they do or do not capture.

Given some theory, consider the preparation of a state on a subspace $\Sigma$. A failure of duality on $\Sigma$ signals an incompleteness of the description of that state that is attainable solely by local observables on $\Sigma$. In particular, this means that there will be distinct states the theory can prepare on $\Sigma$ which will be indistinguishable from the perspective of the algebra of local operators on $\Sigma$. Indistinguishable states on $\Sigma$ will be those differing only by the expectation values of non-local operators in $\mathcal{A}_{\smalltext{max}}(\Sigma)$ which are not in $\mathcal{A}_{\smalltext{min}}(\Sigma)$.

This conditional expectation can be manufactured by partially tracing the state over non-local sectors (thereby removing non-local information) and then lifting it back to an operator in $\mathcal{A}_{\smalltext{max}}(\Sigma)$ with a maximally mixed identity operator in the non-local sector. The resulting state thus retains all information about expectation values of local observables, but is completely ignorant of non-local ones and yields zero for these. Put differently, the two states give identical expectation values for operators in $\mathcal{A}_{\smalltext{min}}(\Sigma)$, and may only differ for those only in $\mathcal{A}_{\smalltext{max}}(\Sigma)$. For instance, given a state on $\Sigma\equiv \Sigma_1\sqcup\Sigma_2$ for the setup on the left of \cref{fig:violations}, one could consider constructing the reduced density operators $\wtilde{\rho}_1$ and $\wtilde{\rho}_2$ on each subspace, and then lifting these back to the product state on $\Sigma$ given by $\wtilde{\rho}_1\otimes\wtilde{\rho}_2$.

More generally, let the state on $\Sigma$ be $\wtilde{\rho}$, and denote the state conditioned on having trivial expectation values with respect to non-local operators on $\Sigma$ by $\wtilde{\rho}_{\smalltext{loc}}$.
With an information theoretic mindset, we would like to consider a situation in which $\wtilde{\rho}$ is the true state on $\Sigma$, and $\wtilde{\rho}_{\smalltext{loc}}$ is a local observer's hypothesis for what the state on $\Sigma$ is. We want to know the number $N$ of copies of $\wtilde{\rho}$ we would need to perform measurements on in order to conclude that the hypothesis $\wtilde{\rho}_{\smalltext{loc}}$ is wrong. This is quantified by requiring $N S(\wtilde{\rho} \, || \, \wtilde{\rho}_{\smalltext{loc}}) \gg 1$, where $S(\wtilde{\rho} \, || \, \wtilde{\rho}_{\smalltext{loc}})$ is the relative entropy of $\wtilde{\rho}$ with respect to $\wtilde{\rho}_{\smalltext{loc}}$, defined by \cite{Witten:2018zva}
\begin{equation}
\label{eq:relent}
    S(\wtilde{\rho} \, || \, \wtilde{\rho}_{\smalltext{loc}}) = \Tr \wtilde{\rho} \left( \log \wtilde{\rho} - \log \wtilde{\rho}_{\smalltext{loc}} \right).
\end{equation}
The relative entropy is a nonnegative quantity by Klein's inequality, and vanishes if and only if the two states are identical.
We emphasize again that this trace shall be taken in $\mathcal{A}_{\smalltext{max}}(\Sigma)$; if the trace were taken in $\mathcal{A}_{\smalltext{min}}(\Sigma)$, the relative entropy of $\wtilde{\rho}$ with respect to $\wtilde{\rho}_{\smalltext{loc}}$ would be identically zero by construction. Equivalently, this means that different states $\wtilde{\rho}_1\neq\wtilde{\rho}_2$ with the same conditional expectation $\wtilde{\rho}_{\smalltext{loc}}$ are perfectly indistinguishable with only access to local operators in $\mathcal{A}_{\smalltext{min}}(\Sigma)$. 
Indeed, the value of the relative entropy in \cref{eq:relent} quantifies the uncertainty of $\wtilde{\rho}$ in $\mathcal{A}_{\smalltext{max}}(\Sigma)$ given the knowledge in $\mathcal{A}_{\smalltext{min}}(\Sigma)$.
Importantly, relative entropies are always well-defined across algebra types in quantum field theory.\footnote{This includes type-III von Neumann algebras, even though for these neither the traces nor the density operators in \cref{eq:relent} actually exist when $\Sigma$ is not a splitting hypersurface.} 
If the conditional expectation that defines $\wtilde{\rho}_{\smalltext{loc}}$ is trace-preserving, \cref{eq:relent} reduces to a simple difference of von Neumann entropies \cite{Casini:2019kex},
\begin{equation}
\label{eq:relvon}
    S(\wtilde{\rho} \, || \, \wtilde{\rho}_{\smalltext{loc}}) = S(\wtilde{\rho}_{\smalltext{loc}}) - S(\wtilde{\rho}),
\end{equation}
an expression which now looks just like a conditional entropy. Here nonnegativity is easily seen to follow from the maximally mixed conditional expectation in $\wtilde{\rho}_{\smalltext{loc}}$. Furthermore, if $\wtilde{\rho}$ is a state on $\Sigma\equiv\Sigma_1\sqcup\Sigma_2$ that reduces to $\wtilde{\rho}_{\smalltext{loc}} = \wtilde{\rho}_1 \otimes \wtilde{\rho}_2$, then \cref{eq:relvon} becomes the mutual information between $\Sigma_1$ and $\Sigma_2$ on the state $\wtilde{\rho}$, i.e.,
\begin{equation}
\label{eq:mii}
    S(\wtilde{\rho} \, || \, \wtilde{\rho}_{\smalltext{loc}}) = I(\Sigma_1:\Sigma_2)_{\wtilde{\rho}}.
\end{equation}
The splitting case in which $\Sigma=\partial X$ is noteworthy. In particular, because in an exact $\QET$ duality holds globally, there would seem to be distinct states of the complete theory on $\Sigma$ which are nonetheless perfectly indistinguishable by exact $\QET$ operators in $\mathcal{A}_{\smalltext{min}}(\partial X)$.
Remarkably, this would mean that there are states which operators on a single copy of the theory cannot possibly distinguish. To distinguish such states, one is necessarily forced to consider operators in multiple copies of the theory (as in \cref{eq:twocopies} for two copies), or quantities like the state replicas in \cref{sec:reps}.

This situation is demystified by thinking of the exact $\QET$ as preparing a canonical purification of $\wtilde{\rho}$ as on the right of \cref{fig:violations}. Then it is clear that $\mathcal{A}_{\smalltext{min}}(\partial X)$ just has access to $\wtilde{\rho}$, the reduced density operator on $\partial X$, but not to any information about the purifying subsystem $\partial \wtilde{X}$. Any state prepared on $\partial X\sqcup\partial \wtilde{X}$ with operator insertions in $\wtilde{X}$ which do not change the state on $\partial X$ will be perfectly indistinguishable by observables in $\mathcal{A}_{\smalltext{min}}(\partial X)$.
In this canonical purification picture, $\wtilde{\rho}_{\smalltext{loc}}$ is the tensor product of the reduced density operators on $\partial X$ and $\partial \wtilde{X}$. This state differs from the canonical purification of $\wtilde{\rho}$ in that quantum wormholes have been traced out, thereby decorrelating the two sides of the system. Correspondingly, \cref{eq:mii} quantifies quantum wormhole correlations, which are given by the mutual information between the state on $\partial X$ and any purification. Because quantum wormholes are the remnant of correlations between $\IR$ and $\UV$ fields in the complete theory, \cref{eq:mii} can also be thought of as measuring the amount of entanglement between these two sectors in pure states of the complete theory. Hence we expect the mutual information in such cases to scale with the interaction term between $\varphi$ and $\Phi$ fields in the complete theory.

The discussion above has concerned distinguishability of states within a single theory. In fact, given how our $\QET$ constructions are related to each other, the relative entropy can also be used to distinguish between states prepared by different theories. This will in particular allow us to determine to what extend the states that an approximate or truncated $\QET$ prepare fail to capture the expectation values of complex operators of the true theory.

To be more precise, let the state of the complete theory on $\Sigma$ be $\rho$. We are just interested in the state of the $\IR$ fields, which can be obtained by tracing out the $\Phi$ sector. By \cref{eq:ffF} the resulting state is precisely $\wtilde{\rho}$, the exact $\QET$ state. Hence the exact $\QET$ perfectly describes the $\varphi$ fields, and should be used as a benchmark. Now, suppose our best description of the $\varphi$ fields is an approximate or truncated $\QET$, which prepares the state $\rho_{\times}$ on $\Sigma$. Theoretical predictions made by computing observables on $\rho_{\times}$ are just approximations to the true results one would obtain by acting with those observables on the actual state $\wtilde{\rho}$ of the system. More precisely, we expect there to be a class of observables $\mathcal{O}_\times$ for which
\begin{equation}
\label{eq:eqoo}
    \Tr (\rho_\times \, \mathcal{O}_\times) \approx \Tr (\wtilde{\rho} \, \mathcal{O}_\times)
\end{equation}
with the approximation holding in some perturbative sense, and some different class of observables $\wtilde{\mathcal{O}}$ for which
\begin{equation}
\label{eq:breakeft}
    {\Tr} (\rho_\times \, \wtilde{\mathcal{O}}) \neq {\Tr} (\wtilde{\rho} \, \wtilde{\mathcal{O}} ),
\end{equation}
with the two sides differing arbitrarily. 
For instance, suppose our observables of interest are correlation functions of a small number of insertions of the $\varphi$ fields.
If the lightest $\Phi$ field that was integrated out from the complete theory had a mass $m_\Phi>0$, one may expect \cref{eq:eqoo} to hold so long as the operator insertions have energies $\Lambda\ll m_\Phi$, whether $\rho_\times$ comes from an approximate or a truncated $\QET$. If instead one considers operators with energies comparable to or larger than $m_\Phi$, the neglect of higher derivative terms will clearly make \cref{eq:eqoo} fail for a truncated $\QET$. For an approximate $\QET$, however, it is not obvious whether or not \cref{eq:eqoo} will hold. 
This is because all that an approximate $\QET$ is missing is quantum wormholes, but otherwise preserves all non-localities within connected components (see \cref{ssec:approx}). There is nonetheless a simple argument for why there must exist observables $\wtilde{\mathcal{O}}$ for which even an approximate $\QET$ must break down in the sense of \cref{eq:breakeft}.

Consider the same logic that led us to introduce the relative entropy in \cref{eq:relent}. Namely, if all we have access to is an approximate $\QET$, our best hypothesis for the state on $\Sigma$ is $\rho_{\times}$ as prepared by the approximate $\QET$. To quantify how hard it is to realize that our hypothesis $\rho_{\times}$ for the state of the system is wrong, we consider the relative entropy
\begin{equation}
\label{eq:kldiv}
    S(\wtilde{\rho} \, || \, \rho_{\times}) = \Tr \wtilde{\rho} \left( \log \wtilde{\rho} - \log \rho_{\times} \right).
\end{equation}
In principle, one may want to take the trace above in the exact $\QET$ algebra $\mathcal{A}_{\smalltext{max}}(\Sigma)$. However, since we are considering states prepared by a single copy of the theory, we can get the same answer by just taking the trace in $\mathcal{A}_{\smalltext{min}}(\Sigma)$. This is also desirable since, as argued at the end of \cref{ssec:qetalg}, we expect the $\mathcal{A}_{\smalltext{min}}(\Sigma)$ algebras of the exact and approximate $\QET$ to agree. The key observation to be made about \cref{eq:kldiv} is sharpest when $\Sigma$ is a splitting hypersurface. Then, as explained in \cref{ssec:appeftst}, the state $\rho_\times$ is pure, while the state $\wtilde{\rho}$ is mixed due to quantum wormholes. As a result, the relative entropy in \cref{eq:kldiv} is in fact formally infinite. This is more easily seen in terms of the underlying probability density functions of each density operator, for which the relative entropy becomes a classical Kullback–Leibler, and the infinity corresponds to the support of $\wtilde{\rho}$ overlapping with the kernel of $\rho_\times$.

The divergence of \cref{eq:kldiv} has operational consequences. From the interpretation of the relative entropy above \cref{eq:relent}, it means that a single measurement on a single copy of the system should suffice to distinguish $\wtilde{\rho}$ from $\rho_{\times}$. In other words, there must exist $\wtilde{\mathcal{O}}$ operators which cause the approximate $\QET$ to break down in the sense of \cref{eq:breakeft}. As pointed out above, it is not clear that this breakdown is necessarily associated to energy scales. Rather, it must correspond to observables which have support on $\wtilde{\rho}$ but not on $\rho_{\times}$, such that their expectation values are zero for the latter but nontrivial for the former. Needless to say, these operators will also cause any truncated $\QET$ to break down.

The kind of observables which an approximate $\QET$ will miss correspond to operators which can tell apart a mixed from a pure state. These must in some sense be complex operators capable of probing the spectrum of the density operator. Intuitively, the more mixed the true state $\wtilde{\rho}$ is, the easier it should be to realize that the hypothesis state $\rho_\times$ is wrong. The mixing of $\wtilde{\rho}$ is caused by quantum wormholes, which arise from the interactions between $\varphi$ and $\Phi$ fields in the complete theory. These quantum wormholes may ultimately be suppressed by the mass $m_\Phi$ of the lightest $\Phi$ field that was integrated out. As a result, the different $\QET$ states will be harder to distinguish the heavier $m_\Phi$ is. Hence one may expect any reasonable notion of operator complexity for the $\wtilde{O}$ observables that break the approximate $\QET$ to grow with $m_\Phi$. \`A priori this does not imply that $\wtilde{O}$ itself must be heavy: it may be possible to scale the complexity of an operator that distinguishes states without scaling its total energy.\footnote{I thank Chris Akers and Netta Engelhardt for very insightful conversations on the breakdown of effective theory with complexity, and also Aidan Herderschee for highlighting the possibility of a separation of scales.}

Let us conclude by pointing out that the failure of approximate $\QET$ states to capture complex operators is akin to how no single projected state $\wtilde{\rho}[\Phi_0]$ in the ensemble that makes up $\wtilde{\rho}$ can possibly make predictions all consistent with $\wtilde{\rho}$ itself (see \cref{ssec:ensemble}). The true state of the system is $\wtilde{\rho}$, and its predictions carry classical uncertainty due to its ensemble nature. This classical uncertainty was in \cref{ssec:theoryensem} formulated in terms of $\wtilde{\rho}$ being prepared by an ensemble of theories, rather than any single one. Correspondingly, we expect a single approximate $\QET$ to fail to capture the true physics of the $\varphi$ fields, but an ensemble thereof to potentially make consistent predictions. Such an ensemble would naturally be over couplings of the theory, since these ultimately come from integrating out genuinely quantum $\Phi$ fields.

\section{Quantum Gravity}
\label{sec:qg}

There are two complementary but qualitatively distinct ways in which the lessons from the previous sections can also be applied to effective theories in the context of quantum gravity.
Holographic duality offers an arena where the gravitational bulk can be understood as an emergent theory that ought to realize every expectation from a boundary field theory, including an exact $\QET$. Alternatively, were one to start from a fully non-perturbative theory of quantum gravity, one may wonder which aspects of the exact $\QET$ story presented here would still apply to a gravitational effective theory (GET). We will discuss the holographic picture first in \cref{ssec:effholo}, and then comment on distinctive ways in which a GET differs from a $\QET$ in \cref{ssec:effgrav}.

\subsection{Effective Holography}
\label{ssec:effholo}

Great progress has been made in understanding quantum gravity within the holographic paradigm by harvesting the connection between entanglement on the boundary and spacetime connectivity in the bulk \cite{VanRaamsdonk:2010pw,Maldacena:2013xja}. The idea that spacetime emerges from quantum correlations has inspired some of the most insightful entries in the holographic dictionary, advancing our understanding of holographic entanglement \cite{Ryu:2006bv,Hubeny:2007xt,Lewkowycz:2013nqa,Dong:2016hjy,Faulkner:2013ana,Engelhardt:2014gca,Rangamani:2016dms,Wall:2012uf,Akers:2019lzs} and bulk reconstruction \cite{Hamilton:2006az,Czech:2012bh,Headrick:2014cta,Jafferis:2014lza,Almheiri:2014lwa,Pastawski:2015qua,Jafferis:2015del,Dong:2016eik,Harlow:2016vwg}.

\subsubsection{Replica Wormholes}
\label{sssec:hawking}

An important development in recent years has in particular brought to the forefront of holography the importance of wormhole contributions in holographic calculations of entropic quantities \cite{Penington:2019kki,Almheiri:2019qdq,Almheiri:2020cfm,Engelhardt:2020qpv,Chandrasekaran:2022asa}. In the context of black hole evaporation, we have come to understand how crucial it is to include connected wormhole topologies in the gravitational path integral in order to obtain a Page curve consistent with unitarity for the entropy of Hawking radiation \cite{Hawking:1975vcx,Almheiri:2019hni,Penington:2019npb,Almheiri:2019psf}. A specific setting where these wormholes make an appearance is in replica calculations in which the path integral is used to prepare the state of an evaporating black hole \cite{Penington:2019kki,Almheiri:2019qdq,Almheiri:2020cfm}. Rather explicitly, one observes how the build up of quantum entanglement between the black hole and the radiation eventually leads to a dominance of wormhole geometries in the replica calculation which connect the replicas, even when these lie on completely disjoint spaces. In other words, quantum correlations across replicas get geometrized in the bulk in the form of replica wormholes. In this section, we would like to understand these replica wormholes from the perspective of the boundary theory.\footnote{A coarse-graining mechanism by which quantum wormholes reproduce the permutation structure of geometric replica wormholes was studied in detail by \cite{Liu:2020jsv}. 
The relation between the quantum wormholes that arise in open quantum systems and those in holographic replica calculations was also explored by \cite{Pelliconi:2023ojb}.}

While the distinction between black hole and radiation degrees of freedom can be rather naturally implemented in the bulk as a spatial splitting along Cauchy surfaces, this is no longer so straightforward from a dual boundary perspective. Indeed, for arguably just technical reasons, one often introduces an auxiliary non-gravitating bath system that couples to both dual theories in order to more easily distinguish degrees of freedom. This way, in the boundary picture, the bath captures most radiation (early radiation only), whereas the genuinely holographic theory captures mostly the black hole (plus late radiation). The need for introducing such an artificial artifact, and nonetheless still getting a rather ambiguous splitting between the two sectors, emphasizes an important point: these two sectors are not naturally localized on any distinct spatial regions on the boundary, but rather simply correspond to distinct sectors of quantum fields of the theory.

The above observations can be easily translated into the language of state preparation employed in this paper. In particular, we have argued that the state of the radiation on the boundary cannot simply be obtained by partially tracing some state of the complete theory over some spatial region. Rather, the state of the radiation corresponds to partially tracing over an internal sector of black hole states of the complete theory. As shown in \cref{sec:prep}, what constructs such a state is precisely an exact $\QET$.

The boundary theory we are interested in here is obtained by integrating out the black hole sector from the complete theory. The result is an exact $\QET$ that holographically describes the radiation, and which is a strongly non-local theory. The fact that the black hole evaporates is a manifestation of the non-unitarity that is characteristic of any exact $\QET$ due to correlations between system and environment. 
As explained in \cref{ssec:approx}, the non-locality and non-unitarity of an exact $\QET$ are not only consistent with unitarity and locality of the complete theory, but required.
Indeed, in the exact $\QET$ the correlations between black hole and radiation are fully captured by quantum wormholes, and would be completely missed by any approximate $\QET$.
In particular, by the arguments in \cref{ssec:relent}, all possible radiation states the exact $\QET$ could prepare differing only by such correlations would be perfectly indistinguishable by an approximate $\QET$. In the exact $\QET$ these states are distinguishable by sufficiently complex operators, but will require multiple copies of the state. In other words, the $\QET$ framework predicts the failure of any local effective theory to distinguish the state of Hawking radiation from a thermal state, and also the need for replicas where the inclusion of wormholes allows one to recover results consistent with unitarity of the complete theory.

Let us restate that in the above paragraph we are arguing for all of these phenomena purely from a boundary perspective. By holographic duality, these predictions must be reproduced by the bulk theory.
Since more than one copy of the theory is needed to make meaningful statements about the state of the radiation, consider the replica path integrals of \cref{sec:reps}.
On the boundary, the quantum wormholes that couple the replicas come to life as soon as the black hole and radiation get coupled, i.e., as soon as the black hole begins to evaporate. Correspondingly, in the bulk one should expect wormholes connecting all replicas to exist as off-shell configurations of the gravitational path integral at all stages of black hole evaporation. Whether replica wormholes dominate or not is a different question that arises in a semiclassical limit. Namely, it refers to the competition between the limit that allows for saddle-point approximations, and the build-up of strongly non-local correlations to overcome local ones, such that the connected saddle dominates.

In line with the old and fruitful intuition that spacetime geometrizes entanglement, we once again are compelled to claim that geometric replica wormholes precisely capture the ways in which replicas are strongly non-locally coupled in any exact $\QET$ by quantum wormholes. The contrapositive of this statement is even more illuminating. As explained in \cref{ssec:appeftst}, one rarely works with an exact $\QET$, and it is customary to reduce such a strongly non-local theory to a local one by performing a derivative expansion and truncating. This procedure may lead to a seemingly unitary theory (if unitarity bounds are respected), but which will make predictions strictly inconsistent with unitarity of the complete theory (cf. \cref{ssec:approx}). Intuitively, this is because a truncated $\QET$ is quite literally obtained by throwing away information about the complete theory. In addition, the destruction of quantum wormholes also means that replica path integrals in any such truncated $\QET$ cannot possibly couple disjoint spaces. In other words, the violation of unitarity in the complete theory comes hand in hand with the absence of quantum wormholes across replicas in an effective theory.
The situation in the bulk side is entirely analogous: the moment one decides to not include contributions from replica wormholes, unitarity manifestly breaks down in the form of information loss during black hole evaporation.\footnote{I thank Marija Toma\v{s}evi\'{c} for emphasizing these similarities.}

More generally, the relation above between wormholes and unitarity is a manifestation of an unavoidable connection between ensembles and unitarity. 
Namely, only an exact $\QET$ is consistent with unitarity of an underlying complete theory, and the way it is so is by producing such ensembles.
Indeed, as explained in \cref{ssec:ensemble}, an exact $\QET$ generally has quantum wormholes which cause mixing in the preparation of states, and the resulting mixed states can be identified with ensembles of pure states of the complete theory.
Equivalently, \cref{ssec:theoryensem} showed how these ensembles of states can also be interpreted as states prepared by an ensemble of complete theories. 
These two interchangeable pictures provide a precise mechanism for and are consistent with the interpretations of \cite{Bousso:2019ykv,Bousso:2020kmy,Pollack:2020gfa,Freivogel:2021ivu,Chandra:2022fwi,deBoer:2023vsm}.

\subsubsection{Non-Replica Wormholes}
\label{sssec:nonrewo}

What follows is a potentially polarizing observation, since it may seem as trivial to some as provocative to others. For quantum wormholes to arise, it was crucial that the way the replica manifold reassembles into disjoint copies of the original manifold is by combining half-spaces from consecutive replicas. This is fundamentally why the eventually disjoint spaces remain coupled: quantum wormholes within each original replica end up straddling across the reassembled copies in the form of quantum wormholes. This is a very specific construction that allows one to compute state replicas, but has nothing to do with the idea of just taking some complete theory, setting it up on different copies of some space, and then constructing an exact $\QET$ by integrating out some sector. Indeed, were one to do this, the partition function of the resulting exact $\QET$ would just factorize across connected components, as shown in \cref{eq:Zeftfact}. Hence the words `performing a path integral on a space with multiple connected components' do not immediately imply factorization or lack thereof even in field theory. In particular, if the theory at hand is strongly non-local, it all depends on how the multi-integrals in the action behave, which is non-optionally determined by which calculation one is doing.\footnote{Alternatively, one may think there is a choice to include or not include quantum wormholes. But as we saw, their inclusion corresponds to a calculation in the exact $\QET$, whereas ignoring them corresponds to using an approximate $\QET$. It may be that the dichotomy as for whether one should include geometric wormholes in the gravitational path integral is analogous.}

If geometric wormhole contributions to the gravitational path integral are understood as geometrizing quantum wormholes of the boundary theory, then the lesson would be: wormholes must be included in holographic replica calculations for an exact $\QET$, but they must not be included when simply evaluating a partition function of a theory on a set of disjoint spaces. Correspondingly, replica calculations should not factorize and holographically they would not, whereas partition functions on disjoint spaces should factorize and holographically they would. Assuming this interpretation makes sense, using the gravitational path integral correctly would not lead to a factorization problem.

The obvious counterargument to this point of view is that calculations such as those in JT gravity and the dual matrix ensembles genuinely seem to be of the partition function of a theory on a set of disjoint spaces. And in these calculations, including wormholes does seem to capture consistent spectral statistics of black hole microstates. There are actually two reasons why this may be reconciled with the statements above. We offer a possible explanation for non-replica wormholes here, and a reason why these wormholes may secretly actually be replica wormholes in \cref{sssec:tfdtmd}.

The first explanation is as follows. In holography, we often write $Z(M)$ to refer to the path integral of a field theory on a boundary manifold $M$, and then $\mathcal{P}(M^n)$ to refer to whatever the gravitational path integral computes with $n$ copies of $M$ as its boundary conditions. Upon the inclusion of wormholes we then obtain the statistical interpretation of non-factorization in $\mathcal{P}(M^n) \equiv \langle Z(M)^n \rangle$ as caused by averaging over complete theories. However, because we are treating gravity as an effective theory, we already see that a more natural interpretation would read $\mathcal{P}(M^n) \equiv \langle \wtilde{Z}(M)^n \rangle$, where the average is over exact $\QET$ partition functions. But then a simple explanation for what happens when we allow for wormholes to connect the $n$ boundaries is that we are simply evaluating $\mathcal{P}(M^n) = \langle \wtilde{Z}(M^n) \rangle$, and allowing for quantum wormholes in the exact $\QET$ to straddle across distinct copies of $M$.
In other words, we would be taking the exact $\QET$ partition function $\wtilde{Z}(M)$ and deciding to define it as a theory on all $n$ boundaries by simply replacing $M\to M^n$ everywhere in the action. This would be very unnatural from the perspective of the exact $\QET$, but would nonetheless give statistical correlations across copies of $M$ which ultimately are nontrivial precisely because the exact $\QET$ does genuinely have quantum wormholes.

This way, $\wtilde{Z}(M)^n$ and $\wtilde{Z}(M^n)$ would already be different from the boundary perspective, and the distinction between them, though unphysical, would be clear. In the factorizing case of $\wtilde{Z}(M)^n$, the exact $\QET$ has strong non-localities only within connected components, whereas in the non-factorizing case of $\wtilde{Z}(M^n)$ these would extend to multiple components in the form of quantum wormholes. One could then speculate that this is also the mechanism at play for gravitational wormholes in the bulk. If so, to compute $\wtilde{Z}(M)^n$ one would be instructed to include all possible topologies anchoring to a single boundary, but to not allow for wormholes connecting multiple boundaries. Notice how this naturally gives single-boundary wormholes and multi-boundary wormholes a qualitatively different status, and a field theory argument for why one should include the former but not the latter in certain calculations. As for the non-factorizing case of $\wtilde{Z}(M^n)$, the artificial quantum wormholes would get geometrized by similarly artificial multi-boundary gravitational wormholes. Any statistical correlations across components of $M^n$ would then seem to be capturing the physics of objects which do exist in the theory (quantum wormholes and gravitational wormholes), but in a setting where they would actually not arise physically.

\subsubsection{Non-Replica Wormholes as Replica Wormholes}
\label{sssec:tfdtmd}

A typical calculation where geometric multi-boundary wormholes are believed to contribute involves providing the gravitational path integral with $n$ copies of a thermal partition function $Z_\beta$ as its boundary conditions \cite{Cotler:2016fpe,Liu:2018hlr,Stanford:2020wkf,Saad:2018bqo,Okuyama:2018gfr,Saad:2019lba,Saad:2019pqd,Winer:2020gdp,Zhou:2023qmk}. The non-factorizing answer that wormholes lead to is statistically interpreted
as corresponding to the calculation of an $n^{\text{th}}$ moment of $Z_\beta$ in some ensemble of theories, which we may denote by $\langle Z_\beta^n \rangle$. Explicitly, writing $M$ for the manifold boundary condition associated to each $Z_\beta$, one identifies
\begin{equation}
\label{eq:multitr}
    \mathcal{P}(M^n) \equiv \langle Z_\beta^n \rangle.
\end{equation}
Because the theories on each copy of $M$ are supposed to be independent from each other, the average in \cref{eq:multitr} seems to be the only reasonable way to make sense of non-factorization. In fact, there are other quantities $\mathcal{P}(M^n)$ could be computing, with identical boundary conditions for the gravitational path integral, and consistent with having a single complete theory on the boundary.

The non-factorizing result $\langle Z_\beta^n \rangle$ when a factorizing $Z_\beta^n$ answer was expected is just a paradigmatic example of a more general concern. Namely, if $Z$ is the partition function of a complete theory, we expect a factorizing result, and not something like $\langle Z^n \rangle$. Actually, as explained in \cref{eq:Zfact}, also the partition function $\wtilde{Z}$ of an exact $\QET$ must factorize. In what follows, we identify replica calculations which are perfectly consistent with both sides of \cref{eq:multitr}, but where the wormholes on the left are replica wormholes, and the average on the right is consistent with having a single, complete theory.

Suppose one has an exact $\QET$ on the boundary and considers preparing a state $\wtilde{\rho}$ on a splitting $\Sigma$. This is done by performing a path integral on the cut-open space $M_\Sigma$. To prepare $n$ independent copies of the state, one can do $n$ independent path integrals on the disjoint space $M_\Sigma^n$. Upon identifications under the trace, as explained in \cref{sec:reps}, the replica path integral for $\Tr\wtilde{\rho}^n$ can be performed on a replica manifold $M_\Sigma^{(n)}$ which turns out to be isomorphic to $M^n$ for $\Sigma$ splitting. Despite being isomorphic to $M^n$, the resulting space connects different replicas as in \cref{eq:repm}. Hence as we have already discussed multiple times the resulting calculation does not factorize due to quantum wormholes, and in addition admits an interpretation as an ensemble average over factorized objects like \cref{eq:ugly}.
To make the connection to the factorization problem explicit, we would like to relate such an ensemble average to one over partition functions of different theories. We now give two general situations in which $\Tr\wtilde{\rho}^n$ can indeed be understood this way.

The first situation follows from \cref{eq:tromewa} and yields
\begin{equation}
\label{eq:ensem1}
    \Tr\wtilde{\rho}^n = \frac{1}{Z^n} \, \left\langle Z_{\Phi_0}^n \right\rangle_{{{{P}}}^n} +  O(\text{RSB}_{\evalat{\Phi}{\Sigma}}),
\end{equation}
where all partition functions are evaluated on $M$. Here, recall that $Z_{\Phi_0}$ are the partition functions of projected theories defined in \cref{eq:theoryzsn}, and which do factorize in the usual sense, i.e., $Z_{\Phi_0}[M^n]=Z_{\Phi_0}[M]^n$. That the RSB corrections in \cref{eq:ensem1} are small relies on the assumption that the projected states from \cref{eq:irstates} are approximately trace-orthogonal. This will generically be the case if the projection from the complete theory does not generate too many null states. More precisely, this assumes that $\dim\mathcal{H}_\Phi$ is not much larger than $\dim\mathcal{H_\varphi}$. Altogether, this means that the replica-symmetric approximation that the first term in \cref{eq:ensem1} gives should hold so long as the Hilbert space of states that was integrated out from the complete theory was not too large. Then, the statement of \cref{eq:ensem1} is that the replica calculation for an exact $\QET$ state takes precisely the form of an ensemble average over partition functions on $M^n$ which do factorize.

The second situation in which this happens does not require assuming anything about the Hilbert space of the $\Phi$ fields in the complete theory. From \cref{eq:rsbphi}, we obtain
\begin{equation}
\label{eq:ensem2}
    \Tr\wtilde{\rho}^n = \frac{1}{Z^n} \left\langle \wtilde{Z}_{\varphi_0}^n \right\rangle_{\wtilde{{{{P}}}}^n} +  O(\text{RSB}_{\evalat{\varphi}{\Sigma}})
\end{equation}
where all partition functions are again evaluated on $M$. The partition functions $Z_{\varphi_0}$ were defined in \cref{eq:zvphi0}, and note that these also factorize due to the projection of $\varphi$ to $\varphi_0$ on $\Sigma$.
Now the RSB term in \cref{eq:ensem2} corresponds to $\varphi$ field configurations across replicas that break replica symmetry on the $\Sigma$ cuts. The assumption that his RSB term is small is the usual assumption of replica symmetry motivated by the cyclic $\mathbb{Z}_n$ symmetry of the overall path integral. Namely, one typically expects replica-symmetric configurations to dominate on the basis of symmetry arguments. This is generically a safe assumption for positive $n\in\mathbb{Z}^+$, particularly when the $\varphi$ path integral admits a saddle-point approximation which respects the $\mathbb{Z}_n$ symmetry.\footnote{Analytic continuations for replica tricks, specifically to $n<1$, are a typical example of when RSB contributions become non-negligible, and in fact oftentimes dominant.} As a result, \cref{eq:ensem2} provides another situation in which a replica calculation takes the form of an ensemble average over theories.

Crucially, note that in both \cref{eq:ensem1,eq:ensem2} we are considering replica calculations for a state $\wtilde{\rho}$ on a splitting $\Sigma$, such that $M_\Sigma^{(n)} \cong M^n$. In a bulk calculation, it would thus be natural to provide the gravitational path integral with $M^n$ as its boundary conditions. The inclusion of geometric wormholes in this case makes sense, since from the boundary perspective there genuinely are quantum wormholes coupling the replicas. This provides an interpretation of $\mathcal{P}(M^n)$ with wormholes included as a bulk calculation of $\Tr\wtilde{\rho}^n$ (where factorization should not happen) rather than a calculation of $\wtilde{Z}^n$ (where factorization should happen).

Alternatively, the constructions above may be thought of as a version of the factorization puzzle which places the focus on the room for ambiguity that exists in how boundary conditions for the gravitational path integral are prescribed. Namely, one lesson from this exercise is that generally $\mathcal{P}(M^n)$ is not a well-defined object: simply feeding the manifold boundary $M^n$ to the gravitational path integral is not enough to single out a specific quantity.\footnote{A related construct which would also gives these ambiguities is the thermo-mixed double of \cite{Verlinde:2020upt,Verlinde:2021kgt}.}

To finish this discussion, let us address the concrete case of thermal partition functions in \cref{eq:multitr} more directly.
Consider a complete theory defined on a space $M$ with a $U(1)$ symmetry around a non-contractible cycle of length $\beta$. Let $\Sigma$ be a hypersurface splitting $M$ into two $\mathbb{Z}_2$-symmetric dual halves $X$ and $X^*$. Each such half-space contains a half-cycle, or segment, of length $\beta/2$. Hence $\partial X$ consists of two disjoint boundary components, one at each end-point of the segment, and similarly for $\partial X^*$. We refer to these as left and right boundaries, respectively $\Sigma_L$ and $\Sigma_R$, and write $\Sigma \equiv \Sigma_L \sqcup \Sigma_R$. The state that the exact path integral prepares on the cut-open space $M_\Sigma$ is the density operator of the thermofield-double (TFD) state. Suppose, however, that we are not using the complete theory on the boundary, but an exact $\QET$. This theory cannot possibly prepare a pure TFD. Instead, it would compute some generally mixed state $\wtilde{\rho}_{\smalltext{TFD}}$ obtained by partially tracing out whichever sector was integrated out from the complete theory. If we further trace out one side of the TFD, what we obtain is a similarly approximate thermal state $\wtilde{\rho}_\beta$ on the other side.

There are two inequivalent calculations one may perform with the states $\wtilde{\rho}_{\smalltext{TFD}}$ and $\wtilde{\rho}_\beta$,
\begin{equation}
\label{eq:wrong}
    (\Tr \wtilde{\rho}_{\beta})^n \neq \Tr (\wtilde{\rho}_{\smalltext{TFD}}^{~n} ).
\end{equation}
Both of these take the form of a boundary path integral on a space isomorphic to $M^n$, but only one of them factorizes. On the left, we have a multi-trace quantity giving $\wtilde{Z}(\beta)^n$ if $\wtilde{\rho}_{\beta}$ is left un-normalized, i.e., the exact $\QET$ computation of $n$ copies of the thermal partition function. This calculation takes place on $M^n$ and obviously factorizes since it would in particular be the same calculation in the complete theory. On the right, we have a single-trace replica calculation for the state $\wtilde{\rho}_{\smalltext{TFD}}$ which obviously cannot factorize because the state is mixed. This calculation takes place on $M_\Sigma^{(n)}\cong M^n$. In other words, both calculations take place on the same manifold. It would thus seem that a gravitational path integral of the form $\mathcal{P}(M^n)$ cannot possibly tell the difference between the distinct quantities on the left and right of \cref{eq:wrong}. This is particularly worrisome because the discrepancy between these quantities is not merely quantitative, but qualitative: the one on the left factorizes, whereas the one on the right does not.

\subsubsection{Quantum Error Correction}
\label{sssec:qec}

The mechanism by which bulk operators are encoded in and can be reconstructed from boundary ones has by now been well understood in terms of operator algebra quantum error correction \cite{Almheiri:2014lwa,Pastawski:2015qua,Jafferis:2015del,Dong:2016eik,Harlow:2016vwg}.
The $\QET$ framework naturally realizes the structure of a quantum error correcting code in terms of the algebras of operators introduced in \cref{sec:algebras}. This can be discussed most transparently by relating the algebras of the approximate $\QET$ and the exact $\QET$ following \cref{ssec:qetalg}, with no need to refer to the complete theory.\footnote{The exact $\QET$ already has all the structure needed to reconstruct operators of the approximate $\QET$, and the additional information the complete theory contains is unnecessary. The extreme case of a complete theory where the sector that is integrated out does not interact with the sector of interest illustrates this.}

The relevant identifications to be made with the ingredients of quantum error correction are as follows. The physical theory is the exact $\QET$, which has a Hilbert space $\mathcal{H}_{\smalltext{exact}}$ of states with quantum wormholes, and an algebra $\mathcal{A}_{\smalltext{exact}}$ which includes both locally generated operators and also non-local ones whose expectation values are nontrivial due to quantum wormholes. This algebra can thus be written
\begin{equation}
\label{eq:aworm}
    \mathcal{A}_{\smalltext{exact}} = \mathcal{A}_{\smalltext{loc}} \vee \mathcal{A}_{\smalltext{worms}}.
\end{equation}
The logical theory is the approximate $\QET$, whose Hilbert space $\mathcal{H}_{\smalltext{approx}}$ is isomorphic to a code subspace $\mathcal{H}_{\smalltext{code}} \subset \mathcal{H}_{\smalltext{exact}}$ of states where quantum wormholes are projected out. The algebra $\mathcal{A}_{\smalltext{approx}}$ of the approximate $\QET$ is locally generated and, by \cref{eq:exapp}, is isomorphic to the local sector of $\mathcal{{A}_{\smalltext{exact}}}$, i.e.,
\begin{equation}
    \mathcal{A}_{\smalltext{approx}} \cong \mathcal{A}_{\smalltext{loc}}.
\end{equation}
That $\mathcal{A}_{\smalltext{approx}}$ is reconstructible from $\mathcal{A}_{\smalltext{exact}}$ is thus obvious at a global level. The more interesting setting corresponds to reconstruction within an arbitrary subspace $\Sigma$. In this case the minimal algebras locally generated within $\Sigma$ of both effective theories are still isomorphic, and so are the maximal algebras if the $\mathcal{A}_{\smalltext{worms}}$ subalgebra of the exact $\QET$ is truncated away.
\footnote{Maximal algebras on $\Sigma$ now both involve non-local operators which are not generated locally. However, these are associate to weak non-localities, and not to quantum wormholes. This non-wormhole sector is what $\mathcal{A}_{\smalltext{loc}}$ labels in this case.}
Reconstruction of operators in the minimal algebra for $\Sigma$ in the approximate $\QET$ will be possible with access to operators in the minimal algebra for $\Sigma$ in the exact $\QET$, and similarly for maximal ones. However, unsurprisingly, it will not be possible to reconstruct non-local operators in the maximal subalgebra for $\Sigma$ in the approximate $\QET$ with just access to the locally generated minimal algebra for $\Sigma$ in the exact $\QET$.

To see how this works, it is instructive to consider the standard theorem addressing the necessary and sufficient conditions for recoverability in operator algebra quantum error correction \cite{B_ny_2007,B_ny_2007b,Almheiri:2014lwa}. The relevant elements for error correction on $\Sigma$ are an arbitrary logical operator $O\in \mathcal{A}_{\smalltext{loc}}$ encoded to act only on $\mathcal{H}_{\smalltext{code}}$, and physical operators in $\mathcal{A}_{\smalltext{exact}}(\Sigma)$ and its commutant. We remain for now agnostic as to whether the algebras assigned to $\Sigma$ are minimal or maximal in either theory. The basic statement of the theorem is that $O$ is recoverable from $\mathcal{A}_{\smalltext{exact}}(\Sigma)$ if and only if\footnote{This theorem is usually stated for $X$ in the algebra of the complementary subspace $\Sigma'$. Such a statement would agree with ours for algebras obeying duality. In our case duality is generally violated, so referring to the algebra of the complementary subspace would be ambiguous.}
\begin{equation}
\label{eq:reconst}
    \bra{\Psi} [ O, X ] \ket{\Psi} = 0, \qquad \forall X\in (\mathcal{A}_{\smalltext{exact}}(\Sigma))', \quad \forall \ket{\Psi}\in\mathcal{H}_{\smalltext{code}},
\end{equation}
It is not hard to figure out for which $O$ this condition holds. Obviously any operator in the bicommutant, $O \in (\mathcal{A}_{\smalltext{exact}}(\Sigma))'' = \mathcal{A}_{\smalltext{exact}}(\Sigma)$, obeys this condition, where we have assumed the algebra is von Neumann for simplicity.
However, we expect more operators to be recoverable given the projection onto $\mathcal{H}_{\smalltext{code}}$ that \cref{eq:reconst} implements.
The algebra $(\mathcal{A}_{\smalltext{exact}}(\Sigma))'$ takes the form of \cref{eq:aworm}, but the projection in \cref{{eq:reconst}} removes any potential contribution from the wormhole sector. Hence if an operator $X$ obeys the commutation relation in \cref{eq:reconst}, any other operator in $(\mathcal{A}_{\smalltext{exact}}(\Sigma))'$ that only differs from $X$ due to quantum wormholes will do so too. This suffices to conclude that the algebra of recoverable operators is given by
\begin{equation}
\label{eq:recono}
    \{ O \in \mathcal{A}_{\smalltext{min}}  \st O \in \mathcal{A}_{\smalltext{exact}}(\Sigma) \vee \mathcal{A}_{\smalltext{worms}} \},
\end{equation}
where recall $\mathcal{A}_{\smalltext{worms}}$ captures globally non-local operators associated to the full theory, and not to any specific subspace $\Sigma$. To diagnose recoverability, it only remains to specify which operators we actually have access to on $\Sigma$, i.e., which algebra we are associating to $\Sigma$ in the exact $\QET$. If it is the minimal one, then the algebra of recoverable operators is simply $\mathcal{A}_{\smalltext{min}}(\Sigma)$, since no operator in $\mathcal{A}_{\smalltext{worms}}$ is in $\mathcal{A}_{\smalltext{min}}$. In this case accessible and recoverable algebras are isomorphic, so the decoding map would also be unique up to isomorphism. If instead we associate the maximal one, then the algebra of recoverable operators consists of those operators in $\mathcal{A}_{\smalltext{max}}(\Sigma) \vee \mathcal{A}_{\smalltext{worms}} = \mathcal{A}_{\smalltext{max}}(\Sigma)$ which are also in the full $\mathcal{A}_{\smalltext{loc}}$. In other words, these are the type of operators that a maximal algebra on $\Sigma$ would include in the approximate $\QET$, but which in the exact $\QET$ have to be restricted to being additively generated within the full theory so that they have no support on $\mathcal{A}_{\smalltext{worms}}$. This case is more interesting, since now the accessible algebra includes $\mathcal{A}_{\smalltext{worms}}$, whereas the recoverable algebra does not. 

The last situation above is more reminiscent of holography, where there is a large amount of non-local boundary freedom in reconstructing local bulk operators. In other words, there exist decoding maps which allow to reconstruct bulk operators from many different boundary regions. This freedom here comes from the quantum wormholes inherited from the complete theory, and which respect \cref{eq:recono} by virtue of only acting nontrivially outside the code subspace. States in the code subspace of the exact $\QET$ can be perfectly represented by the approximate $\QET$, much like semiclassical bulk gravity is dual to holographic boundary states in the code subspace. On the opposite extreme one has wormhole states of the exact $\QET$ with no support in the code subspace, which cannot be represented at all by the approximate $\QET$. In holography, these are presumably highly quantum gravitational states with no geometric interpretation. 

Continuing with this line of reasoning, one may expect that states of the exact $\QET$ which are mostly but not completely in the code subspace will be very accurately described by the approximate $\QET$. The description of the latter will however only be approximate, with errors corresponding to its inability to describe quantum wormholes. These errors can be quantified by the relative entropy measures introduced in \cref{ssec:relent}, which diagnose how distinguishable states of the two theories are. As explained therein, operationally these are distinguishable by sufficiently complex operators. In holography, this nicely connects with the idea that highly complex and non-local observables are needed in order to capture any departure of bulk physics from semiclassical gravity \cite{Harlow:2013tf,Susskind:2014rva,Engelhardt:2018kcs,Brown:2019rox,Kim:2020cds,Engelhardt:2021mue,Engelhardt:2021qjs,Akers:2022qdl,Yang:2023zic}.

\subsubsection{Baby Universes}
\label{sssec:babies}

Another set of ideas that is precisely captured by our constructs in \cref{ssec:ensemble,ssec:theoryensem} are the baby universes and $\alpha$-states of \cite{Coleman:1988cy,Coleman:1988tj,Giddings:1988cx,Giddings:1988wv,Preskill:1988na,Klebanov:1988eh,Lyons:1991im,Hawking:1991vs,Marolf:2020xie,McNamara:2020uza,Blommaert:2021fob,Blommaert:2022ucs}. 
In our language, consider a complete theory defined on a manifold $M$, and a splitting hypersurface $\Sigma$ consisting of arbitrarily many connected components $\{\Sigma_k\}$. 
The state one obtains on $\Sigma$ by evaluating the path integral on the cut-open space $M_\Sigma$ is $\rho$, which is pure. If instead one uses the path integral of an exact $\QET$ to prepare a state on $M_\Sigma$, one obtains the mixed state $\wtilde{\rho}$. By virtue of being mixed, $\wtilde{\rho}$ can be thought of as an ensemble of pure states.

Our boundaries for baby universes are the distinct connected components $\{\Sigma_k\}$, and the choice of sources on each corresponds to the specification of field configurations ${\Phi_0}_k$ for the $\Phi$ fields. In the path integral for the complete theory on $M_\Sigma$, we may thus consider inserting projections of the form $\delta(\evalat{\Phi}{\Sigma_k} - {\Phi_0}_k)$. In the quantum mechanical language these would be the operators
\begin{equation}
\label{eq:sourcezj}
    \widehat{B[{\Phi_0}_k]} \equiv \ket{{\Phi_0}_k}\bra{{\Phi_0}_k}.
\end{equation}
These operators commute, as is most obviously seen from the fact that they are associated to delta functions in the path integral.
The state one obtains by inserting no operators in the path integral on $M_\Sigma$ is the pure state $\rho$ above. We identify this as our Hartle-Haking state, and write $\rho \equiv \ket{\text{HH}}\bra{\text{HH}}$. Generally, $\ket{\text{HH}}$ is acted on nontrivially by all possible operators in \cref{eq:sourcezj}.
Using this state, one can thus think of the path integral on $M_\Sigma$ with source insertions as actions of the operators in \cref{eq:sourcezj} on $\ket{\text{HH}}$.
Similarly, the path integral with insertions on $M$ without cuts corresponds to computing expectation values of these operators in the Hartle-Hawking state. 

The states the operators in \cref{eq:sourcezj} give rise to are those in \cref{eq:liftrho}, which correspond to lifts of the projected states from \cref{eq:irstates} to the Hilbert space $\mathcal{H}$ of the complete theory. 
The Hilbert space of baby universes is thus more naturally associated to the projected states, which here we denote by $\wtilde{\rho}[\{{\Phi_0}_k\}]$. Recall that these states are pure only on splitting hypersurfaces, and are otherwise mixed. Therefore here purity corresponds to having choices of sources $\{{\Phi_0}_k\}$ specified on every $\Sigma_k$ boundary. Otherwise, for any finite number of sources, these $\wtilde{\rho}[\{{\Phi_0}_k\}]$ states are mixed.
The span of all possible such states defines a (not-necessarily proper) subspace of $\mathcal{H}_\varphi$. Because the $\Phi$ fields define the space of sources for baby universes, it follows that there will be null state combinations of the $\wtilde{\rho}[\{{\Phi_0}_k\}]$ states if $\dim\mathcal{H}_\Phi$ exceeds the dimensionality of the Hilbert space of baby universes, as observed in \cref{eq:nulls}.

Because the operators in \cref{eq:sourcezj} all commute, they can be simultaneously diagonalized. The orthonormal eigenstates to all of them, known as $\alpha$-states, are here given by states of the form $\Ket{\{{\Phi_0}_k\}}$ with delta sources specified on every single $\Sigma_k$ boundary by a projection operator $\alpha[\{{\Phi_0}_k\}]$. These were constructed as combinations of projected states in \cref{eq:chista}, but here can be more easily seen to come from specifications of infinite combinations of sources.
That these are orthonormal is just the statement of \cref{eq:orthoiruv}. The overlaps of $\alpha$-states with the Hartle-Hawking state give precisely the probability density functions introduced in \cref{eq:decomprho,eq:palpha},
\begin{equation}
    P_*[\alpha] = \frac{\left|\langle {\text{HH}| \{{\Phi_0}_k\}} \rangle \right|^2}{\braket{\text{HH}}}.
\end{equation}
This is the probability of projecting the state of the complete theory onto an $\alpha$-sector. In terms of the exact $\QET$, the state one obtains without projection is $\wtilde{\rho}$, which by \cref{eq:ensemrho} is an ensemble of baby universes. By projecting onto an $\alpha$-sector, one obtains the state $\wtilde{\rho}[\{{\Phi_0}_k\}]$, which is a pure state corresponding to a definite baby universe state.

There is, however, a fundamental difference between the usual narrative for $\alpha$-states in the literature and our construction. The picture the $\QET$ framework provides for $\alpha$-states more closely parallels the perspective of \cite{Hawking:1991vs}. In particular, different $\alpha$-states correspond to different states of the $\Phi$ fields of the complete theory, which are not merely global parameters but genuine quantum fields with nontrivial position-dependent configurations. Correspondingly, an average over $\alpha$-sectors does not take the form of a integral over ad-hoc global $\alpha$-parameters that generate strong non-localities that do not decay with distance. Instead, our average over $\alpha$-sectors is a functional integral over physical degrees of freedom that were already there in the complete theory. In \cref{sssec:trunc} we provide an example of a (wrong) way of truncating a complete theory that leads to an exact $\QET$ with the features of a traditional $\alpha$-parameter integral where strong non-localities do not decay with distance.

\subsection{Effective Gravity}
\label{ssec:effgrav}

Although holographic duality provides a non-perturbative definition of quantum gravity in terms of a non-gravitational quantum field theory, one would ultimately hope there to also exist an equally complete but inherently gravitational description of quantum gravity. Whether this is string theoretic or not, dual to a field theory or not, for such a non-perturbative formulation to have a gravitational flavor, one would at least expect it to reduce in some appropriate limit to a geometric theory of metric manifolds. In other words, a theory of quantum gravity should reproduce semiclassical gravity as we know it: a GET where spacetime is dynamical, with all its dimensions manifest and solely governed by a (potentially effective rather than fundamental) metric field.

Given the explorations in this paper on $\QET$, one would hope to be able to draw some lessons about GET as well. However, to say that a GET should somehow behave like a $\QET$ just because both are effective theories begs the questions: which aspects of the derivation of a $\QET$ from a complete field theory do still makes sense in quantum gravity? While answering this question reliably would require actually having a complete theory of quantum gravity, a few distinguishing features the gravity framework exhibits which are generically absent in field theory and their expected consequences can already be pointed out. In what follows, we will first explain how IR/UV mixing in gravity may naturally cause a GET to exhibit more dramatic non-localities than a $\QET$ obtained by renormalization does, and also touch on the ways in which Kaluza-Klein renormalization intrinsically truncates gravity in ways that Wilsonian renormalization in field theory does not.

\subsubsection{IR/UV Mixing}
\label{sssec:iruv}

As emphasized at the beginning of \cref{ssec:eftnonlo}, the construction of a $\QET$ does not rely in any way on the idea of renormalization \`a la Wilson, and simply requires a splitting of degrees of freedom into two sectors. For instance, in \cref{ssec:effholo} these two sectors were the black hole and the radiation. The notion of a GET does however encompass the idea of working with gravity at low energies or, more precisely, low curvatures. But the distinction between high and low energies in gravity is a very subtle one, as it turns out, due to the phenomenon of IR/UV mixing.\footnote{I thank Jake McNamara for pointing out the importance of IR/UV mixing in effective gravity.} The paradigmatic thought experiment illustrating IR/UV mixing in quantum gravity proceeds as follows.

Consider a particle physics collision at some center of mass energy $E$, and let $E_P$ denote the Planck energy. In the IR, at energies $E$ many orders of magnitude below $E_P$, gravity plays no role and the Standard Model applies. At higher energies but still with $E\ll E_P$ gravity begins becoming important, but it can still be treated semiclassically as quantum fields propagating on a fixed curved background. A perturbative treatment of quantum gravity is also possible, and allows one to account for graviton contributions to particle scattering. 
However, as soon as scales $E \sim E_P$ are reached, one loses perturbative control and this framework breaks down: gravity becomes genuinely quantum. At such energies we expect the collision to involve the formation of a Planck-sized black hole, the physics of which is unknown. In the context of string theory, this is the regime in which Horowitz-Polchinski describes a continuous transition as the energy is raised between a state of highly excited strings and D-branes, and a string-size black hole \cite{Horowitz:1996nw}. Remarkably though, as one continues to raise the energies towards $E\gg E_P$, semiclassical gravity takes over again. The outcome of such an ultra-high energy collision is again well understood and takes the form of a black hole whose Hawking radiation can be well described semiclassically. And if $E$ is many orders of magnitude larger than $E_P$, the black hole will be sufficiently large that quantum effects in fact become negligible and the end state of the collision is just a classical black hole.

The fact that the deep UV is controlled by IR physics is a characteristic feature of gravity that is rarely seen in field theory. Conversely, in gravity one also observes that the very IR energy scale of the Hawking radiation a large black hole emits is governed inversely by the very UV scale of the mass of the black hole. These are clear manifestations of the phenomenon of IR/UV mixing, which makes assuming a decoupling between IR and UV sectors in a GET na\"ive.
Another reason for expecting significant IR/UV mixing effects in gravity which explicitly also leads to non-local physics is diffeomorphism invariance. The very fact that there exist no local observables in quantum gravity and that any operator has to be gravitationally dressed gives rise in and on itself to a mixing between length scales with an intrinsically non-local flavor \cite{Giddings:2005id,Donnelly:2015hta}.

These IR/UV mixing features do not mean a GET does not make sense. In fact, as we emphasized multiple times, effective theories can be generally defined irrespective of any notion of energy scales. All it means is that the mechanisms that suppress or enhance certain phenomena in an effective theory are not the ones na\"ively expected from the perspective of renormalization in field theory. 
In particular, the non-localities the effective theory exhibits may not merely be exponentially decaying in distance scales determined by some mass of the lightest field that was integrated out (if massive at all). Instead, the scale of non-localities will be governed by IR/UV mixing effects which may actually be dynamically determined and not as simple as some parameter of the theory.

This intuition comes from somewhat exotic field theory constructions which accomplish IR/UV mixing effects by virtue of being defined on non-commutative spaces \cite{Minwalla:1999px,Douglas:2001ba,Szabo:2001kg,Craig:2019zbn}. These non-commutative field theories exhibit momentum-dependent non-localities of arbitrarily long range which actually resemble the physics of strings in many ways \cite{Rajaraman:2000dw,Kiem:2000wt}. Such non-commutative spaces also arise when studying string theory itself with a background magnetic field \cite{Witten:1985cc,Seiberg:1999vs}. In string theory, a simple argument for IR/UV mixing arises simply from worldsheet modular invariance. In particular, this symmetry relates the low and high energy spectrum of the string, suggesting that the string tension may set a scale for IR/UV mixing.

All in all, one may expect that the natural consequence of IR/UV mixing for a GET would be to further exacerbate the non-localities relative to those expected for a $\QET$. This would lend further support to the idea put forward in \cref{ssec:effholo} that gravitational wormholes may be a manifestation in a GET of what we have been calling quantum wormholes in a $\QET$. This is compelling from a holographic perspective since, as we have seen, gravitational wormholes behave in very much the same way as and capture similar physics to quantum wormholes. But from an intrinsically quantum gravity perspective it also suggests an intriguing possibility. Generally, gravitational wormholes can act as mediators of non-local interactions between quantum fields propagating on them, and thus integrating over gravitational wormholes induces non-localities on otherwise local quantum fields.
If gravitational wormholes really are on the same footing as quantum wormholes, it would be interesting to understand if the end result of integrating out gravity wormholes can be fully captured by a strongly non-local field theory on a fixed background akin to an exact $\QET$.\footnote{I thank Daniel Harlow for bringing up this possibility.}

\subsubsection{Kaluza-Klein Renormalization}
\label{sssec:kkr}

Here we comment on some standard procedures within GET interpreted as a gravitational path integral, and the ways in which they are comparable to very coarse truncations in $\QET$, and nowhere close to exact or even approximate $\QET$ manipulations.

We will take the gravitational path integral as our starting point. Such an object is formally understood as path integrating not only over all possible metrics one can endow a given manifold with, but also over manifolds of all possible topologies. In the semiclassical limit of string theory though, it is customary to begin by discriminating large, extended dimensions from small, compact ones, and fixing the topology of the latter.
From the point of view of a gravity, this is already in tension with diffeomorphism invariance of the theory. Moreover, if the complete theory is to allow for all possible metrics and topologies, this also constitutes a dramatic truncation of the phase space of quantum gravity. 

After having fixed a topology for the compact space, Kaluza-Klein dimensional reduction then allows one to fully encode all possible metric fluctuations of the small dimensions into matter fields propagating on the extended ones. If the resulting tower of Kaluza-Klein modes admits a consistent truncation, then one generally also throws away all but the lightest ones, effectively freezing the compact dimensions or even getting rid of them entirely.
This procedure again does not consist of integrating out some sector of heavy Kaluza-Klein modes, but rather of drastically truncating away infinitely many quantum fields.
While this may seem natural from the perspective of decoupling in Kaluza-Klein renormalization, it is a rather radical operation from a spacetime point of view: this truncation effectively removes regions of spacetime (entire dimensions, actually). Were one to actually integrate out the tower of Kaluza-Klein modes, this would correspond to tracing out spacetime dimensions, a procedure which from a field theory perspective should lead to very significant mixing in the states that the resulting GET would prepares. Put differently, an actual trace over Kaluza-Klein modes would result in a strongly non-local $\QET$, but this is not what is often done. We exemplify in \cref{sssec:trunc} why the mass hierarchy of the tower of Kaluza-Klein modes is not obviously helpful for performing this integration.

All truncations above are mainly motivated by the intuition from Wilsonian renormalization to low energies, and correspondingly one typically assumes that energy scales such as that of the lowest Kaluza-Klein mode that was truncated away will rule the breakdown of the resulting GET. But as reviewed in \cref{sssec:iruv}, the phenomenon of IR/UV mixing should make one wary of such expectations. After all, it may not be so surprising if a GET constructed this way exhibits non-localities in certain regimes which are a lot more dramatic than expected.

\subsubsection{A Truncation Gone Wrong}
\label{sssec:trunc}

This section presents a simple example of how a na\"ive mode truncation inside a path integral can lead to a completely wrong result as compared to what exact integration actually gives.

Consider a setting like in \cref{ssec:eg}, involving a theory on a space $M$ with an action given by \cref{eq:iii,eq:Zeftfact}, and with the interaction term in \cref{eq:simint}.
We would like to construct a $\QET$ for the $\varphi$ fields. We already know the form of the exact $\QET$ action one obtains by exact integration, which is given in \cref{eq:fstqet}. From \cref{eq:fstqetapp} and below, we also know how to obtain approximations to this result in a reliable way. Here we will attempt (and fail) to find a shortcut to these results by trying to avoid exact integration of a full tower of modes.

Suppose the Laplacian on $M$ admits a complete set of orthonormal eigenfunctions $\{\Phi_k\}$ with nonnegative eigenvalues $\{\lambda_k^2\}$, i.e.,\footnote{Similar expressions apply to cases in which this set is a continuous family of eigenfunctions, for which sums become integrals. We are using a discrete notation just for the sake of clarity.}
\begin{equation}
    -\nabla^2 \Phi_k = \lambda_k^2 \Phi_k,
\end{equation}
where completeness and orthonormality respectively mean
\begin{equation}
\label{eq:comport}
    \sum_{k=0}^\infty \Phi^\dagger_k(x) \Phi_k(y) = \delta(x-y), \qquad \int\displaylimits_M d^dx \, \Phi_k^\dagger(x) \Phi_{l}(x) = \delta_{kl}.
\end{equation}
One can decompose $\Phi$ into eigenfunctions of the Laplacian (cf. a momentum representation),
\begin{equation}
\label{eq:PhiN}
    \Phi(x) = \sum_{k=0}^{\infty} \alpha_k \Phi_k(x), \qquad \alpha_k \equiv \int d^dx \, \Phi_k^\dagger(x) \Phi(x),
\end{equation}
and also easily construct the propagator for $\Phi$ satisfying \cref{eq:Ggreen} as
\begin{equation}
\label{eq:greeneig}
    G(x,y) = \sum_{k=0}^\infty \frac{\Phi^\dagger_k(x) \Phi_k(y)}{\lambda_k^2+m_\Phi^2}.
\end{equation}
In the $\QET$, suppose we are only interested in probing $\varphi$ at energies much smaller than some scale $\Lambda$. In Wilsonian renormalization this allows for some nice simplifications if $\Lambda\ll m_{\Phi}$, as carried out in \cref{eq:fstqetapp} and below, which eventually take one from a local complete theory to a local $\QET$. Alternatively, one could also do this using perturbation theory, but we would like to proceed within the framework of exact integration.

Unfortunately though, usually the path integral over $\Phi$ is too hard to compute in the first place. If this is the case, an exact $\QET$ expression like that in \cref{eq:nonloeg} is no longer an available starting point. Instead, one has to start with the complete theory, and hope to be able to perform some simplifications first. Inspired by Wilsonian renormalization, an approach that seems reasonable is the following.\footnote{
The reader is encouraged to read what follows with skepticism; we are going to illustrate an argument that leads to a catastrophically wrong result.}

Without loss of generality, let $\lambda_{k+1}^2\geq\lambda_{k}^2 $ for all $k$. Because we are only interested in probing $\varphi$ at scales much smaller than some energy $\Lambda$, one may expect the dynamics of heavy modes of the $\Phi$ field with $\lambda_k^2$ exceeding $\Lambda^2$ to play a negligible role. Hence consider truncating the mode expansion in \cref{eq:PhiN} to some $k=N$ such that $\lambda_N^2\approx\Lambda^2$. This truncation may be as low as to $N=0$ if already $\lambda_1 \gg \lambda_0$, as one can easily engineer by making $M$ a parametrically small compact space.
Then the free action of $\Phi$ in terms of the $\alpha_k$ coefficients is simply
\begin{equation}
    I_\Phi[\Phi] = \frac{1}{2} \sum_{k=0}^{N} \alpha_k^2 (\lambda_k^2 + m_\Phi^2).
\end{equation}
The path integral over $\Phi$ becomes a multi-integral over every mode $\alpha_k$, the measure of which will for convenience be simply denoted by $\mathcal{D}\alpha$. The resulting expression takes the form of a standard Hubbard-Stratonovich transformation,
\begin{align}
\label{eq:hubbardd}
    \int\displaylimits_M \mathcal{D}\Phi \, e^{- I[\varphi,\Phi]} 
    & \reprel{?}{\propto}
    e^{-I_\varphi[\varphi]} \int\displaylimits \mathcal{D}\alpha \, \exp[- \sum_{k=0}^{N} \, \left( \frac{\alpha_k^2}{2} (\lambda_k^2 + m_\Phi^2)  + \lambda \, \alpha_k  \int\displaylimits_M  d^dx f_\varphi(x) \Phi_k(x) \right)] \\
\label{eq:hubbard2d}
    & \propto
    \exp[-I_\varphi[\varphi] + \frac{\lambda^2}{2} 
    \sum_{k=0}^{N}
    \int\displaylimits_{M^2} d^dx \, d^dy \, \frac{f_\varphi(x) \Phi_k(x) \, f_\varphi(y) \Phi_k^\dagger(y)}{\lambda_k^2 + m_\Phi^2} ].
\end{align}
In the limit $N\to\infty$ this consistently reproduces \cref{eq:nonloeg} after the mode expansion is resummed. However, at any finite $N$, this calculation emphasizes what a grave error it would have been to truncate the mode expansion of $\Phi$: non-localities in \cref{eq:hubbard2d} do not die off with distance, but exhibit a periodicity of order $\Lambda^{-1}$.\footnote{This is akin to a Fourier series where the highest frequency is of order $\Lambda$.} An extreme case corresponds to truncating to the lowest $k=0$ mode only, which would have seemed reasonable if $\lambda_1 \gg \lambda_0$, whether or not $m_\Phi$ is large. This corresponds to constant eigenfunctions for which $\lambda_0=0$ and one gets
\begin{align}
\label{eq:hubbarddbad}
    \int\displaylimits_M \mathcal{D}\Phi \, e^{- I[\varphi,\Phi]} 
    & \reprel{?}{\propto}
    \exp[-I_\varphi[\varphi] + \frac{\lambda^2}{2} \frac{\Phi_0^2}{m_\Phi^2}
    \int\displaylimits_{M^2} d^dx \, d^dy \, f_\varphi(x) f_\varphi(y) ].
\end{align}
This bilocal interaction is completely independent of the separation between the two points, in contrast with the exact one in \cref{eq:nonloeg} which always exhibits a decay dictated by the Green's function. The reason the correlations induced by \cref{eq:hubbarddbad} are insensitive to separation is that, by truncating $\Phi$ to its lowest, position-independent mode, one is effectively correlating the value of the field $\Phi$ everywhere. The correlations $\varphi$ inherits are thus equally scale-independent, and would be able to break factorization at the level of partition functions. This is the same effect that a dynamical cosmological constant would have, as discussed at the end of \cref{ssec:factor}. 

In conclusion, a truncation of a field to its lowest energy modes at the level of the path integral has catastrophically non-local and unphysical consequences for an effective theory on the same space. Interestingly, the bilocal coupling that \cref{eq:hubbarddbad} gives is the type of interaction $\alpha$-parameters are often considered to give when integrated out in the context of gravitational wormholes.

\section*{Acknowledgements}

It is a pleasure to acknowledge very insightful conversations and invaluable feedback from
Netta Engelhardt,
Daniel Harlow,
Tom Hartman,
Mukund Rangamani,
Marija Toma\v{s}evi\'{c}, and
Wayne Weng.
I would also like to thank
Ning Bao,
Mike Blake,
Nikolay Bobev,
Elliott Gesteau,
Aidan Herderschee,
Gary Horowitz,
Veronika Hubeny,
Hong Liu,
Alex Maloney,
Don Marolf,
Jake McNamara,
Steve Shenker,
Jon Sorce,
and
Evita Verheijden
for stimulating discussions.
Finally, I am grateful to Elliott Gesteau, Mukund Rangamani, and Marija Toma\v{s}evi\'{c} for very helpful comments on an earlier draft of this paper

This research was supported by the U.S. Department of Energy (award DE-SC0021886), by the Templeton Foundation via the Black Hole Initiative (award 62286), and by the MIT Center for Theoretical Physics.
I also thank the Kavli Institute for Theoretical Physics (KITP), supported by grant NSF PHY-2309135, for its hospitality during the completion of this work.
The opinions expressed in this publication are those of the author and do not necessarily reflect the views of the John Templeton Foundation.

\addcontentsline{toc}{section}{References}
\bibliographystyle{JHEP}
\bibliography{references.bib}

\providecommand{\href}[2]{#2}\begingroup\raggedright\begin{thebibliography}{100}

\bibitem{Maldacena:1997re}
J.~M. Maldacena, \emph{{The Large N limit of superconformal field theories and
  supergravity}}, \href{https://doi.org/10.4310/ATMP.1998.v2.n2.a1}{\emph{Adv.
  Theor. Math. Phys.} {\bfseries 2} (1998) 231}
  [\href{https://arxiv.org/abs/hep-th/9711200}{{\ttfamily hep-th/9711200}}].

\bibitem{Witten:1998qj}
E.~Witten, \emph{{Anti-de Sitter space and holography}},
  \href{https://doi.org/10.4310/ATMP.1998.v2.n2.a2}{\emph{Adv. Theor. Math.
  Phys.} {\bfseries 2} (1998) 253}
  [\href{https://arxiv.org/abs/hep-th/9802150}{{\ttfamily hep-th/9802150}}].

\bibitem{Gubser:1998bc}
S.~S. Gubser, I.~R. Klebanov and A.~M. Polyakov, \emph{{Gauge theory
  correlators from noncritical string theory}},
  \href{https://doi.org/10.1016/S0370-2693(98)00377-3}{\emph{Phys. Lett. B}
  {\bfseries 428} (1998) 105}
  [\href{https://arxiv.org/abs/hep-th/9802109}{{\ttfamily hep-th/9802109}}].

\bibitem{Aharony:1999ti}
O.~Aharony, S.~S. Gubser, J.~M. Maldacena, H.~Ooguri and Y.~Oz, \emph{{Large N
  field theories, string theory and gravity}},
  \href{https://doi.org/10.1016/S0370-1573(99)00083-6}{\emph{Phys. Rept.}
  {\bfseries 323} (2000) 183}
  [\href{https://arxiv.org/abs/hep-th/9905111}{{\ttfamily hep-th/9905111}}].

\bibitem{Maldacena:2004rf}
J.~M. Maldacena and L.~Maoz, \emph{{Wormholes in AdS}},
  \href{https://doi.org/10.1088/1126-6708/2004/02/053}{\emph{JHEP} {\bfseries
  02} (2004) 053} [\href{https://arxiv.org/abs/hep-th/0401024}{{\ttfamily
  hep-th/0401024}}].

\bibitem{Saad:2018bqo}
P.~Saad, S.~H. Shenker and D.~Stanford, \emph{{A semiclassical ramp in SYK and
  in gravity}},  \href{https://arxiv.org/abs/1806.06840}{{\ttfamily
  1806.06840}}.

\bibitem{Harlow:2018tqv}
D.~Harlow and D.~Jafferis, \emph{{The Factorization Problem in
  Jackiw-Teitelboim Gravity}},
  \href{https://doi.org/10.1007/JHEP02(2020)177}{\emph{JHEP} {\bfseries 02}
  (2020) 177} [\href{https://arxiv.org/abs/1804.01081}{{\ttfamily
  1804.01081}}].

\bibitem{Chandra:2022fwi}
J.~Chandra and T.~Hartman, \emph{{Coarse graining pure states in AdS/CFT}},
  \href{https://doi.org/10.1007/JHEP10(2023)030}{\emph{JHEP} {\bfseries 10}
  (2023) 030} [\href{https://arxiv.org/abs/2206.03414}{{\ttfamily
  2206.03414}}].

\bibitem{VanRiet:2020pcn}
T.~Van~Riet, \emph{{Instantons, Euclidean wormholes and AdS/CFT}},
  \href{https://doi.org/10.22323/1.376.0121}{\emph{PoS} {\bfseries CORFU2019}
  (2020) 121} [\href{https://arxiv.org/abs/2004.08956}{{\ttfamily
  2004.08956}}].

\bibitem{Astesiano:2023iql}
D.~Astesiano and F.~F. Gautason, \emph{{Supersymmetric wormholes in String
  theory}},  \href{https://arxiv.org/abs/2309.02481}{{\ttfamily 2309.02481}}.

\bibitem{Hebecker:2018ofv}
A.~Hebecker, T.~Mikhail and P.~Soler, \emph{{Euclidean wormholes, baby
  universes, and their impact on particle physics and cosmology}},
  \href{https://doi.org/10.3389/fspas.2018.00035}{\emph{Front. Astron. Space
  Sci.} {\bfseries 5} (2018) 35}
  [\href{https://arxiv.org/abs/1807.00824}{{\ttfamily 1807.00824}}].

\bibitem{Arkani-Hamed:2007cpn}
N.~Arkani-Hamed, J.~Orgera and J.~Polchinski, \emph{{Euclidean wormholes in
  string theory}},
  \href{https://doi.org/10.1088/1126-6708/2007/12/018}{\emph{JHEP} {\bfseries
  12} (2007) 018} [\href{https://arxiv.org/abs/0705.2768}{{\ttfamily
  0705.2768}}].

\bibitem{Marolf:2021kjc}
D.~Marolf and J.~E. Santos, \emph{{AdS Euclidean wormholes}},
  \href{https://doi.org/10.1088/1361-6382/ac2cb7}{\emph{Class. Quant. Grav.}
  {\bfseries 38} (2021) 224002}
  [\href{https://arxiv.org/abs/2101.08875}{{\ttfamily 2101.08875}}].

\bibitem{Belin:2020hea}
A.~Belin and J.~de~Boer, \emph{{Random statistics of OPE coefficients and
  Euclidean wormholes}},
  \href{https://doi.org/10.1088/1361-6382/ac1082}{\emph{Class. Quant. Grav.}
  {\bfseries 38} (2021) 164001}
  [\href{https://arxiv.org/abs/2006.05499}{{\ttfamily 2006.05499}}].

\bibitem{Afkhami-Jeddi:2020ezh}
N.~Afkhami-Jeddi, H.~Cohn, T.~Hartman and A.~Tajdini, \emph{{Free partition
  functions and an averaged holographic duality}},
  \href{https://doi.org/10.1007/JHEP01(2021)130}{\emph{JHEP} {\bfseries 01}
  (2021) 130} [\href{https://arxiv.org/abs/2006.04839}{{\ttfamily
  2006.04839}}].

\bibitem{Maloney:2020nni}
A.~Maloney and E.~Witten, \emph{{Averaging over Narain moduli space}},
  \href{https://doi.org/10.1007/JHEP10(2020)187}{\emph{JHEP} {\bfseries 10}
  (2020) 187} [\href{https://arxiv.org/abs/2006.04855}{{\ttfamily
  2006.04855}}].

\bibitem{Cotler:2020ugk}
J.~Cotler and K.~Jensen, \emph{{AdS$_{3}$ gravity and random CFT}},
  \href{https://doi.org/10.1007/JHEP04(2021)033}{\emph{JHEP} {\bfseries 04}
  (2021) 033} [\href{https://arxiv.org/abs/2006.08648}{{\ttfamily
  2006.08648}}].

\bibitem{Benjamin:2021wzr}
N.~Benjamin, C.~A. Keller, H.~Ooguri and I.~G. Zadeh, \emph{{Narain to
  Narnia}}, \href{https://doi.org/10.1007/s00220-021-04211-x}{\emph{Commun.
  Math. Phys.} {\bfseries 390} (2022) 425}
  [\href{https://arxiv.org/abs/2103.15826}{{\ttfamily 2103.15826}}].

\bibitem{Collier:2021rsn}
S.~Collier and A.~Maloney, \emph{{Wormholes and spectral statistics in the
  Narain ensemble}}, \href{https://doi.org/10.1007/JHEP03(2022)004}{\emph{JHEP}
  {\bfseries 03} (2022) 004}
  [\href{https://arxiv.org/abs/2106.12760}{{\ttfamily 2106.12760}}].

\bibitem{Chandra:2022bqq}
J.~Chandra, S.~Collier, T.~Hartman and A.~Maloney, \emph{{Semiclassical 3D
  gravity as an average of large-c CFTs}},
  \href{https://doi.org/10.1007/JHEP12(2022)069}{\emph{JHEP} {\bfseries 12}
  (2022) 069} [\href{https://arxiv.org/abs/2203.06511}{{\ttfamily
  2203.06511}}].

\bibitem{Collier:2023fwi}
S.~Collier, L.~Eberhardt and M.~Zhang, \emph{{Solving 3d gravity with Virasoro
  TQFT}}, \href{https://doi.org/10.21468/SciPostPhys.15.4.151}{\emph{SciPost
  Phys.} {\bfseries 15} (2023) 151}
  [\href{https://arxiv.org/abs/2304.13650}{{\ttfamily 2304.13650}}].

\bibitem{Collier:2024mgv}
S.~Collier, L.~Eberhardt and M.~Zhang, \emph{{3d gravity from Virasoro TQFT:
  Holography, wormholes and knots}},
  \href{https://arxiv.org/abs/2401.13900}{{\ttfamily 2401.13900}}.

\bibitem{Jackiw:1984je}
R.~Jackiw, \emph{{Lower Dimensional Gravity}},
  \href{https://doi.org/10.1016/0550-3213(85)90448-1}{\emph{Nucl. Phys. B}
  {\bfseries 252} (1985) 343}.

\bibitem{Teitelboim:1983ux}
C.~Teitelboim, \emph{{Gravitation and Hamiltonian Structure in Two Space-Time
  Dimensions}}, \href{https://doi.org/10.1016/0370-2693(83)90012-6}{\emph{Phys.
  Lett. B} {\bfseries 126} (1983) 41}.

\bibitem{Almheiri:2014cka}
A.~Almheiri and J.~Polchinski, \emph{{Models of AdS$_{2}$ backreaction and
  holography}}, \href{https://doi.org/10.1007/JHEP11(2015)014}{\emph{JHEP}
  {\bfseries 11} (2015) 014} [\href{https://arxiv.org/abs/1402.6334}{{\ttfamily
  1402.6334}}].

\bibitem{Saad:2019lba}
P.~Saad, S.~H. Shenker and D.~Stanford, \emph{{JT gravity as a matrix
  integral}},  \href{https://arxiv.org/abs/1903.11115}{{\ttfamily 1903.11115}}.

\bibitem{Stanford:2019vob}
D.~Stanford and E.~Witten, \emph{{JT gravity and the ensembles of random matrix
  theory}}, \href{https://doi.org/10.4310/ATMP.2020.v24.n6.a4}{\emph{Adv.
  Theor. Math. Phys.} {\bfseries 24} (2020) 1475}
  [\href{https://arxiv.org/abs/1907.03363}{{\ttfamily 1907.03363}}].

\bibitem{Johnson:2019eik}
C.~V. Johnson, \emph{{Nonperturbative Jackiw-Teitelboim gravity}},
  \href{https://doi.org/10.1103/PhysRevD.101.106023}{\emph{Phys. Rev. D}
  {\bfseries 101} (2020) 106023}
  [\href{https://arxiv.org/abs/1912.03637}{{\ttfamily 1912.03637}}].

\bibitem{Almheiri:2019hni}
A.~Almheiri, R.~Mahajan, J.~Maldacena and Y.~Zhao, \emph{{The Page curve of
  Hawking radiation from semiclassical geometry}},
  \href{https://doi.org/10.1007/JHEP03(2020)149}{\emph{JHEP} {\bfseries 03}
  (2020) 149} [\href{https://arxiv.org/abs/1908.10996}{{\ttfamily
  1908.10996}}].

\bibitem{Penington:2019npb}
G.~Penington, \emph{{Entanglement Wedge Reconstruction and the Information
  Paradox}}, \href{https://doi.org/10.1007/JHEP09(2020)002}{\emph{JHEP}
  {\bfseries 09} (2020) 002}
  [\href{https://arxiv.org/abs/1905.08255}{{\ttfamily 1905.08255}}].

\bibitem{Almheiri:2019psf}
A.~Almheiri, N.~Engelhardt, D.~Marolf and H.~Maxfield, \emph{{The entropy of
  bulk quantum fields and the entanglement wedge of an evaporating black
  hole}}, \href{https://doi.org/10.1007/JHEP12(2019)063}{\emph{JHEP} {\bfseries
  12} (2019) 063} [\href{https://arxiv.org/abs/1905.08762}{{\ttfamily
  1905.08762}}].

\bibitem{Penington:2019kki}
G.~Penington, S.~H. Shenker, D.~Stanford and Z.~Yang, \emph{{Replica wormholes
  and the black hole interior}},
  \href{https://doi.org/10.1007/JHEP03(2022)205}{\emph{JHEP} {\bfseries 03}
  (2022) 205} [\href{https://arxiv.org/abs/1911.11977}{{\ttfamily
  1911.11977}}].

\bibitem{Almheiri:2019qdq}
A.~Almheiri, T.~Hartman, J.~Maldacena, E.~Shaghoulian and A.~Tajdini,
  \emph{{Replica Wormholes and the Entropy of Hawking Radiation}},
  \href{https://doi.org/10.1007/JHEP05(2020)013}{\emph{JHEP} {\bfseries 05}
  (2020) 013} [\href{https://arxiv.org/abs/1911.12333}{{\ttfamily
  1911.12333}}].

\bibitem{Almheiri:2020cfm}
A.~Almheiri, T.~Hartman, J.~Maldacena, E.~Shaghoulian and A.~Tajdini,
  \emph{{The entropy of Hawking radiation}},
  \href{https://doi.org/10.1103/RevModPhys.93.035002}{\emph{Rev. Mod. Phys.}
  {\bfseries 93} (2021) 035002}
  [\href{https://arxiv.org/abs/2006.06872}{{\ttfamily 2006.06872}}].

\bibitem{Page:1993df}
D.~N. Page, \emph{{Average entropy of a subsystem}},
  \href{https://doi.org/10.1103/PhysRevLett.71.1291}{\emph{Phys. Rev. Lett.}
  {\bfseries 71} (1993) 1291}
  [\href{https://arxiv.org/abs/gr-qc/9305007}{{\ttfamily gr-qc/9305007}}].

\bibitem{Page:1993up}
D.~N. Page, \emph{{Black hole information}},  in \emph{{5th Canadian Conference
  on General Relativity and Relativistic Astrophysics (5CCGRRA)}}, 5, 1993,
  \href{https://arxiv.org/abs/hep-th/9305040}{{\ttfamily hep-th/9305040}}.

\bibitem{Mathur:2009hf}
S.~D. Mathur, \emph{{The Information paradox: A Pedagogical introduction}},
  \href{https://doi.org/10.1088/0264-9381/26/22/224001}{\emph{Class. Quant.
  Grav.} {\bfseries 26} (2009) 224001}
  [\href{https://arxiv.org/abs/0909.1038}{{\ttfamily 0909.1038}}].

\bibitem{Almheiri:2012rt}
A.~Almheiri, D.~Marolf, J.~Polchinski and J.~Sully, \emph{{Black Holes:
  Complementarity or Firewalls?}},
  \href{https://doi.org/10.1007/JHEP02(2013)062}{\emph{JHEP} {\bfseries 02}
  (2013) 062} [\href{https://arxiv.org/abs/1207.3123}{{\ttfamily 1207.3123}}].

\bibitem{Harlow:2014yka}
D.~Harlow, \emph{{Jerusalem Lectures on Black Holes and Quantum Information}},
  \href{https://doi.org/10.1103/RevModPhys.88.015002}{\emph{Rev. Mod. Phys.}
  {\bfseries 88} (2016) 015002}
  [\href{https://arxiv.org/abs/1409.1231}{{\ttfamily 1409.1231}}].

\bibitem{Marolf:2017jkr}
D.~Marolf, \emph{{The Black Hole information problem: past, present, and
  future}}, \href{https://doi.org/10.1088/1361-6633/aa77cc}{\emph{Rept. Prog.
  Phys.} {\bfseries 80} (2017) 092001}
  [\href{https://arxiv.org/abs/1703.02143}{{\ttfamily 1703.02143}}].

\bibitem{Cotler:2016fpe}
J.~S. Cotler, G.~Gur-Ari, M.~Hanada, J.~Polchinski, P.~Saad, S.~H. Shenker
  et~al., \emph{{Black Holes and Random Matrices}},
  \href{https://doi.org/10.1007/JHEP05(2017)118}{\emph{JHEP} {\bfseries 05}
  (2017) 118} [\href{https://arxiv.org/abs/1611.04650}{{\ttfamily
  1611.04650}}].

\bibitem{Liu:2018hlr}
J.~Liu, \emph{{Spectral form factors and late time quantum chaos}},
  \href{https://doi.org/10.1103/PhysRevD.98.086026}{\emph{Phys. Rev. D}
  {\bfseries 98} (2018) 086026}
  [\href{https://arxiv.org/abs/1806.05316}{{\ttfamily 1806.05316}}].

\bibitem{Stanford:2020wkf}
D.~Stanford, \emph{{More quantum noise from wormholes}},
  \href{https://arxiv.org/abs/2008.08570}{{\ttfamily 2008.08570}}.

\bibitem{Okuyama:2018gfr}
K.~Okuyama, \emph{{Eigenvalue instantons in the spectral form factor of random
  matrix model}}, \href{https://doi.org/10.1007/JHEP03(2019)147}{\emph{JHEP}
  {\bfseries 03} (2019) 147}
  [\href{https://arxiv.org/abs/1812.09469}{{\ttfamily 1812.09469}}].

\bibitem{Saad:2019pqd}
P.~Saad, \emph{{Late Time Correlation Functions, Baby Universes, and ETH in JT
  Gravity}},  \href{https://arxiv.org/abs/1910.10311}{{\ttfamily 1910.10311}}.

\bibitem{Winer:2020gdp}
M.~Winer and B.~Swingle, \emph{{Hydrodynamic Theory of the Connected Spectral
  form Factor}}, \href{https://doi.org/10.1103/PhysRevX.12.021009}{\emph{Phys.
  Rev. X} {\bfseries 12} (2022) 021009}
  [\href{https://arxiv.org/abs/2012.01436}{{\ttfamily 2012.01436}}].

\bibitem{Zhou:2023qmk}
Y.-N. Zhou, T.-G. Zhou and P.~Zhang, \emph{{Universal Properties of the
  Spectral Form Factor in Open Quantum Systems}},
  \href{https://arxiv.org/abs/2303.14352}{{\ttfamily 2303.14352}}.

\bibitem{Schwinger:1960qe}
J.~S. Schwinger, \emph{{Brownian motion of a quantum oscillator}},
  \href{https://doi.org/10.1063/1.1703727}{\emph{J. Math. Phys.} {\bfseries 2}
  (1961) 407}.

\bibitem{Keldysh:1964ud}
L.~V. Keldysh, \emph{{Diagram technique for nonequilibrium processes}},
  {\emph{Zh. Eksp. Teor. Fiz.} {\bfseries 47} (1964) 1515}.

\bibitem{Haehl:2016pec}
F.~M. Haehl, R.~Loganayagam and M.~Rangamani, \emph{{Schwinger-Keldysh
  formalism. Part I: BRST symmetries and superspace}},
  \href{https://doi.org/10.1007/JHEP06(2017)069}{\emph{JHEP} {\bfseries 06}
  (2017) 069} [\href{https://arxiv.org/abs/1610.01940}{{\ttfamily
  1610.01940}}].

\bibitem{Haehl:2016uah}
F.~M. Haehl, R.~Loganayagam and M.~Rangamani, \emph{{Schwinger-Keldysh
  formalism. Part II: thermal equivariant cohomology}},
  \href{https://doi.org/10.1007/JHEP06(2017)070}{\emph{JHEP} {\bfseries 06}
  (2017) 070} [\href{https://arxiv.org/abs/1610.01941}{{\ttfamily
  1610.01941}}].

\bibitem{Pelliconi:2023ojb}
P.~Pelliconi and J.~Sonner, \emph{{The influence functional in open holography:
  entanglement and R\'enyi entropies}},
  \href{https://doi.org/10.1007/JHEP06(2024)185}{\emph{JHEP} {\bfseries 06}
  (2024) 185} [\href{https://arxiv.org/abs/2310.13047}{{\ttfamily
  2310.13047}}].

\bibitem{Feynman:1963fq}
R.~P. Feynman and F.~L. Vernon, Jr., \emph{{The Theory of a general quantum
  system interacting with a linear dissipative system}},
  \href{https://doi.org/10.1016/0003-4916(63)90068-X}{\emph{Annals Phys.}
  {\bfseries 24} (1963) 118}.

\bibitem{Lombardo:1995fg}
F.~Lombardo and F.~D. Mazzitelli, \emph{{Coarse graining and decoherence in
  quantum field theory}},
  \href{https://doi.org/10.1103/PhysRevD.53.2001}{\emph{Phys. Rev. D}
  {\bfseries 53} (1996) 2001}
  [\href{https://arxiv.org/abs/hep-th/9508052}{{\ttfamily hep-th/9508052}}].

\bibitem{weiss2008quantum}
U.~Weiss, \emph{Quantum Dissipative Systems}, Series in modern condensed matter
  physics. World Scientific, 2008.

\bibitem{Lindblad1976OnTG}
G.~Lindblad, \emph{On the generators of quantum dynamical semigroups},
  {\emph{Communications in Mathematical Physics} {\bfseries 48} (1976) 119}.

\bibitem{breuer2002theory}
H.-P. Breuer and F.~Petruccione, \emph{The theory of open quantum systems}. OUP
  Oxford, 2002.

\bibitem{GRABERT1988115}
H.~Grabert, P.~Schramm and G.-L. Ingold, \emph{Quantum brownian motion: The
  functional integral approach},
  \href{https://doi.org/https://doi.org/10.1016/0370-1573(88)90023-3}{\emph{Physics
  Reports} {\bfseries 168} (1988) 115}.

\bibitem{aurell2020operator}
E.~Aurell, R.~Kawai and K.~Goyal, \emph{An operator derivation of the
  feynman--vernon theory, with applications to the generating function of bath
  energy changes and to an-harmonic baths}, {\emph{Journal of Physics A:
  Mathematical and Theoretical} {\bfseries 53} (2020) 275303}.

\bibitem{LaFlamme:1990kd}
R.~LaFlamme and J.~Louko, \emph{{Reduced density matrices and decoherence in
  quantum cosmology}},
  \href{https://doi.org/10.1103/PhysRevD.43.3317}{\emph{Phys. Rev. D}
  {\bfseries 43} (1991) 3317}.

\bibitem{Barvinsky:2023jkl}
A.~O. Barvinsky and N.~Kolganov, \emph{{Nonequilibrium Schwinger-Keldysh
  formalism for density matrix states: Analytic properties and implications in
  cosmology}}, \href{https://doi.org/10.1103/PhysRevD.109.025004}{\emph{Phys.
  Rev. D} {\bfseries 109} (2024) 025004}
  [\href{https://arxiv.org/abs/2309.03687}{{\ttfamily 2309.03687}}].

\bibitem{Caldeira:1981rx}
A.~O. Caldeira and A.~J. Leggett, \emph{{Influence of dissipation on quantum
  tunneling in macroscopic systems}},
  \href{https://doi.org/10.1103/PhysRevLett.46.211}{\emph{Phys. Rev. Lett.}
  {\bfseries 46} (1981) 211}.

\bibitem{Betzios:2019rds}
P.~Betzios, E.~Kiritsis and O.~Papadoulaki, \emph{{Euclidean Wormholes and
  Holography}}, \href{https://doi.org/10.1007/JHEP06(2019)042}{\emph{JHEP}
  {\bfseries 06} (2019) 042}
  [\href{https://arxiv.org/abs/1903.05658}{{\ttfamily 1903.05658}}].

\bibitem{Eberhardt:2020bgq}
L.~Eberhardt, \emph{{Partition functions of the tensionless string}},
  \href{https://doi.org/10.1007/JHEP03(2021)176}{\emph{JHEP} {\bfseries 03}
  (2021) 176} [\href{https://arxiv.org/abs/2008.07533}{{\ttfamily
  2008.07533}}].

\bibitem{Eberhardt:2021jvj}
L.~Eberhardt, \emph{{Summing over Geometries in String Theory}},
  \href{https://doi.org/10.1007/JHEP05(2021)233}{\emph{JHEP} {\bfseries 05}
  (2021) 233} [\href{https://arxiv.org/abs/2102.12355}{{\ttfamily
  2102.12355}}].

\bibitem{Gesteau:2024gzf}
E.~Gesteau, M.~Marcolli and J.~McNamara, \emph{{Wormhole Renormalization: The
  gravitational path integral, holography, and a gauge group for topology
  change}},  \href{https://arxiv.org/abs/2407.20324}{{\ttfamily 2407.20324}}.

\bibitem{Polchinski:1983gv}
J.~Polchinski, \emph{{Renormalization and Effective Lagrangians}},
  \href{https://doi.org/10.1016/0550-3213(84)90287-6}{\emph{Nucl. Phys. B}
  {\bfseries 231} (1984) 269}.

\bibitem{Morris:1993qb}
T.~R. Morris, \emph{{The Exact renormalization group and approximate
  solutions}}, \href{https://doi.org/10.1142/S0217751X94000972}{\emph{Int. J.
  Mod. Phys. A} {\bfseries 9} (1994) 2411}
  [\href{https://arxiv.org/abs/hep-ph/9308265}{{\ttfamily hep-ph/9308265}}].

\bibitem{Wetterich:1992yh}
C.~Wetterich, \emph{{Exact evolution equation for the effective potential}},
  \href{https://doi.org/10.1016/0370-2693(93)90726-X}{\emph{Phys. Lett. B}
  {\bfseries 301} (1993) 90}
  [\href{https://arxiv.org/abs/1710.05815}{{\ttfamily 1710.05815}}].

\bibitem{Rosten:2010vm}
O.~J. Rosten, \emph{{Fundamentals of the Exact Renormalization Group}},
  \href{https://doi.org/10.1016/j.physrep.2011.12.003}{\emph{Phys. Rept.}
  {\bfseries 511} (2012) 177}
  [\href{https://arxiv.org/abs/1003.1366}{{\ttfamily 1003.1366}}].

\bibitem{skinnernotes}
D.~Skinner, \emph{{Part III Lecture Notes on Advanced Quantum Field Theory}},
  2017.

\bibitem{Balasubramanian:2011wt}
V.~Balasubramanian, M.~B. McDermott and M.~Van~Raamsdonk, \emph{{Momentum-space
  entanglement and renormalization in quantum field theory}},
  \href{https://doi.org/10.1103/PhysRevD.86.045014}{\emph{Phys. Rev. D}
  {\bfseries 86} (2012) 045014}
  [\href{https://arxiv.org/abs/1108.3568}{{\ttfamily 1108.3568}}].

\bibitem{Hartle:1983ai}
J.~B. Hartle and S.~W. Hawking, \emph{{Wave Function of the Universe}},
  \href{https://doi.org/10.1103/PhysRevD.28.2960}{\emph{Phys. Rev. D}
  {\bfseries 28} (1983) 2960}.

\bibitem{Hawking:1983hj}
S.~W. Hawking, \emph{{The Quantum State of the Universe}},
  \href{https://doi.org/10.1016/0550-3213(84)90093-2}{\emph{Nucl. Phys. B}
  {\bfseries 239} (1984) 257}.

\bibitem{Page:1986vw}
D.~N. Page, \emph{{Density Matrix of the Universe}},
  \href{https://doi.org/10.1103/PhysRevD.34.2267}{\emph{Phys. Rev. D}
  {\bfseries 34} (1986) 2267}.

\bibitem{Anous:2020lka}
T.~Anous, J.~Kruthoff and R.~Mahajan, \emph{{Density matrices in quantum
  gravity}}, \href{https://doi.org/10.21468/SciPostPhys.9.4.045}{\emph{SciPost
  Phys.} {\bfseries 9} (2020) 045}
  [\href{https://arxiv.org/abs/2006.17000}{{\ttfamily 2006.17000}}].

\bibitem{Chen:2020tes}
Y.~Chen, V.~Gorbenko and J.~Maldacena, \emph{{Bra-ket wormholes in
  gravitationally prepared states}},
  \href{https://doi.org/10.1007/JHEP02(2021)009}{\emph{JHEP} {\bfseries 02}
  (2021) 009} [\href{https://arxiv.org/abs/2007.16091}{{\ttfamily
  2007.16091}}].

\bibitem{Hawking:1982dj}
S.~W. Hawking, \emph{{The Unpredictability of Quantum Gravity}},
  \href{https://doi.org/10.1007/BF01206031}{\emph{Commun. Math. Phys.}
  {\bfseries 87} (1982) 395}.

\bibitem{Giddings:1988cx}
S.~B. Giddings and A.~Strominger, \emph{{Loss of incoherence and determination
  of coupling constants in quantum gravity}},
  \href{https://doi.org/10.1016/0550-3213(88)90109-5}{\emph{Nucl. Phys. B}
  {\bfseries 307} (1988) 854}.

\bibitem{Coleman:1988cy}
S.~R. Coleman, \emph{{Black holes as red herrings: Topological fluctuations and
  the loss of quantum coherence}},
  \href{https://doi.org/10.1016/0550-3213(88)90110-1}{\emph{Nucl. Phys. B}
  {\bfseries 307} (1988) 867}.

\bibitem{Hawking:1991vs}
S.~W. Hawking, \emph{{The Alpha parameters of wormholes}},
  \href{https://doi.org/10.1088/0031-8949/1991/T36/023}{\emph{Phys. Scripta T}
  {\bfseries 36} (1991) 222}.

\bibitem{Balasubramanian:2020lux}
V.~Balasubramanian, J.~J. Heckman, E.~Lipeles and A.~P. Turner,
  \emph{{Statistical Coupling Constants from Hidden Sector Entanglement}},
  \href{https://doi.org/10.1103/PhysRevD.103.066024}{\emph{Phys. Rev. D}
  {\bfseries 103} (2021) 066024}
  [\href{https://arxiv.org/abs/2012.09182}{{\ttfamily 2012.09182}}].

\bibitem{Giddings:2020yes}
S.~B. Giddings and G.~J. Turiaci, \emph{{Wormhole calculus, replicas, and
  entropies}}, \href{https://doi.org/10.1007/JHEP09(2020)194}{\emph{JHEP}
  {\bfseries 09} (2020) 194}
  [\href{https://arxiv.org/abs/2004.02900}{{\ttfamily 2004.02900}}].

\bibitem{Pollack:2020gfa}
J.~Pollack, M.~Rozali, J.~Sully and D.~Wakeham, \emph{{Eigenstate
  Thermalization and Disorder Averaging in Gravity}},
  \href{https://doi.org/10.1103/PhysRevLett.125.021601}{\emph{Phys. Rev. Lett.}
  {\bfseries 125} (2020) 021601}
  [\href{https://arxiv.org/abs/2002.02971}{{\ttfamily 2002.02971}}].

\bibitem{Freivogel:2021ivu}
B.~Freivogel, D.~Nikolakopoulou and A.~F. Rotundo, \emph{{Wormholes from
  averaging over states}},
  \href{https://doi.org/10.21468/SciPostPhys.14.3.026}{\emph{SciPost Phys.}
  {\bfseries 14} (2023) 026}
  [\href{https://arxiv.org/abs/2105.12771}{{\ttfamily 2105.12771}}].

\bibitem{Marolf:2020rpm}
D.~Marolf and H.~Maxfield, \emph{{Observations of Hawking radiation: the Page
  curve and baby universes}},
  \href{https://doi.org/10.1007/JHEP04(2021)272}{\emph{JHEP} {\bfseries 04}
  (2021) 272} [\href{https://arxiv.org/abs/2010.06602}{{\ttfamily
  2010.06602}}].

\bibitem{Marolf:2021ghr}
D.~Marolf and H.~Maxfield, \emph{{The page curve and baby universes}},
  \href{https://doi.org/10.1142/S021827182142027X}{\emph{Int. J. Mod. Phys. D}
  {\bfseries 30} (2021) 2142027}
  [\href{https://arxiv.org/abs/2105.12211}{{\ttfamily 2105.12211}}].

\bibitem{Casini:2019kex}
H.~Casini, M.~Huerta, J.~M. Mag\'an and D.~Pontello, \emph{{Entanglement
  entropy and superselection sectors. Part I. Global symmetries}},
  \href{https://doi.org/10.1007/JHEP02(2020)014}{\emph{JHEP} {\bfseries 02}
  (2020) 014} [\href{https://arxiv.org/abs/1905.10487}{{\ttfamily
  1905.10487}}].

\bibitem{Casini:2020rgj}
H.~Casini, M.~Huerta, J.~M. Magan and D.~Pontello, \emph{{Entropic order
  parameters for the phases of QFT}},
  \href{https://doi.org/10.1007/JHEP04(2021)277}{\emph{JHEP} {\bfseries 04}
  (2021) 277} [\href{https://arxiv.org/abs/2008.11748}{{\ttfamily
  2008.11748}}].

\bibitem{Casini:2021zgr}
H.~Casini and J.~M. Magan, \emph{{On completeness and generalized symmetries in
  quantum field theory}},
  \href{https://doi.org/10.1142/S0217732321300251}{\emph{Mod. Phys. Lett. A}
  {\bfseries 36} (2021) 2130025}
  [\href{https://arxiv.org/abs/2110.11358}{{\ttfamily 2110.11358}}].

\bibitem{Arveson1976-wp}
W.~Arveson, \emph{An Invitation to {C*-Algebras}}. Springer New York, New York,
  NY, 1976.

\bibitem{horuzhy2012}
S.~S. Horuzhy, \emph{Introduction to Algebraic Quantum Field Theory}, vol.~19.
  Springer Science \& Business Media, 2012.

\bibitem{Haag_1996}
R.~Haag, \emph{Local Quantum Physics: Fields, Particles, Algebras}. Springer,
  Berlin, Germany, 1996.

\bibitem{Witten:2018zva}
E.~Witten, \emph{{A Mini-Introduction To Information Theory}},
  \href{https://doi.org/10.1007/s40766-020-00004-5}{\emph{Riv. Nuovo Cim.}
  {\bfseries 43} (2020) 187}
  [\href{https://arxiv.org/abs/1805.11965}{{\ttfamily 1805.11965}}].

\bibitem{VanRaamsdonk:2010pw}
M.~Van~Raamsdonk, \emph{{Building up spacetime with quantum entanglement}},
  \href{https://doi.org/10.1142/S0218271810018529}{\emph{Gen. Rel. Grav.}
  {\bfseries 42} (2010) 2323}
  [\href{https://arxiv.org/abs/1005.3035}{{\ttfamily 1005.3035}}].

\bibitem{Maldacena:2013xja}
J.~Maldacena and L.~Susskind, \emph{{Cool horizons for entangled black holes}},
  \href{https://doi.org/10.1002/prop.201300020}{\emph{Fortsch. Phys.}
  {\bfseries 61} (2013) 781} [\href{https://arxiv.org/abs/1306.0533}{{\ttfamily
  1306.0533}}].

\bibitem{Ryu:2006bv}
S.~Ryu and T.~Takayanagi, \emph{{Holographic derivation of entanglement entropy
  from AdS/CFT}},
  \href{https://doi.org/10.1103/PhysRevLett.96.181602}{\emph{Phys. Rev. Lett.}
  {\bfseries 96} (2006) 181602}
  [\href{https://arxiv.org/abs/hep-th/0603001}{{\ttfamily hep-th/0603001}}].

\bibitem{Hubeny:2007xt}
V.~E. Hubeny, M.~Rangamani and T.~Takayanagi, \emph{{A Covariant holographic
  entanglement entropy proposal}},
  \href{https://doi.org/10.1088/1126-6708/2007/07/062}{\emph{JHEP} {\bfseries
  07} (2007) 062} [\href{https://arxiv.org/abs/0705.0016}{{\ttfamily
  0705.0016}}].

\bibitem{Lewkowycz:2013nqa}
A.~Lewkowycz and J.~Maldacena, \emph{{Generalized gravitational entropy}},
  \href{https://doi.org/10.1007/JHEP08(2013)090}{\emph{JHEP} {\bfseries 08}
  (2013) 090} [\href{https://arxiv.org/abs/1304.4926}{{\ttfamily 1304.4926}}].

\bibitem{Dong:2016hjy}
X.~Dong, A.~Lewkowycz and M.~Rangamani, \emph{{Deriving covariant holographic
  entanglement}}, \href{https://doi.org/10.1007/JHEP11(2016)028}{\emph{JHEP}
  {\bfseries 11} (2016) 028}
  [\href{https://arxiv.org/abs/1607.07506}{{\ttfamily 1607.07506}}].

\bibitem{Faulkner:2013ana}
T.~Faulkner, A.~Lewkowycz and J.~Maldacena, \emph{{Quantum corrections to
  holographic entanglement entropy}},
  \href{https://doi.org/10.1007/JHEP11(2013)074}{\emph{JHEP} {\bfseries 11}
  (2013) 074} [\href{https://arxiv.org/abs/1307.2892}{{\ttfamily 1307.2892}}].

\bibitem{Engelhardt:2014gca}
N.~Engelhardt and A.~C. Wall, \emph{{Quantum Extremal Surfaces: Holographic
  Entanglement Entropy beyond the Classical Regime}},
  \href{https://doi.org/10.1007/JHEP01(2015)073}{\emph{JHEP} {\bfseries 01}
  (2015) 073} [\href{https://arxiv.org/abs/1408.3203}{{\ttfamily 1408.3203}}].

\bibitem{Rangamani:2016dms}
M.~Rangamani and T.~Takayanagi, \emph{{Holographic Entanglement Entropy}},
  vol.~931. Springer, 2017,
  \href{https://doi.org/10.1007/978-3-319-52573-0}{10.1007/978-3-319-52573-0},
  [\href{https://arxiv.org/abs/1609.01287}{{\ttfamily 1609.01287}}].

\bibitem{Wall:2012uf}
A.~C. Wall, \emph{{Maximin Surfaces, and the Strong Subadditivity of the
  Covariant Holographic Entanglement Entropy}},
  \href{https://doi.org/10.1088/0264-9381/31/22/225007}{\emph{Class. Quant.
  Grav.} {\bfseries 31} (2014) 225007}
  [\href{https://arxiv.org/abs/1211.3494}{{\ttfamily 1211.3494}}].

\bibitem{Akers:2019lzs}
C.~Akers, N.~Engelhardt, G.~Penington and M.~Usatyuk, \emph{{Quantum Maximin
  Surfaces}}, \href{https://doi.org/10.1007/JHEP08(2020)140}{\emph{JHEP}
  {\bfseries 08} (2020) 140}
  [\href{https://arxiv.org/abs/1912.02799}{{\ttfamily 1912.02799}}].

\bibitem{Hamilton:2006az}
A.~Hamilton, D.~N. Kabat, G.~Lifschytz and D.~A. Lowe, \emph{{Holographic
  representation of local bulk operators}},
  \href{https://doi.org/10.1103/PhysRevD.74.066009}{\emph{Phys. Rev. D}
  {\bfseries 74} (2006) 066009}
  [\href{https://arxiv.org/abs/hep-th/0606141}{{\ttfamily hep-th/0606141}}].

\bibitem{Czech:2012bh}
B.~Czech, J.~L. Karczmarek, F.~Nogueira and M.~Van~Raamsdonk, \emph{{The
  Gravity Dual of a Density Matrix}},
  \href{https://doi.org/10.1088/0264-9381/29/15/155009}{\emph{Class. Quant.
  Grav.} {\bfseries 29} (2012) 155009}
  [\href{https://arxiv.org/abs/1204.1330}{{\ttfamily 1204.1330}}].

\bibitem{Headrick:2014cta}
M.~Headrick, V.~E. Hubeny, A.~Lawrence and M.~Rangamani, \emph{{Causality \&
  holographic entanglement entropy}},
  \href{https://doi.org/10.1007/JHEP12(2014)162}{\emph{JHEP} {\bfseries 12}
  (2014) 162} [\href{https://arxiv.org/abs/1408.6300}{{\ttfamily 1408.6300}}].

\bibitem{Jafferis:2014lza}
D.~L. Jafferis and S.~J. Suh, \emph{{The Gravity Duals of Modular
  Hamiltonians}}, \href{https://doi.org/10.1007/JHEP09(2016)068}{\emph{JHEP}
  {\bfseries 09} (2016) 068} [\href{https://arxiv.org/abs/1412.8465}{{\ttfamily
  1412.8465}}].

\bibitem{Almheiri:2014lwa}
A.~Almheiri, X.~Dong and D.~Harlow, \emph{{Bulk Locality and Quantum Error
  Correction in AdS/CFT}},
  \href{https://doi.org/10.1007/JHEP04(2015)163}{\emph{JHEP} {\bfseries 04}
  (2015) 163} [\href{https://arxiv.org/abs/1411.7041}{{\ttfamily 1411.7041}}].

\bibitem{Pastawski:2015qua}
F.~Pastawski, B.~Yoshida, D.~Harlow and J.~Preskill, \emph{{Holographic quantum
  error-correcting codes: Toy models for the bulk/boundary correspondence}},
  \href{https://doi.org/10.1007/JHEP06(2015)149}{\emph{JHEP} {\bfseries 06}
  (2015) 149} [\href{https://arxiv.org/abs/1503.06237}{{\ttfamily
  1503.06237}}].

\bibitem{Jafferis:2015del}
D.~L. Jafferis, A.~Lewkowycz, J.~Maldacena and S.~J. Suh, \emph{{Relative
  entropy equals bulk relative entropy}},
  \href{https://doi.org/10.1007/JHEP06(2016)004}{\emph{JHEP} {\bfseries 06}
  (2016) 004} [\href{https://arxiv.org/abs/1512.06431}{{\ttfamily
  1512.06431}}].

\bibitem{Dong:2016eik}
X.~Dong, D.~Harlow and A.~C. Wall, \emph{{Reconstruction of Bulk Operators
  within the Entanglement Wedge in Gauge-Gravity Duality}},
  \href{https://doi.org/10.1103/PhysRevLett.117.021601}{\emph{Phys. Rev. Lett.}
  {\bfseries 117} (2016) 021601}
  [\href{https://arxiv.org/abs/1601.05416}{{\ttfamily 1601.05416}}].

\bibitem{Harlow:2016vwg}
D.~Harlow, \emph{{The Ryu\textendash{}Takayanagi Formula from Quantum Error
  Correction}}, \href{https://doi.org/10.1007/s00220-017-2904-z}{\emph{Commun.
  Math. Phys.} {\bfseries 354} (2017) 865}
  [\href{https://arxiv.org/abs/1607.03901}{{\ttfamily 1607.03901}}].

\bibitem{Engelhardt:2020qpv}
N.~Engelhardt, S.~Fischetti and A.~Maloney, \emph{{Free energy from replica
  wormholes}}, \href{https://doi.org/10.1103/PhysRevD.103.046021}{\emph{Phys.
  Rev. D} {\bfseries 103} (2021) 046021}
  [\href{https://arxiv.org/abs/2007.07444}{{\ttfamily 2007.07444}}].

\bibitem{Chandrasekaran:2022asa}
V.~Chandrasekaran, N.~Engelhardt, S.~Fischetti and S.~Hern\'andez-Cuenca,
  \emph{{A tale of two saddles}},
  \href{https://doi.org/10.1007/JHEP11(2022)110}{\emph{JHEP} {\bfseries 11}
  (2022) 110} [\href{https://arxiv.org/abs/2207.09472}{{\ttfamily
  2207.09472}}].

\bibitem{Hawking:1975vcx}
S.~W. Hawking, \emph{{Particle Creation by Black Holes}},
  \href{https://doi.org/10.1007/BF02345020}{\emph{Commun. Math. Phys.}
  {\bfseries 43} (1975) 199}.

\bibitem{Liu:2020jsv}
H.~Liu and S.~Vardhan, \emph{{Entanglement Entropies of Equilibrated Pure
  States in Quantum Many-Body Systems and Gravity}},
  \href{https://doi.org/10.1103/PRXQuantum.2.010344}{\emph{PRX Quantum}
  {\bfseries 2} (2021) 010344}
  [\href{https://arxiv.org/abs/2008.01089}{{\ttfamily 2008.01089}}].

\bibitem{Bousso:2019ykv}
R.~Bousso and M.~Toma\v{s}evi\'c, \emph{{Unitarity From a Smooth Horizon?}},
  \href{https://doi.org/10.1103/PhysRevD.102.106019}{\emph{Phys. Rev. D}
  {\bfseries 102} (2020) 106019}
  [\href{https://arxiv.org/abs/1911.06305}{{\ttfamily 1911.06305}}].

\bibitem{Bousso:2020kmy}
R.~Bousso and E.~Wildenhain, \emph{{Gravity/ensemble duality}},
  \href{https://doi.org/10.1103/PhysRevD.102.066005}{\emph{Phys. Rev. D}
  {\bfseries 102} (2020) 066005}
  [\href{https://arxiv.org/abs/2006.16289}{{\ttfamily 2006.16289}}].

\bibitem{deBoer:2023vsm}
J.~de~Boer, D.~Liska, B.~Post and M.~Sasieta, \emph{{A principle of maximum
  ignorance for semiclassical gravity}},
  \href{https://doi.org/10.1007/JHEP02(2024)003}{\emph{JHEP} {\bfseries 02}
  (2024) 003} [\href{https://arxiv.org/abs/2311.08132}{{\ttfamily
  2311.08132}}].

\bibitem{Verlinde:2020upt}
H.~Verlinde, \emph{{ER = EPR revisited: On the Entropy of an Einstein-Rosen
  Bridge}},  \href{https://arxiv.org/abs/2003.13117}{{\ttfamily 2003.13117}}.

\bibitem{Verlinde:2021kgt}
H.~Verlinde, \emph{{Wormholes in Quantum Mechanics}},
  \href{https://arxiv.org/abs/2105.02129}{{\ttfamily 2105.02129}}.

\bibitem{B_ny_2007}
C.~Bény, A.~Kempf and D.~W. Kribs, \emph{Generalization of quantum error
  correction via the heisenberg picture},
  \href{https://doi.org/10.1103/physrevlett.98.100502}{\emph{Physical Review
  Letters} {\bfseries 98} (2007) }.

\bibitem{B_ny_2007b}
C.~Bény, A.~Kempf and D.~W. Kribs, \emph{Quantum error correction of
  observables},
  \href{https://doi.org/10.1103/physreva.76.042303}{\emph{Physical Review A}
  {\bfseries 76} (2007) }.

\bibitem{Harlow:2013tf}
D.~Harlow and P.~Hayden, \emph{{Quantum Computation vs. Firewalls}},
  \href{https://doi.org/10.1007/JHEP06(2013)085}{\emph{JHEP} {\bfseries 06}
  (2013) 085} [\href{https://arxiv.org/abs/1301.4504}{{\ttfamily 1301.4504}}].

\bibitem{Susskind:2014rva}
L.~Susskind, \emph{{Computational Complexity and Black Hole Horizons}},
  \href{https://doi.org/10.1002/prop.201500092}{\emph{Fortsch. Phys.}
  {\bfseries 64} (2016) 24} [\href{https://arxiv.org/abs/1403.5695}{{\ttfamily
  1403.5695}}].

\bibitem{Engelhardt:2018kcs}
N.~Engelhardt and A.~C. Wall, \emph{{Coarse Graining Holographic Black Holes}},
  \href{https://doi.org/10.1007/JHEP05(2019)160}{\emph{JHEP} {\bfseries 05}
  (2019) 160} [\href{https://arxiv.org/abs/1806.01281}{{\ttfamily
  1806.01281}}].

\bibitem{Brown:2019rox}
A.~R. Brown, H.~Gharibyan, G.~Penington and L.~Susskind, \emph{{The
  Python\textquoteright{}s Lunch: geometric obstructions to decoding Hawking
  radiation}}, \href{https://doi.org/10.1007/JHEP08(2020)121}{\emph{JHEP}
  {\bfseries 08} (2020) 121}
  [\href{https://arxiv.org/abs/1912.00228}{{\ttfamily 1912.00228}}].

\bibitem{Kim:2020cds}
I.~H. Kim, E.~Tang and J.~Preskill, \emph{{The ghost in the radiation: robust
  encodings of the black hole interior (invited paper)}},
  \href{https://doi.org/10.1145/3406325.3465357}{\emph{JHEP} {\bfseries 06}
  (2020) 031} [\href{https://arxiv.org/abs/2003.05451}{{\ttfamily
  2003.05451}}].

\bibitem{Engelhardt:2021mue}
N.~Engelhardt, G.~Penington and A.~Shahbazi-Moghaddam, \emph{{A world without
  pythons would be so simple}},
  \href{https://doi.org/10.1088/1361-6382/ac2de5}{\emph{Class. Quant. Grav.}
  {\bfseries 38} (2021) 234001}
  [\href{https://arxiv.org/abs/2102.07774}{{\ttfamily 2102.07774}}].

\bibitem{Engelhardt:2021qjs}
N.~Engelhardt, G.~Penington and A.~Shahbazi-Moghaddam, \emph{{Finding pythons
  in unexpected places}},
  \href{https://doi.org/10.1088/1361-6382/ac3e75}{\emph{Class. Quant. Grav.}
  {\bfseries 39} (2022) 094002}
  [\href{https://arxiv.org/abs/2105.09316}{{\ttfamily 2105.09316}}].

\bibitem{Akers:2022qdl}
C.~Akers, N.~Engelhardt, D.~Harlow, G.~Penington and S.~Vardhan, \emph{{The
  black hole interior from non-isometric codes and complexity}},
  \href{https://arxiv.org/abs/2207.06536}{{\ttfamily 2207.06536}}.

\bibitem{Yang:2023zic}
L.~Yang and N.~Engelhardt, \emph{{The Complexity of Learning (Pseudo)random
  Dynamics of Black Holes and Other Chaotic Systems}},
  \href{https://arxiv.org/abs/2302.11013}{{\ttfamily 2302.11013}}.

\bibitem{Coleman:1988tj}
S.~R. Coleman, \emph{{Why There Is Nothing Rather Than Something: A Theory of
  the Cosmological Constant}},
  \href{https://doi.org/10.1016/0550-3213(88)90097-1}{\emph{Nucl. Phys. B}
  {\bfseries 310} (1988) 643}.

\bibitem{Giddings:1988wv}
S.~B. Giddings and A.~Strominger, \emph{{Baby Universes, Third Quantization and
  the Cosmological Constant}},
  \href{https://doi.org/10.1016/0550-3213(89)90353-2}{\emph{Nucl. Phys. B}
  {\bfseries 321} (1989) 481}.

\bibitem{Preskill:1988na}
J.~Preskill, \emph{{Wormholes in Space-time and the Constants of Nature}},
  \href{https://doi.org/10.1016/0550-3213(89)90592-0}{\emph{Nucl. Phys. B}
  {\bfseries 323} (1989) 141}.

\bibitem{Klebanov:1988eh}
I.~R. Klebanov, L.~Susskind and T.~Banks, \emph{{Wormholes and the Cosmological
  Constant}}, \href{https://doi.org/10.1016/0550-3213(89)90538-5}{\emph{Nucl.
  Phys. B} {\bfseries 317} (1989) 665}.

\bibitem{Lyons:1991im}
A.~Lyons and S.~W. Hawking, \emph{{Wormholes in string theory}},
  \href{https://doi.org/10.1103/PhysRevD.44.3802}{\emph{Phys. Rev. D}
  {\bfseries 44} (1991) 3802}.

\bibitem{Marolf:2020xie}
D.~Marolf and H.~Maxfield, \emph{{Transcending the ensemble: baby universes,
  spacetime wormholes, and the order and disorder of black hole information}},
  \href{https://doi.org/10.1007/JHEP08(2020)044}{\emph{JHEP} {\bfseries 08}
  (2020) 044} [\href{https://arxiv.org/abs/2002.08950}{{\ttfamily
  2002.08950}}].

\bibitem{McNamara:2020uza}
J.~McNamara and C.~Vafa, \emph{{Baby Universes, Holography, and the
  Swampland}},  \href{https://arxiv.org/abs/2004.06738}{{\ttfamily
  2004.06738}}.

\bibitem{Blommaert:2021fob}
A.~Blommaert, L.~V. Iliesiu and J.~Kruthoff, \emph{{Gravity factorized}},
  \href{https://doi.org/10.1007/JHEP09(2022)080}{\emph{JHEP} {\bfseries 09}
  (2022) 080} [\href{https://arxiv.org/abs/2111.07863}{{\ttfamily
  2111.07863}}].

\bibitem{Blommaert:2022ucs}
A.~Blommaert, L.~V. Iliesiu and J.~Kruthoff, \emph{{Alpha states demystified
  \textemdash{} towards microscopic models of AdS$_{2}$ holography}},
  \href{https://doi.org/10.1007/JHEP08(2022)071}{\emph{JHEP} {\bfseries 08}
  (2022) 071} [\href{https://arxiv.org/abs/2203.07384}{{\ttfamily
  2203.07384}}].

\bibitem{Horowitz:1996nw}
G.~T. Horowitz and J.~Polchinski, \emph{{A Correspondence principle for black
  holes and strings}},
  \href{https://doi.org/10.1103/PhysRevD.55.6189}{\emph{Phys. Rev. D}
  {\bfseries 55} (1997) 6189}
  [\href{https://arxiv.org/abs/hep-th/9612146}{{\ttfamily hep-th/9612146}}].

\bibitem{Giddings:2005id}
S.~B. Giddings, D.~Marolf and J.~B. Hartle, \emph{{Observables in effective
  gravity}}, \href{https://doi.org/10.1103/PhysRevD.74.064018}{\emph{Phys. Rev.
  D} {\bfseries 74} (2006) 064018}
  [\href{https://arxiv.org/abs/hep-th/0512200}{{\ttfamily hep-th/0512200}}].

\bibitem{Donnelly:2015hta}
W.~Donnelly and S.~B. Giddings, \emph{{Diffeomorphism-invariant observables and
  their nonlocal algebra}},
  \href{https://doi.org/10.1103/PhysRevD.93.024030}{\emph{Phys. Rev. D}
  {\bfseries 93} (2016) 024030}
  [\href{https://arxiv.org/abs/1507.07921}{{\ttfamily 1507.07921}}].

\bibitem{Minwalla:1999px}
S.~Minwalla, M.~Van~Raamsdonk and N.~Seiberg, \emph{{Noncommutative
  perturbative dynamics}},
  \href{https://doi.org/10.1088/1126-6708/2000/02/020}{\emph{JHEP} {\bfseries
  02} (2000) 020} [\href{https://arxiv.org/abs/hep-th/9912072}{{\ttfamily
  hep-th/9912072}}].

\bibitem{Douglas:2001ba}
M.~R. Douglas and N.~A. Nekrasov, \emph{{Noncommutative field theory}},
  \href{https://doi.org/10.1103/RevModPhys.73.977}{\emph{Rev. Mod. Phys.}
  {\bfseries 73} (2001) 977}
  [\href{https://arxiv.org/abs/hep-th/0106048}{{\ttfamily hep-th/0106048}}].

\bibitem{Szabo:2001kg}
R.~J. Szabo, \emph{{Quantum field theory on noncommutative spaces}},
  \href{https://doi.org/10.1016/S0370-1573(03)00059-0}{\emph{Phys. Rept.}
  {\bfseries 378} (2003) 207}
  [\href{https://arxiv.org/abs/hep-th/0109162}{{\ttfamily hep-th/0109162}}].

\bibitem{Craig:2019zbn}
N.~Craig and S.~Koren, \emph{{IR Dynamics from UV Divergences: UV/IR Mixing,
  NCFT, and the Hierarchy Problem}},
  \href{https://doi.org/10.1007/JHEP03(2020)037}{\emph{JHEP} {\bfseries 03}
  (2020) 037} [\href{https://arxiv.org/abs/1909.01365}{{\ttfamily
  1909.01365}}].

\bibitem{Rajaraman:2000dw}
A.~Rajaraman and M.~Rozali, \emph{{Noncommutative gauge theory, divergences and
  closed strings}},
  \href{https://doi.org/10.1088/1126-6708/2000/04/033}{\emph{JHEP} {\bfseries
  04} (2000) 033} [\href{https://arxiv.org/abs/hep-th/0003227}{{\ttfamily
  hep-th/0003227}}].

\bibitem{Kiem:2000wt}
Y.~Kiem and S.~Lee, \emph{{UV / IR mixing in noncommutative field theory via
  open string loops}},
  \href{https://doi.org/10.1016/S0550-3213(00)00430-2}{\emph{Nucl. Phys. B}
  {\bfseries 586} (2000) 303}
  [\href{https://arxiv.org/abs/hep-th/0003145}{{\ttfamily hep-th/0003145}}].

\bibitem{Witten:1985cc}
E.~Witten, \emph{{Noncommutative Geometry and String Field Theory}},
  \href{https://doi.org/10.1016/0550-3213(86)90155-0}{\emph{Nucl. Phys. B}
  {\bfseries 268} (1986) 253}.

\bibitem{Seiberg:1999vs}
N.~Seiberg and E.~Witten, \emph{{String theory and noncommutative geometry}},
  \href{https://doi.org/10.1088/1126-6708/1999/09/032}{\emph{JHEP} {\bfseries
  09} (1999) 032} [\href{https://arxiv.org/abs/hep-th/9908142}{{\ttfamily
  hep-th/9908142}}].

\end{thebibliography}\endgroup

\end{document}